%% file: main.tex
\documentclass[12pt, english, letterpaper]{article}

\usepackage{jheppub}
\usepackage{lineno}
\usepackage[T1]{fontenc}
\usepackage[english]{babel}
\usepackage{amsmath}
\usepackage{amssymb}
\usepackage{amsfonts}
\usepackage{bbold}
\usepackage{mathtools}
\usepackage{graphicx}
\usepackage{enumerate}
\usepackage{csquotes}
\usepackage{xcolor}
\usepackage{svg}
\usepackage{caption}
\usepackage{subcaption}
\usepackage{makecell}

\DeclareMathOperator{\tr}{tr}
\DeclareMathOperator{\pf}{pf}
\DeclareMathOperator{\sgn}{sgn}
\DeclareMathOperator{\adj}{adj}

\DeclareMathOperator{\Ortho}{O}

\DeclareMathOperator{\action}{\mathcal{I}}

\newcommand{\bra}[1]{\langle #1 \rvert}
\newcommand{\ket}[1]{\lvert #1 \rangle}

\newcommand{\ketbra}[1]{\ket{#1}\bra{#1}}

\DeclareMathOperator{\majo}{\mu}

\DeclareMathOperator{\prob}{\pi}

\DeclareMathOperator{\diag}{diag}

\newcommand{\Partial}{\mathcal{D}}

\arxivnumber{2602.12339}

\title{Magic and Wormholes in the Sachdev-Ye-Kitaev Model}

\author[1]{Val\'erie Bettaque\note{Corresponding author.}}
\author{and Brian Swingle}
\affiliation{Martin A. Fisher School of Physics \\ Brandeis University \\ Waltham, MA 02453, USA}

\emailAdd{vbettaque@brandeis.edu}
\emailAdd{bswingle@brandeis.edu}

\abstract{
Any quantum state is fully specified by the expectation values of a complete set of Hermitian operators. For a system of Majorana fermions, such as the Sachdev-Ye-Kitaev (SYK) model, this set of observables can be taken to be all possible strings of Majorana fermion operators. The expectation values of these fermion strings in a thermal state depend erratically on the microscopic couplings that specify the SYK Hamiltonian, and we study their statistical properties directly in the thermodynamic limit using path integral techniques. When the underlying SYK Hamiltonian is chaotic, we find that these expectation values are well-modeled as real Gaussian random variables with zero mean and a variance that we compute. In contrast, for the integrable variant of SYK, we find that the expectation values are actually non-Gaussian. We then use these results to study measures of magic in the SYK thermal state, including the robustness of magic and the stabilizer R\'enyi entropy. We also show that our results can be quantitatively reproduced with a dual gravity calculation in the chaotic case at sufficiently low temperature. In this dual gravity model the variance of a given microscopic operator string is related to a wormhole geometry stabilized by a massive particle which is dual to the operator string. Our results thus provide a concrete and quantitative setting in which to study the relationship between randomness, wormholes and closed universes, as well as a holographic dual of quantum magic.
}

\begin{document}

\maketitle

\newpage

\input{sections/introduction.tex}

\input{sections/path_integral.tex}

\input{sections/solutions.tex}
\input{sections/applications.tex}

\input{sections/outlook.tex}

\bibliographystyle{JHEP}
\bibliography{refs.bib}

\appendix

\input{appendices/numerics.tex}
\input{appendices/ensembles.tex}

\end{document}

%% file: sections/introduction.tex
\section{Introduction}

Any quantum state is fully determined by the expectation values of a complete set of Hermitian operators. For example, the density matrix of a single qubit is fully determined by the expectation values of the three Pauli matrices. In a many-qubit system, the generalization requires the expectation values of exponentially many Hermitian operators, essentially a choice of identity or Pauli matrix for each qubit. This is a large amount of data, which goes far beyond what is normally accessed in experiments that probe few-body observables. Nevertheless, various statistical properties of these data are of considerable interest. For example, they are used to track the dynamics of information scrambling \cite{xu_scrambling_dynamics_out_2024}, they help quantify the amount of non-stabilizerness or magic in a state \cite{bravyi_universal_quantum_computation_2005,leone_stabilizer_renyi_entropy_2022,leone_stabilizer_entropies_are_2024}, they are related to classical shadows \cite{aaronson_shadow_tomography_quantum_2018, huang_predicting_many_properties_2020}, and they feature in recent models of wormholes and closed universes in quantum gravity \cite{stanford_more_quantum_noise_2020,antonini_cosmology_random_entanglement_2023, usatyuk_closed_universes_two_2024}. 

Here we study the analog of this data in a many-body system of fermions. We consider $N$ Majorana fermions in a thermal state of the Sachdev-Ye-Kitaev (SYK) model \cite{sachdev_gapless_spin_fluid_1993, kitaev_simple_model_quantum_2015, maldacena_remarks_sachdev_ye_2016, polchinski_spectrum_sachdev_ye_2016, kitaev_soft_mode_sachdev_2018} and study expectation values of a complete set of Majorana string operators. These \emph{thermal 1-point functions} completely specify the state and depend erratically on the couplings in the SYK Hamiltonian. Moreover, the ensemble of SYK Hamiltonians produces an ensemble of thermal 1-point functions for a given operator. We study the statistical properties of this ensemble and show that the thermal 1-point functions are well-modeled as Gaussian random variables when the system is chaotic. These random variables are zero on average and have a non-trivial variance which depends on the operator.

In addition to providing a natural ensemble of expectation values, we focus on the Sachdev-Ye-Kitaev model for our study because it is a widely-used toy model in many-body physics \cite{chowdhury_sachdev_ye_kitaev_2022}, it connects to simple models of holographic quantum gravity \cite{kitaev_simple_model_quantum_2015, maldacena_remarks_sachdev_ye_2016, polchinski_spectrum_sachdev_ye_2016, kitaev_soft_mode_sachdev_2018}, it has an integrable variant, and -- as we show -- the data of interest is tractable at large $N$ via path integral methods. Indeed, our main results are obtained by computing statistical moments of thermal 1-point functions using path integrals over multiple copies of the system.

Building on this characterization of the ensemble of expectation values, we also discuss the application of our results to two problems of current interest. First, we analyze the robustness of magic \cite{howard_application_resource_theory_2017} and stabilizer R\'enyi entropy  \cite{leone_stabilizer_renyi_entropy_2022}. We show that thermal states of SYK have a lower bound on their robustness of magic that implies they are very magical at sufficiently low temperature. We also show that the stabilizer R\'enyi entropy of the thermal state is consistent with high magic at low temperature but does not necessarily imply it. Magic in the SYK model has previously been studied in several works, including one that uses similar path integral methods \cite{bera_non_stabilizerness_sachdev_2025, jasser_stabilizer_entropy_entanglement_2025, zhang_stabilizer_renyi_entropy_2026, malvimat_multipartite_non_local_2026}.

Second, we show how our results can be interpreted holographically in terms of a dual quantum gravity system when the temperature is low enough. In this duality, the SYK model is the boundary and the quantum gravity system is the bulk. Concretely, we compare our SYK results at low temperature to a simple bulk model built from fermionic fields propagating on a \enquote{double trumpet} wormhole geometry sourced by matter \cite{stanford_more_quantum_noise_2020} in Jackiw-Teitelboim (JT) gravity \cite{teitelboim_gravitation_hamiltonian_structure_1983, jackiw_lower_dimensional_gravity_1985, almheiri_models_ads2_backreaction_2015}, and we show that the SYK results can be well fit with this gravitational model. The wormhole solutions are also related to models of closed universes \cite{maldacena_wormholes_ads_2004} and our results amount to a microscopic verification of an operator randomness hypothesis put forward in that literature \cite{antonini_cosmology_random_entanglement_2023, antonini_baby_universe_fine_2025, sasieta_baby_universes_thermal_2025}. One may also view these wormholes as providing a holographic dual for magic which is related to gravitational back-reaction \cite{cao_gravitational_backreaction_magical_2025}.

\subsection{Preliminaries}

The system of interest consists of $N$ Majorana fermions which are represented by Hermitian operators $\chi_i = \chi_i^\dagger$ that obey the algebra
\begin{equation}
    \{ \chi_i , \chi_j \} = \delta_{ij}.
\end{equation} 
Note that this is the \enquote{physics} normalization convention, which is more convenient for the path integral analysis that is central to our results. The dimension of the Hilbert space on which these Majorana modes form an irreducible representation is
\begin{equation}
    D= 2^{N/2}.
\end{equation}

A complete basis of Hermitian operators acting on states in said Hilbert space is provided by the \emph{Majorana strings},
\begin{equation}
    \majo(a) \coloneqq 2^{W/2} \, i^{\lfloor W/2 \rfloor} \, \chi_1^{a_1} \chi_2^{a_2} \cdots \chi_N^{a_N}, \quad a \equiv (a_1, a_2, \ldots, a_N) \in \{0, 1\}^N.
\end{equation}
Here, $W \equiv \sum_i a_i$ is the \emph{(operator) weight} of the string $\majo(a)$, and $\lfloor W/2 \rfloor$ is the nearest integer less than or equal to $W/2$. The prefactor ensures that $\majo$ is Hermitian and squares to the identity. There are $2^N$ such (unique) strings, including the identity operator. A special role is played by the Majorana string with $a_i=1$ for all $1 \leq i \leq N$, which is known as the fermion parity operator,
\begin{equation}
\label{eq:fermion_parity}
    (-1)^F = (2i)^{N/2} \, \chi_1 \chi_2 \cdots \chi_N.
\end{equation}
Physical processes can only create or destroy fermions in pairs, so any Hamiltonian must therefore commute with fermion parity.

We consider an ensemble of Hamiltonians known as the $q$-body SYK model. An instance of this ensemble of Hamiltonians is defined by an instance of the Gaussian random couplings $J_{i_1 \cdots i_q}$ via
\begin{equation}
\label{eq:syk_hamiltonian}
    H_\mathrm{SYK} = (i)^{q/2} \sum_{1 \leq i_1 < \cdots < i_q \leq N} J_{i_1 \ldots i_q} \, \chi_{i_1} \cdots \chi_{i_q}.
\end{equation}
Throughout we use an overline to denote an average over the ensemble of Hamiltonians, which is the same as an average over the $J$ couplings. In this notation, these couplings have zero mean and variance
\begin{equation}
    \overline{J^2_{i_1 \ldots i_q}} = \frac{J^2 (q-1)!}{N^{q-1}}.
\end{equation}
The $J$ couplings for different sets of indices are uncorrelated. Note that $[H_\mathrm{SYK},(-1)^F]=0$, provided $q$ is even. 

The case of $q=2$ is special because it reduces to an integrable model of non-interacting fermions. The cases with $q\geq 4$ are all qualitatively similar to each other in that they are quantum chaotic models which feature a holographic duality with a simple model of quantum gravity at low temperature. We focus primarily on $q\geq 4$ in this paper; we briefly discuss the $q=2$ case whenever relevant, but postpone a full discussion to future work. 

Throughout we consider a thermal (Gibbs) state corresponding to $H_\mathrm{SYK}$ at inverse temperature $\beta$:
\begin{equation}
    \rho = \frac{e^{-\beta H_\mathrm{SYK}}}{Z(\beta)},
\end{equation}
where $Z(\beta)$ is the partition function,
\begin{equation}
    Z(\beta) = \tr( e^{-\beta H_\mathrm{SYK}}).
\end{equation}
The central object of interest is the thermal expectation value or thermal 1-point function,
\begin{equation}
    \xi(a) \coloneqq \tr(\rho \majo(a)).
\end{equation}
Because $\majo$ is Hermitian, these expectation values are real. They depend erratically on the couplings $J$ defining the SYK instance.  Our goal is to compute disorder-averaged statistical moments of $\xi(a)$, 
\begin{equation}
    \overline{ \xi(a)^R},
\end{equation}
where $R$ is usually taken to be even. These quantities can be obtained at small $N$ from a full diagonalization of the Hamiltonian. Our contribution is to study them at large $N$ using a path integral formulation.

\subsection{Summary}
\label{sec:summary}

With these preliminaries in place and before diving into the technical details, we discuss the main results in more detail and place them in context with prior work.

\subsubsection*{Gaussianity}

The core technical result of our study is that, provided $q \geq 4$, the $\xi(a)$s (corresponding to string with even parity\footnote{Odd-parity operator expectation values vanish exactly due to parity superselection rules.})are well-modeled as Gaussian random variables with mean zero and variance parameterized by
\begin{equation}
    \overline{ \xi(a)^2 } \equiv e^{- 2 N \phi(w)},
\end{equation}
where $w \coloneqq W/N$. As this notation implies, the variance of $\xi(a)$ (and indeed any moment) can only depend on the relative weight $w$ of $\majo(a)$. This is because the ensemble of SYK Hamiltonians has a large $O(N)$ statistical symmetry, which means that the statistical moments of $\xi(a)$ cannot depend on which fermions make up $\majo(a)$. We show how to compute the function $\phi(w)$ from a path integral representation of the variance in Equation \eqref{eq:phi_w}, and we report results for $\phi$ for a representative collection of $\beta$s in Section \ref{sec:solutions}.

Looking at higher moments, we find that the first four moments of $\xi(a)$ are
\begin{equation}
    \overline{\xi(a)} = 0, \quad \overline{\xi(a)^2}=e^{- 2 N \phi(w)}, \quad \overline{\xi(a)^3} = 0, \quad \overline{\xi(a)^4} \approx 3 \left[\overline{\xi(a)^2}\right]^2 = 3 \, e^{- 4 N \phi(w)},
\end{equation}
where the last equality for the fourth moment is the expected result for a real Gaussian random variable. 

The $q=2$ case turns out to be special, and $\xi(a)$ is not Gaussian random in that case. We can still compute the important exponential parts of $\overline{\xi(a)^R}$ using our path integral approach, but computations that depend in detail on the Gaussianity have to be revisited. We will return to study the $q=2$ case in more detail in future work. 

The Gaussianity of the moments for $q = 4$ was previously observed in \cite{bera_non_stabilizerness_sachdev_2025} based on finite $N$ results. They also noted that the $q=2$ case was non-Gaussian. The case of fermions in a general Gaussian state (not to be confused with the statistics of expectation values) was studied in \cite{collura_non_stabilizerness_fermionic_2026}, the results of which can also be used to argue that $q=2$ thermal 1-point functions are not Gaussian distributed. Our results establish the approximate Gaussianity of the $q\geq 4$ thermal 1-point functions in the thermodynamic limit of large $N$. We also comment on how the non-Gaussianity of $q=2$ emerges from our path integral analysis, although we defer a full discussion to future work.

\subsubsection*{Fermionic Magic}

With the Gaussian characterization of the moments in hand, we consider several applications. First, we study a fermionic version of the robustness of magic \cite{howard_application_resource_theory_2017, pashayan_estimating_outcome_probabilities_2015, heinrich_robustness_magic_symmetries_2019}, $\mathcal{R}$, and stabilizer Renyi entropy $M_\alpha$ \cite{leone_stabilizer_renyi_entropy_2022, leone_stabilizer_entropies_are_2024}. We emphasize that we are considering magic in the sense of non-stabilizerness, for suitably defined fermionic stabilizer states \cite{bravyi_majorana_fermion_codes_2010} and fermionic Clifford operations \cite{mclauchlan_fermion_parity_based_2022, mudassar_encoding_majorana_codes_2024, bettaque_structure_majorana_clifford_2025}. We are not considering deviations from fermionic Gaussian states or matchgate tensors; indeed, the $q=2$ thermal state is a fermionic Gaussian state but has magic in the non-stabilizer sense.

We don't directly compute the robustness, which involves a non-trivial optimization problem. Instead, we invoke a lower bound from the 1-norm of the expectation value data, 
\begin{equation}
    \mathcal{R}(\rho) \geq \frac{F_1(\rho)}{D},
\end{equation}
where
\begin{equation}
    F_1(\rho) = \sum_a |\xi(a)|.
\end{equation}
Using the standard result for the averaged absolute value of a Gaussian random variable with zero mean:
\begin{equation}
    \overline{ |\xi(a)|} = \sqrt{\frac{2}{\pi}} \, \sqrt{\overline{\xi(a)^2}} = \sqrt{\frac{2}{\pi}} \, e^{- N \phi(w)},
\end{equation}
we get a lower bound on $\mathcal{R}$ of
\begin{equation}
    \overline{\mathcal{R}(\rho)} \geq \sqrt{\frac{2}{\pi}} \, \sum_w \binom{N}{Nw} \, e^{- N \left[ \phi(w) + \ln(2)/2\right]}.
\end{equation}

As we review in Section \ref{sec:robustness}, any state $\rho$ in the fermion stabilizer polytope has $\mathcal{R}(\rho) = 1$, so $\mathcal{R}(\rho) > 1$ is sufficient to show that the thermal state is not in the stabilizer polytope. We also discuss the evaluation of this lower bound based on our path integral results. Once the thermal purity is low enough, the lower bound indeed exceeds unity by an exponentially large amount.

We also study the stabilizer R\'enyi entropy (SRE) in Section \ref{sec:sre}. For the mixed states considered here, there are a few possible definitions of the SRE. One is derived using a convex roof construction, which involves choosing a pure state decomposition of the mixed state, computing the average pure state SRE over the decomposition, and then minimizing over all possible such decompositions \cite{leone_stabilizer_entropies_are_2024}. Alternatively, one may directly generalize the SRE formula to mixed states without invoking any pure state decomposition. Within our framework, only the second option is accessible.

Our choice of mixed state SRE is defined in terms of a normalized probability distribution
\begin{equation}
    \prob_\rho(a) = \frac{ |\xi(a)|^2}{D \tr(\rho^2)}, 
\end{equation}
via the associated classical R\'enyi entropy
\begin{equation}
    M_{\alpha} (\rho) \coloneqq \frac{1}{1-\alpha} \log \left( \sum_a \prob_\rho^{\alpha}(a) \right) - \log(D \tr(\rho^2)).
\end{equation}
Applied to the SYK thermal state, we find that the SRE is dominated by the identity operator when $q\geq4$ and $\alpha=2$. This means the sum over $a$ can be replaced with just the identity term ($a_i=0$). This situation is consistent with a high degree of magic but, as we review in Appendix \ref{app:ensembles}, it can be mimicked with a random mixture of stabilizer states when the thermal entropy is large enough. This is not to say that the SYK mixed state isn't magical, only that the SRE cannot guarantee it in some regimes. In the $q=2$ case, we also find a (subleading) dependence on terms beyond just the identity, and can reach a regime where said non-identity contributions cannot be reproduced by a random mixture of stabilizer states. This (albeit only marginal) difference in behavior can be captured by a filtered version of the SRE, which has appeared before in the literature to address similar issues in other systems that admit a large-$N$ limit \cite{turkeshi_pauli_spectrum_nonstabilizerness_2025, haug_efficient_witnessing_testing_2026, collura_non_stabilizerness_fermionic_2026}.

Our results come in the context of a flurry of activity studying magic in the SYK model, with prior work focusing primarily on finite-$N$ exact numerical approaches \cite{bera_non_stabilizerness_sachdev_2025, jasser_stabilizer_entropy_entanglement_2025, malvimat_multipartite_non_local_2026}. In particular, a series of works defined and studied the stabilizer R\'enyi entropy which we used in our study. One feature of our approach is that it applies directly in the thermodynamic limit rather than relying on finite $N$. Shortly before this paper was completed, another work appeared that discussed the stabilizer R\'enyi entropy in SYK using a similar large $N$ path integral approach \cite{zhang_stabilizer_renyi_entropy_2026}. Our results are consistent with theirs where there is overlap. The ability to work directly at large $N$ is also desirable given the intractability of measuring magic in totally generic situations \cite{garcia_hardness_measuring_magic_2025}.

This activity is also part of a broader recent interest in magic in quantum many-body systems that includes early studies \cite{white_conformal_field_theories_2021, sarkar_characterization_operational_quantum_2020, liu_many_body_quantum_2022}, the notion of SRE \cite{leone_stabilizer_renyi_entropy_2022, oliviero_measuring_magic_quantum_2022}, and many other works, of which a sampling is \cite{oliviero_magic_state_resource_2022, niroula_phase_transition_magic_2024, zhang_quantum_magic_dynamics_2026, turkeshi_magic_spreading_random_2025, dowling_bridging_entanglement_magic_2025, russomanno_nonstabilizerness_unitary_monitored_2025}.

\subsubsection*{Holographic Wormholes}

The second application is to holographic wormhole spacetimes which we study in Section \ref{sec:wormholes}. To explain this, it is useful to begin with the now well-established fact that the $q=4$ SYK model has a precise correspondence with a gravitational system when the temperature satisfies $N \gg \beta J \gg 1$ \cite{sachdev_gapless_spin_fluid_1993, kitaev_simple_model_quantum_2015, maldacena_remarks_sachdev_ye_2016, polchinski_spectrum_sachdev_ye_2016, kitaev_soft_mode_sachdev_2018}. In this regime, the thermodynamics of the SYK model is captured by the so-called Schwarzian effective theory, and precisely the same effective theory arises from a model known as Jackiw-Teitelboim (JT) gravity that lives in two spacetime dimensions \cite{teitelboim_gravitation_hamiltonian_structure_1983, jackiw_lower_dimensional_gravity_1985, almheiri_models_ads2_backreaction_2015}. This correspondence is holographic in the sense that the physics of SYK, which has just a time coordinate, is dual to the physics of gravity in a higher-dimensional geometry. We refer to SYK as the boundary and the gravitational system as the bulk.  

\begin{figure}[h!tb]
    \centering
    \includegraphics[width=0.9\linewidth]{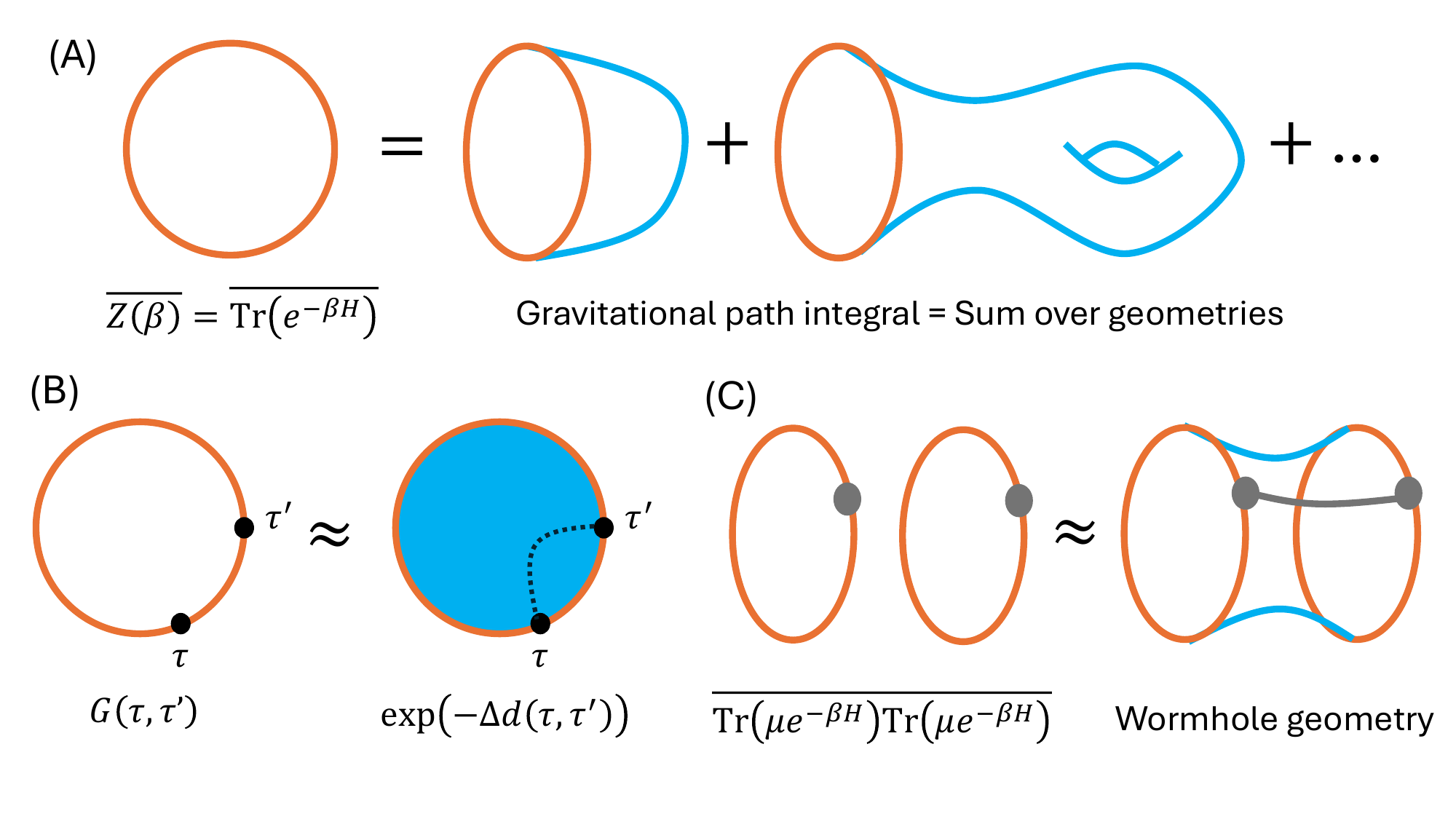}
    \caption{(A) The basic entry in the holographic dictionary states that the ensemble averaged SYK partition function is equal to a gravitational path integral that includes a sum over geometries and different topologies. The bulk theory entering this path integral reduces to JT gravity for simple observables but in general includes a rich and not well understood collection of bulk degrees of freedom. Below, we will model this bulk theory as JT gravity coupled to $N$ free fermions. (B) Boundary correlations of the fermion operators are related to correlations of the bulk fermion fields propagating in the gravitational geometry. For ease of analysis, we will treat these bulk correlations within a geodesic approximation. (C) For some observables, the sum over topologies is important in that there is no analog of the disk contribution and the first contribution instead comes from a geometry with nontrivial topology.. In our case, a wormhole geometry serves to capture the variance of the erratic thermal 1-point functions.}
    \label{fig:syk_jt}
\end{figure}

These elements of the holographic dictionary are sketched in Figure \ref{fig:syk_jt}(A), where the first term in the gravitational sum is the ``disk'' contribution which captures the above physics. Concretely, the thermal time circle in the SYK model is interpreted as the boundary of the bulk two-dimensional spacetime, and -- for the thermodynamic free energy -- the bulk spacetime is simply a hyperbolic disk which is cutoff at large radius. In general, the bulk dual of SYK is not fully understood and does not contain just JT gravity \cite{rosenhaus_introduction_syk_model_2019, sarosi_ads_2_holography_syk_2018}. In this work, we will model the bulk in terms of JT gravity coupled to $N$ bulk fermionic fields. Figure \ref{fig:syk_jt}(B) sketches how boundary correlations can be related to these bulk fields propagating in the disk geometry. In Figure \ref{fig:syk_jt}(C) we sketch how more elaborate bulk spacetimes can be obtained by considering more elaborate SYK observables. Thermal 1-point functions are zero on average, but have a non-trivial variance in the SYK ensemble. In the gravitational language, this variance corresponds to a bulk spacetime with two boundaries which are connected in a \enquote{wormhole} geometry. There are two boundaries because the variance $\overline{\xi^2(a)}$ has two copies of the thermal 1-point function $\xi(a)$. The precise form of this wormhole geometry depends on the details of the theory.

In this paper we use a simple model in which the wormhole is mostly empty except for a single very massive particle which stabilizes it. The massive particle is the bulk dual of the $\majo(a)$ operator, and it is \enquote{heavy} when the weight $W$ is large, scaling with $N$. We show that this simple bulk theory can capture a number of qualitative features of our microscopic results, and comment on how it might be further improved. Our result shows that there are concrete statistical observables in SYK that are parsimoniously described using a higher dimensional wormhole geometry, similar to prior results for the entropy \cite{almheiri_replica_wormholes_entropy_2020, penington_replica_wormholes_black_2022} and the spectral form factor \cite{saad_semiclassical_ramp_syk_2019}. This connection opens the door to concrete quantitative explorations of a number of interesting topics in quantum gravity, including the physics of closed universes \cite{antonini_cosmology_random_entanglement_2023, usatyuk_closed_universes_two_2024, sasieta_baby_universes_thermal_2025}.

The gravity solutions we consider were previously obtained in \cite{stanford_more_quantum_noise_2020, usatyuk_closed_universes_two_2024} but those works did not compare the bulk results with SYK results on the boundary. By considering a simple extension of the bulk theory to include $N$ free fermions, we show that gravity calculations based on these wormhole geometries provide a quantitatively accurate model of the SYK results at large $\beta$. We also note several prior studies of magic and related quantities in holography \cite{cao_gravitational_backreaction_magical_2025, basu_stabilizer_complexity_hawking_2025}. Our results are particularly synergistic with those of \cite{cao_gravitational_backreaction_magical_2025} since the gravitational back-reaction of the heavy particle is what stabilizes the wormhole geometry, and the action of the heavy particle encodes the weight dependence of the variance. Thus we conclude along with \cite{cao_gravitational_backreaction_magical_2025} that gravitational back-reaction is crucially related to magic, and our results provide quantitative and non-perturbative bulk duals for a variety of magic measures based on heavy operator data.

%% file: sections/path_integral.tex
\section{Path Integral for Statistical Moments}
\label{sec:path_integral}

We begin our technical analysis by formulating a path integral approach for the statistical moments $\overline{\xi(a)^R}$, which enables us to work directly at large $N$. We also review how the path integral may be evaluated using saddle point methods and derive the Schwinger-Dyson equations that such saddle points must obey. In Section \ref{sec:solutions} we then discuss explicit solutions of the Schwinger-Dyson equations and thereby obtain the behavior of the moments.

\subsection{Derivation}
\label{sec:path_integral_derivation}

Our goal is to evaluate the $R$th moment of $\xi(a) = \tr(\majo(a) \, e^{-\beta H_\mathrm{SYK}}) / Z$. This is difficult a priori since the numerator and denominator both contain factors of $e^{-\beta H}$ with the same instance of the couplings. However, the partition function $Z$ is self-averaging (i.e.\ $\overline{Z^R} \approx \overline{Z}^R$) when $\beta J \ll N$, so one can replace $Z$ in the denominator with its average as separately evaluated using a single SYK replica (the standard thermodynamic calculation).\footnote{For example, the covariance between $Z$ and $\tr(\majo(a) e^{-\beta H})$ vanishes at large $N$. This follows from the emergent $O(N)$ symmetry which leaves $Z$invariant but not $\majo(a)$.}

Focusing therefore on the numerator, we may express it as a single trace over $R$ replicas of the state and rearrange the modes such that they are sorted by mode index:
\begin{align}
\begin{split}
\label{eq:syk_stat_moment_trace}
    \tr\left[\majo(a) \, e^{-\beta H_\mathrm{SYK}}\right]^R &= \tr\left[\majo(a)^{\otimes_f R} \exp(-\beta H_\mathrm{SYK})^{\otimes_f R}\right] \\
    &= \tr\left[\majo^{(1)}(a) \majo^{(2)}(a) \cdots \majo^{(R)}(a) \exp\left(-\beta \sum_{r = 1}^R H_\mathrm{SYK}^{(r)}\right)\right] \\
    &\equiv \tr\left[(-1)^{a_1 F_{R,1}} \cdots (-1)^{a_N F_{R,N}} \exp\left(-\beta \sum_{r = 1}^R H_\mathrm{SYK}^{(r)}\right)\right].
\end{split}
\end{align}
Here we introduced the multi-replica fermion parity operators
\begin{equation}
\label{eq:replica_parity_operator}
    (-1)^{F_{R,i}} \coloneqq (2i)^{R/2} \, \chi^{(1)}_i \chi^{(2)}_i \cdots \chi^{(R)}_i,
\end{equation}
of which $W$ many are present in the expression. We also suppress potential leading signs resulting from this rearrangement, as they can be absorbed into the path integral measure.

After expressing \eqref{eq:syk_stat_moment_trace} as a path integral and disorder-averaging as usual \cite{maldacena_remarks_sachdev_ye_2016, kitaev_soft_mode_sachdev_2018}, the introduction of the Green's function
\begin{equation}
\label{eq:G_definition}
    G_{rs}(\tau, \tau') \coloneqq \frac{1}{N} \sum_{i = 1}^N \chi_i^{(r)}(\tau) \, \chi_i^{(s)}(\tau'), \quad G_{rs}(\tau, \tau') = - G_{sr}(\tau', \tau),
\end{equation}
and the Lagrange multiplier $\Sigma_{rs}(\tau, \tau')$ enforcing this equality (and which turns out to correspond to the self-energy) causes the Majorana integrals to decouple with regard to the modes, and to couple with regard to the replicas:
\begin{align}
\begin{split}
\label{eq:syk_path_integral_decoupled}
    &\overline{\tr(\majo(a) \, e^{-\beta H_\mathrm{SYK}})^R} \\
    \propto{}& \prod_{i=1}^N \int  \left[\Partial\chi_i\right] (-1)^{a_i F_{R,i}} \exp\left(-\frac{1}{2} \sum_{r,s=1}^{R} \int_0^{\beta} d\tau d\tau' \chi^{(r)}_i(\tau) \left[\delta_{rs} \Partial(\tau, \tau') - \Sigma_{rs}(\tau, \tau') \right] \chi^{(s)}_i(\tau') \right).
\end{split}
\end{align}
Each insertion of $(-1)^{F_{R,i}}$ therefore determines the parity of all modes involved in the corresponding $W$ Majorana path integrals, which causes them to flip the boundary conditions of the thermal circle from antiperiodic to periodic. The other $N - W$ integrals are left untouched by this, which overall leads to
\begin{align}
\begin{split}
\label{eq:syk_path_integral_integrated_out}
    &\overline{\tr(\majo(a) \, e^{-\beta H_\mathrm{SYK}})^R} \\
    ={}& \int \left[\Partial\Sigma\right] \left[\Partial G\right] \pf[I_R \otimes \Partial^- - \Sigma]^{N-W} \pf[I_R \otimes \Partial^+ -\Sigma]^{W} \\
    &\hspace{2cm} \exp \left(- \frac{N}{2} \sum_{r,s=1}^{R} \int_0^{\beta} d\tau d\tau' \left(\Sigma_{rs}(\tau, \tau') \, G_{rs}(\tau, \tau') - \frac{J^2}{q} \, G_{rs}(\tau, \tau')^q \right) \right) \\
    \equiv{}& \int \left[\Partial\Sigma\right] \left[\Partial G\right] \exp\left( - N \action^{(R, w)}_\mathrm{SYK}[G, \Sigma] \right).
\end{split}
\end{align}
The corresponding effective action $\action^{(R, w)}_\mathrm{SYK}[G, \Sigma]$ now only depends on the relative weight $w \coloneqq W/N \in [0, 1]$ of $\majo(a)$ and is given by
\begin{align}
\begin{split}
\label{eq:syk_w_action}
    \action^{(R, w)}_\mathrm{SYK}[G, \Sigma] ={}& - (1-w) \ln \pf[I_R \otimes \Partial^- - \Sigma] - w \ln \pf[I_R \otimes \Partial^+ - \Sigma] \\
    & + \frac{1}{2} \sum_{r,s=1}^{R} \int_0^{\beta} d\tau d\tau' \left(\Sigma_{rs}(\tau, \tau') \, G_{rs}(\tau, \tau') - \frac{J^2}{q} \, G_{rs}(\tau, \tau')^q \right).
\end{split}
\end{align}
The $\mp$ superscript of $\Partial^\mp(\tau, \tau') \sim \delta(\tau - \tau') \, \partial_{\tau'}$ indicates if the associated expression is to be evaluated with periodic or antiperiodic boundary conditions.

By varying $\action^{(R, w)}_\mathrm{SYK}$ with regard to $G$ and $\Sigma$, the associated Schwinger-Dyson equations of the large-$N$ saddle can then be found to be
\begin{align}
\begin{split}
\label{eq:syk_w_schwinger_dyson}
    G_{rs}(\tau, \tau') &= (1-w) \, [I_R \otimes \Partial^- - \Sigma]^{-1}_{rs}(\tau, \tau') + w \, [I_R \otimes \Partial^+ - \Sigma]^{-1}_{rs}(\tau, \tau'), \\
    \Sigma_{rs}(\tau, \tau') &= J^2 \, G_{rs}(\tau, \tau')^{q-1},
\end{split}
\end{align}
which in the case of $w = 0$ (i.e.\ non-heavy operators) reduce to those of ordinary multi-replica SYK, as expected. Overall, the saddle point approximation for the averaged statistical moment $\overline{\xi(a)^R} \equiv \overline{\xi(w)^R}$ is therefore
\begin{equation}
\label{eq:averaged_moment_saddle}
    \overline{\xi(w)^R} \approx \exp\left(-N \left[\action^{(R,w)}_\mathrm{SYK} - R \action^{(1,0)}_\mathrm{SYK} \right]\right),
\end{equation}
where $\action^{(R,w)}_\mathrm{SYK}$ and $\action^{(1,0)}_\mathrm{SYK} = - \ln(Z)$ are evaluated at their respective minimizing saddles. In the case of $R = 2$, this means that $\phi(w)$ as introduced in Section \ref{sec:summary} will turn out to be equivalent to
\begin{equation}
\label{eq:phi_w}
    \phi(w) \equiv \frac{1}{2} \action^{(2,w)}_\mathrm{SYK} - \action^{(1,0)}_\mathrm{SYK}.
\end{equation}

\subsection{Discretization}

When computing just the free energy with a single replica, the saddle point value of $G(\tau_1,\tau_2)$ is time-translation invariant, depending only on $\tau_1 - \tau_2$. This enormously simplifies the solution of the Schwinger-Dyson equations. In contrast, the insertion of $\majo(a)$ at $\tau = 0$ means that the saddle point value of $G_{rs}(\tau_1,\tau_2)$ which computes $\overline{\xi(a)^R}$ won't generically be time-translation invariant. To solve \eqref{eq:syk_w_schwinger_dyson}, one therefore has to resort to discretizing the setup in terms of $R L \times R L$ matrices describing $G$ and $\Sigma$, with $L$ ideally being much larger than $\beta J$. This is accomplished by dividing each of the imaginary time integrals in the action into $L$ smaller integrals and applying the trapezoid rule to those. The resulting discretized action then is
\begin{align}
\begin{split}
\label{eq:syk_w_action_discrete}
    \action^{(R, w)}_\mathrm{SYK}[G, \Sigma] \approx{}& - \frac{1-w}{2} \ln \det\left(I_R \otimes \Partial^- - (\Delta\tau)^2 \, \Sigma\right) - \frac{w}{2} \ln\det\left(I_R \otimes \Partial^+ - (\Delta\tau)^2 \, \Sigma\right) \\
    & + \frac{(\Delta\tau)^2}{2} \left( \tr \left( \Sigma \, G^T\right) - \frac{J^2}{q} \sum_{i,j=1}^{RL} (G_{ij})^q \right),
\end{split}
\end{align}
where $\Delta\tau = \beta / L$, and the derivatives $\Partial^-$ and $\Partial^+$ corresponding to the two possible boundary conditions are also now described by $L \times L$ matrices:
\begin{equation}
\label{eq:discrete_derivative}
    \Partial^\mp  = \begin{pmatrix} 
        1 & 0 & 0 & \cdots & 0 & \pm 1 \\
        -1 & 1 & 0 & \cdots & 0 & 0 \\
        0 & -1 & 1 & \cdots & 0 & 0 \\
        \vdots &\vdots &\vdots & \ddots & \vdots & \vdots \\
        0 & 0 & 0 & \cdots & 1 & 0 \\
        0 & 0 & 0 & \cdots & -1 & 1
    \end{pmatrix}.
\end{equation}
The associated discrete Schwinger-Dyson equations can then be found to be
\begin{align}
\begin{split}
\label{eq:syk_w_schwinger_dyson_discrete}
    G^T &= - (1-w) \, (I_R \otimes \Partial^- - (\Delta\tau)^2 \, \Sigma)^{-1} - w \, (I_R \otimes \Partial^+ - (\Delta\tau)^2 \, \Sigma)^{-1}, \\
    \Sigma_{ij} &= J^2 \, (G_{ij})^{q-1},
\end{split}
\end{align}

Note that while $G$ (and by extension $\Sigma$) were originally antisymmetric fields \eqref{eq:G_definition}, we cannot assume that $G^T = -G$ and $\Sigma^T = -\Sigma$ when deriving these equations due to their discretized versions only being antisymmetric up to non-zero diagonal entries. This is a consequence of the derivative discretization intentionally not being antisymmetric either, thus breaking Hermiticity to avoid fermion doubling as predicted by the Nielsen-Ninomiya theorem \cite{nielsen_absence_neutrinos_lattice_1981}. That is also the reason why the Pfaffians in the action had to be replaced by (square roots of) determinants. For more information see Appendix \ref{sec:discretization}.

\subsection{Corrections to the Saddle Point Evaluation}
\label{sec:saddle_point_corrections}

We also briefly review how these solutions are used to compute the quantities of interest and comment on a subtlety. To illustrate the procedure in the simplest setting, consider a one-dimensional integral of the form 
\begin{equation}
    Q = \int_{-\infty}^\infty dx \, e^{- N q(x)}.
\end{equation}
Suppose for simplicity that the action $q(x)$ has a minimum at $x_0$. Rewriting the integral in terms of $y=x-x_0$ gives
\begin{equation}
    Q = \int dy \, e^{- N \left( q(x_0) + \frac{1}{2} q''(x_0) y^2 + \frac{1}{6} q'''(x_0) y^3 + \cdots \right)}.
\end{equation}
At large $N$ this rewrite is useful because the size of the fluctuation $y$, as determined by the leading quadratic term in the action,  is suppressed by $N$ according to
\begin{equation}
    y \sim \frac{1}{\sqrt{N q''(x_0)}}.
\end{equation}

Plugging this typical value of $y$ into the $\ell$th order term in the exponent gives a typical value of those terms of order
\begin{equation}
    \frac{1}{\ell!} N^{1 - \ell/2} \frac{q^{(\ell)}(x_0)}{(q''(x_0))^{\ell/2}},
\end{equation}
which is $1/N$-suppressed for $\ell>2$. Intuitively, one may treat these higher order terms perturbatively and write
\begin{equation}
    Q = e^{- N q(x_0)} \sqrt{\frac{2\pi}{N q''(x_0)}} \left(1 + \mathbb{E}\left[ - N \frac{q'''(x_0) y^3}{6} \right] + \cdots \right)
\end{equation}
where $\mathbb{E}$ denotes an expectation with respect to the normalized Gaussian distribution obtained from the quadratic term in the exponent.

Despite the $\ell$th term in the series scaling as $N^{1-\ell/2}$, the series is not necessarily convergent. Instead, it is typically an asymptotic series or more generally a \emph{transseries} when we include other saddle points~\cite{aniceto_primer_resurgent_transseries_2019}. Nevertheless, one expects the leading term in the $1/N$-expansion to accurately compute the leading behavior of the $Q$ integral at large $N$,
\begin{equation}
    Q \approx e^{- N q(x_0)} \sqrt{\frac{2\pi}{N q''(x_0)}} .
\end{equation}

We now consider the case of interest, in which we have a path integral or functional integral over $G_{rs}(\tau,\tau')$ and $\Sigma_{rs}(\tau,\tau')$,
\begin{equation}
    \mathcal{Z} = \int \Partial G \, \Partial \Sigma \, e^{- N \action[G,\Sigma]}.
\end{equation}
Here $\action$ is a generic proxy for any of the one-or-several-replica actions considered above. The leading large $N$ result for the functional integral is structurally similar to the single-variable case just reviewed,
\begin{equation}
    \mathcal{Z} \approx \frac{e^{- N \action(G_0,\Sigma_0)}}{ \sqrt{\det{\frac{N \action''}{2\pi}}}}
\end{equation}
where $\action''$ is a schematic indicating the matrix of second derivatives of the action and one takes the determinant of this matrix. One may also generalize this expression to include the possibility of multiple saddles, $G_{0,(i)}$ and $\Sigma_{0,(i)}$,
\begin{equation}
    \mathcal{Z} \approx \sum_i \frac{e^{- N \action(G_{0,(i)},\Sigma_{0,(i)})}}{ \sqrt{\det{\frac{N \action''_{(i)}}{2\pi}}}} \left( 1 + \dots \right)
\end{equation}

For most of the applications in Section \ref{sec:applications}, the most important part of this formula is the $N$-dependent exponent, however, in some situations we also care about the prefactor. When assessing the Gaussianity of $\xi$, we must study ratios of moments, for example, the fourth moment must be precisely three times the second moment squared. For this to be true, the exponents must match on both sides, but the prefactors must also be consistent.

As we will see below, the factor of three will come from summing over distinct saddles with the same action. However, we must also address the functional determinant that appears. The way Gaussianity arises in the analysis below is that the relevant saddle for $R=4$ consists of two copies of the $R=2$ saddle, meaning the collective fields take a block diagonal form in the $R=4$ saddle. What does the functional determinant look like in this case?

Let $A$ and $B$ denote the replica indices involved in the two blocks of the $R=4$ saddle. The fluctuation variables, $\delta G$ and $\delta \Sigma$, can be divided into fluctuations with both indices in $A$, both in $B$, or one in $A$ and one in $B$. We denote the latter type collectively as $\delta G_{AB}$ and $\delta \Sigma_{AB}$. Because the $(G_{rs})^q$ part of the action is ultra-local in replica indices, it follows that there is no $\delta G_{AB}^2$ term in the quadratic part of the action so long as $q>2$,
\begin{equation}
    \left[(G_0)_{rs} + \delta G_{rs} \right]^q = (\delta G_{rs})^q \quad \text{for $r\in A$, $s\in B$}.
\end{equation}
The only place $\delta G_{AB}$ appears to quadratic order is in the $\delta G_{AB} \delta \Sigma_{AB}$ term. Hence, carrying out the $\delta G_{AB}$ integral at quadratic order produces a delta function for $\delta \Sigma_{AB}$ that sets it equal to zero. By carefully defining the integration measure and the delta function, one can thus show that the contribution of these between-block fluctuations to the functional determinant is unity. Hence, the $R=4$ functional determinant in this block-diagonal setting is the square of the $R=2$ functional determinant. All that remains is the aforementioned numerical factor of three which counts distinct saddles.

As noted, this discussion is specific to the case $q>2$. When $q=2$, there is a term in the action containing $\delta G_{AB}^2$ and the situation is more complex. This is one of several reasons why the thermal 1-point functions turn out to be non-Gaussian for $q=2$.

%% file: sections/solutions.tex
\section{Solutions to the Schwinger-Dyson Equations}
\label{sec:solutions}

We next enumerate and discuss the solutions to the Schwinger-Dyson saddle point equations obtained in Section \ref{sec:path_integral} for different values of $R$. 

A little reflection suggests that the simplest possible case that can be considered ($R = 1$) should be trivial as it corresponds to the (signed) coefficients $\xi(a)$ arising in the expansion of $\rho$ in terms of the complete basis of Majorana strings $\majo(a)$. The coefficients associated to non-identity strings are expected to follow some to be determined random distribution with mean zero, and should therefore vanish when averaged over. That the path integral therefore evaluates to zero for $w > 0$ is shown both analytically and numerically in Section \ref{sec:single_replica_case}. Similarly, all averaged odd moments of these coefficients are also expected to be zero.

Therefore, the primary focus of this section is finding solutions of the Schwinger-Dyson equation corresponding to even moments $R \geq 2$, which necessarily have to exhibit cross-replica correlations as to not reduce to the vanishing single-replica case. Unfortunately, the theory does not tell us which form these off-diagonal replica entries will take. In fact, there can be many possible options, each contributing in some way to the full path integral result. Here we are only interested in the dominating large-$N$ saddle though, which means we have to make an educated guess based on the limited information available to us. This guess takes the form of an initial value we have to choose when iteratively solving \eqref{eq:syk_w_schwinger_dyson_discrete}, and which we assume to be independent of $\beta$ and $w > 0$. This in itself is also an assumption, but one that the numerical results in Section \ref{sec:applications} largely support.

\subsection{The Need for Cross-Replica Correlations}

In this section, we incrementally build towards a reasonable replica ansatz for arbitrary (even) $R$ by first considering the reason for the $R = 1$ solution vanishing, and then using the gained intuition to inform an initial value for $R = 2$ that is later shown to give the correct result for the second R\'enyi entropy (see Section \ref{sec:purity}). 

\subsubsection*{The Vanishing of the Single-Replica Case}
\label{sec:single_replica_case}

To understand why the path integral vanishes in the case of $R = 1$ and $w > 0$, it suffices to consider the determinant associated to $\Partial^+$ (i.e.\ periodic boundary conditions) in the discretized action \eqref{eq:syk_w_action_discrete} with $\Delta\tau = \beta / L$ arbitrarily small for fixed $\beta$ (and $J$). Using Jacobi's formula for non-invertible matrices ($\Partial^+$ has a zero eigenvalue), one can expand said determinant $\det(\Partial^+ -(\Delta\tau)^2 \, \Sigma)$ to first order in $(\Delta\tau)^2$:
\begin{align}
\begin{split}
\label{eq:determinant_expansion}
    \det(\Partial^+ - (\Delta\tau)^2 \, \Sigma) &= \det(\Partial^+) -(\Delta\tau)^2 \tr(\adj(\Partial^+) \, \Sigma) + \mathcal{O}((\Delta\tau)^4) \\
    &\equiv - \left( \beta J / L \right)^2 \tr\left(\adj(\Partial^+) \, \widetilde{\Sigma}\right) + \mathcal{O}\left(\left(\beta J / L \right)^4\right),
\end{split}
\end{align}
where we identified $\det(\Partial^+) = 0$ and $\widetilde{\Sigma} \coloneqq \Sigma / J^2$. $\adj(\Partial^+)$ refers to the \emph{adjugate matrix} of $\Partial^+$, which can be shown to be the $L \times L$ matrix $\mathbb{1}_L$ whose entries are all equal to 1. It is straightforward to verify that the trace of $\mathbb{1}_L$ and $\Sigma$ (or any other matrix) is equal to the sum over all entries of $\Sigma$. In the continuum case, this sum therefore corresponds to the zero mode of $\Sigma$ (up to a sign), meaning:
\begin{equation}
        \det(\Partial^+ - \Sigma) \propto \int_0^\beta d\tau d\tau' \, \Sigma(\tau, \tau') \approx (\Delta\tau)^2\tr\left(\mathbb{1}_L \, \Sigma\right).
\end{equation}
But since $\Sigma(\tau, \tau')$ is necessarily antisymmetric on-shell \eqref{eq:G_definition}, the continuous expression for both the determinant and the path integral as a whole should therefore vanish exactly due to both $\Partial^+$ and $\Sigma$ having a shared eigenvalue of zero.

\begin{figure}[h!tb]
    \centering
    \begin{subfigure}[b]{0.49\textwidth}
         \centering
         \includegraphics[width=\textwidth]{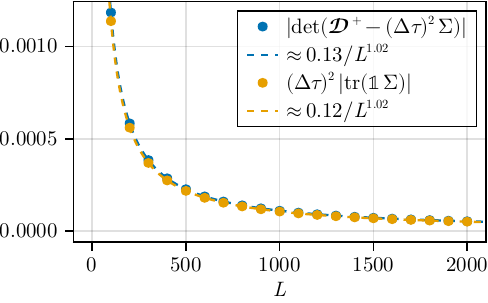}
         \caption{$\beta = 1$}
     \end{subfigure}
     \hfill
     \begin{subfigure}[b]{0.49\textwidth}
         \centering
         \includegraphics[width=\textwidth]{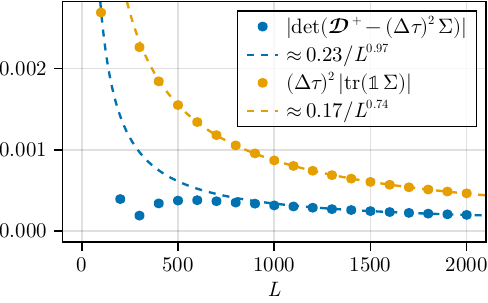}
         \caption{$\beta = 5$}
     \end{subfigure}
     \par\bigskip
     \begin{subfigure}[b]{0.49\textwidth}
         \centering
         \includegraphics[width=\textwidth]{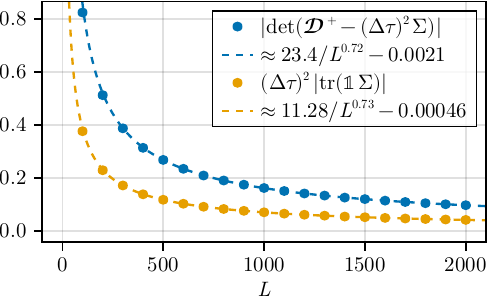}
         \caption{$\beta = 10$}
     \end{subfigure}
     \hfill
     \begin{subfigure}[b]{0.49\textwidth}
         \centering
         \includegraphics[width=\textwidth]{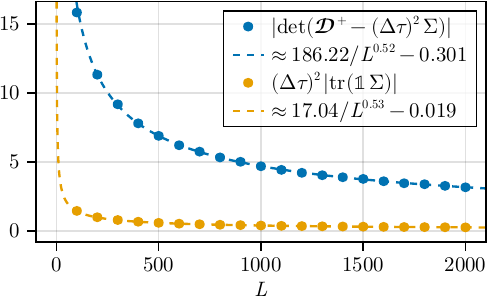}
         \caption{$\beta = 20$}
     \end{subfigure}
    \caption{$L$ dependence of the single-replica determinant $\det(\Partial^+ - (\Delta \tau)^2 \, \Sigma) \approx (\Delta\tau)^2 \tr(\mathbb{1} \, \Sigma)$ evaluated at $w = 0$ and for $q = 4$. In all cases a fit of the values (offset rounded to five decimals) indicates that both the determinant and its zero mode approximation decrease with about the same inverse power of $L$ towards 0, at least approximately. The only outlier for this trend is the case of $\beta = 5$, where no good fit can be made because the determinant is initially negative.}
    \label{fig:single_replica_L_dependence}
\end{figure}

Things are a bit more subtle in the discretized case, as there $\widetilde{\Sigma}$ is expected to only be antisymmetric up to some $\mathcal{O}(1)$ diagonal entries arising from the choice of discretization. This implies that the trace should scale approximately linearly with the matrix size:
\begin{equation}
    \tr(\mathbb{1}_L \, \widetilde{\Sigma}) \propto c L.
\end{equation}
However, this anomalous effect is not enough to counteract the suppression by the $\mathcal{O}(L^{-2})$ prefactor in \eqref{eq:determinant_expansion}, which means that it must vanish in the limit of $L \gg (\beta J)^2$. That this is indeed the case in the numerics is shown in Figure \ref{fig:single_replica_L_dependence}, however the rate of convergence only has the predicted $\mathcal{O}(L^{-1})$ scaling for small $\beta J$, and seems to approach a $\mathcal{O}(L^{-1/2})$ scaling at larger $\beta J$ (here $J = 1$).

As alluded to in the introduction of this section, the physical reason for the path integral vanishing at $R = 1$ (and larger odd $R$) and $w > 0$ lies in the fact that the sign of $\xi(a)$ varies erratically, so that $\overline{\xi(a)}$ averages to zero. By contrast, the average of the absolute value of $\xi(a)$ does not vanish, but it cannot be straightforwardly represented using replicas. Below, we instead use the Gaussianity of $\xi(a)$ to indirectly evaluate the average of its absolute value, and even when this approach is not available, as in the $q=2$ case, there is still a lower bound on the average of the absolute value in terms of the second and fourth moments that gives the same conclusion up to $\mathcal{O}(1)$ numerical prefactors.

However, note that the moment at $w = 0$ must always be strictly 1 for any $R$ and $N$ because it is equal to the trace of the state itself. This is indeed true even in the large-$N$ limit because at this point \eqref{eq:syk_w_action} reduces to the action of ordinary SYK, and so $\overline{\xi(0)} = \overline{Z/Z} = 1$. This implies that there must be a discontinuity in the first moments at large $N$, which can be interpreted as being a consequence of the limits $N \rightarrow \infty$ and $w \rightarrow 0$ not commuting. We will revisit the differences between $w = 0$ and $w > 0$ for higher moments in Section \ref{sec:w_0_behavior}.

\subsubsection*{Purity-Inspired Correlations for \textit{R} = 2}
\label{sec:purity_correlations}

The single-replica solution vanishes almost everywhere because the on-shell conditions \eqref{eq:G_definition} forbid the presence of a zero-mode in the bilocal fields. Even in the multi-replica case, taking $G$ and $\Sigma$ to be symmetric in the replica indices and antisymmetric in imaginary time \cite{arefeva_replica_nondiagonal_solutions_2019} would lead to the same outcome. The other possible option\footnote{One could in principle also consider mixtures of these two extremes, but anecdotally those either converged to the same case considered here, or didn't converge at all.} therefore is to make the off-diagonal replica correlations \emph{symmetric} in imaginary time, with the overall replica structure being antisymmetric. The symmetry of the correlations then allows for the existence of a non-vanishing zero mode, which has the effect of stabilizing the determinant with periodic boundary conditions.

For $R = 2$, this proposal can be made concrete using the fact that we can relate the evaluation of individual second moments of a state to the evaluation of its purity:
\begin{equation}
\label{eq:purity_connection}
    \sum_{a} \tr \left(\majo(a) \, \rho \right)^2 = D \tr(\rho^2).
\end{equation}
Computing the purity saddle for ordinary disorder-averaged SYK is a straightforward endeavor as it essentially just means setting $\beta \rightarrow 2\beta$ when solving the Schwinger-Dyson equations. Up to some constant, the (not replica-diagonal) saddle corresponding to $w$-deformed SYK at inverse temperature $\beta$ and some \enquote{critical} relative weight $w = w_\mathrm{crit}$ should therefore somehow be equivalent to the single-replica saddle of ordinary SYK at inverse temperature $2\beta$. That this is indeed true is proven explicitly in terms of the second R\'enyi entropy in Section \ref{sec:purity}, but for now we only care about the implication that the saddle points for $G$ (and $\Sigma$) resulting from both approaches should also match in some way.

To understand how a single-replica solution can be equivalent to a two-replica solution, it is illustrative to look at the Green's function $G_\mathrm{free}$ of a free Majorana fermion, which is also usually chosen as the initial value when iteratively solving the ordinary Schwinger-Dyson equations. In the discretized setting, it can be retrieved as the inverse of the anti-periodic derivative matrix $\Partial^-$ \eqref{eq:discrete_derivative}, up to the usual anomalous diagonal elements\footnote{Subtracting the diagonal elements of $(\Partial^-)^{-1}$  has shown to improve numerical convergence.}:
\begin{equation}
\label{eq:G_free}
    G_\mathrm{free} \sim (\Partial^-)^{-1} - \frac{1}{2} \, I = \frac{1}{2}\begin{pmatrix}
        0 & -1 & -1 & \cdots & -1 & -1 \\
        1 & 0 & -1 & \cdots & -1 & -1 \\
        1 & 1 & 0 & \cdots & -1 & -1 \\
        \vdots & \vdots & \vdots & \ddots & \vdots & \vdots \\
        1 & 1 & 1 & \cdots & 0 & -1 \\
        1 & 1 & 1 & \cdots & 1 & 0
    \end{pmatrix} \equiv \frac{1}{2} \, [\sgn(i - j)]_{ij}.
\end{equation}
Assuming the number of rows and columns to be $2L$, one can interpret $G_\mathrm{free}$ as both the aforementioned $2L \times 2L$ matrix for the single-replica case and a $2 \times 2$ replica matrix containing $L \times L$ replicas for the two-replica case:
\begin{equation}
\label{eq:G_free_decomposition}
    (G_\mathrm{free})_{2L} = \begin{pmatrix} 1 & 0 \\ 0 & 1 \end{pmatrix} \otimes (G_\mathrm{free})_L + \begin{pmatrix} 0 & -1 \\ 1 & 0 \end{pmatrix} \otimes \frac{1}{2} \, \mathbb{1}_L,
\end{equation}
where $\mathbb{1}_L$ again denotes the (symmetric) $L \times L$ matrix with all ones. Expressing it like that, clearly this matrix satisfies the requirements outlined before, namely having the cross-replica correlations be symmetric in (discretized) imaginary time but antisymmetric with regard to the replica indices. Identifying
\begin{equation}
    T \coloneqq \begin{pmatrix} 0 & -1 \\ 1 & 0 \end{pmatrix} \equiv I_2 - \Partial^-_2,
\end{equation}
we therefore make the ansatz that any (not necessarily discretized) two-replica saddle point solution $M \in \{G, \Sigma, \ldots\}$ of $w$-deformed SYK can be written as
\begin{equation}
    M \equiv \sum_{r = 0}^1 T^r \otimes M_r = I \otimes M_0 + T \otimes M_1,
\end{equation}
where $M_0$ and $M_1$ correspond to the diagonal and off-diagonal elements of $M$ respectively. It is straightforward to check that this \emph{replica symmetry} generated by $T$ is left invariant under the Schwinger Dyson equations \eqref{eq:syk_w_schwinger_dyson}, in particular the second due to $q-1$ being odd, which causes the signs of $T$ to be preserved.

\begin{figure}[h!tb]
    \centering
    \begin{subfigure}[b]{0.49\textwidth}
         \centering
         \includegraphics[width=\textwidth]{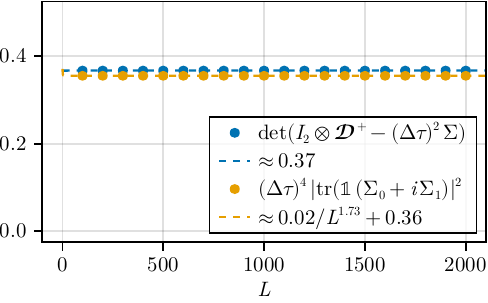}
         \caption{$\beta = 1$}
     \end{subfigure}
     \hfill
     \begin{subfigure}[b]{0.49\textwidth}
         \centering
         \includegraphics[width=\textwidth]{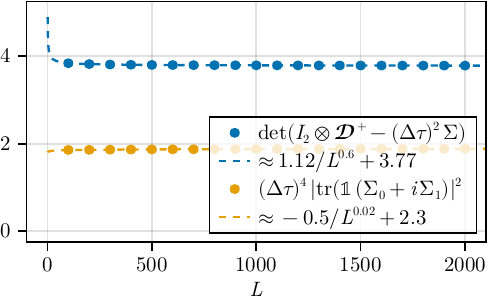}
         \caption{$\beta = 5$}
     \end{subfigure}
     \par\bigskip
     \begin{subfigure}[b]{0.49\textwidth}
         \centering
         \includegraphics[width=\textwidth]{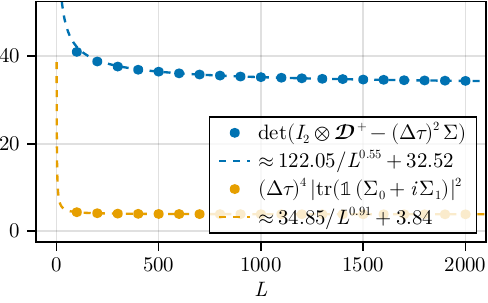}
         \caption{$\beta = 10$}
     \end{subfigure}
     \hfill
     \begin{subfigure}[b]{0.49\textwidth}
         \centering
         \includegraphics[width=\textwidth]{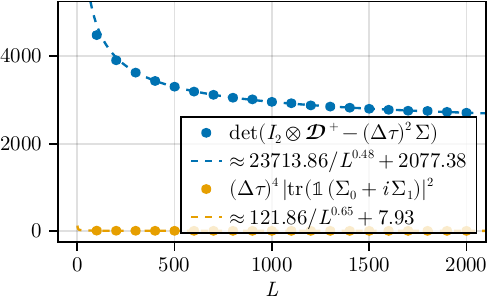}
         \caption{$\beta = 20$}
     \end{subfigure}
    \caption{$L$ dependence of the two-replica determinant $\det(I_2 \otimes \Partial^+ - (\Delta \tau)^2 \, \Sigma) \approx (\Delta\tau)^4 |\tr(\mathbb{1} \, (\Sigma_0 + i \, \Sigma_1))|^2$ evaluated at $w = 0.5$ and for $q = 4$. In all cases a fit of the values (rounded to two decimals) indicates that both the determinant and its zero mode approximation -- while not matching at larger $\beta J$ -- are either essentially independent of $L$ (up to small corrections) or converge towards a non-zero value at large $L$.}
    \label{fig:two_replica_L_dependence}
\end{figure}

We can now explicitly check that a two-replica solution with initial value \eqref{eq:G_free_decomposition} leads to non-singular periodic determinant. Since $T$ has eigenvalues $\pm i$, the determinant of any matrix endowed with that replica structure is
\begin{align}
\begin{split}
    \det(M) &= \det(M_0 + i M_1) \det(M_0 - i M_1) \\
    &= \lvert \det(M_0 + i M_1) \rvert^2.
\end{split}
\end{align}
Using the zero-mode approximation \eqref{eq:determinant_expansion} on each of the terms, we get the following:
\begin{align}
\begin{split}
    \det(I_2 \otimes \Partial^+ - (\Delta\tau)^2 \, \Sigma) &=  \lvert\det(\Partial^+ - (\Delta\tau)^2 \, (\Sigma_0 + i \Sigma_1))\rvert^2 \\
    &= (\Delta\tau)^4 \, \lvert\tr\left(\mathbb{1}_L \, (\Sigma_0 + i \Sigma_1)\right)\rvert^2 + \mathcal{O}\left((\Delta\tau)^6\right) \\
    &\equiv \left( \beta J / L \right)^4 \left[\tr(\mathbb{1}_L \, \widetilde{\Sigma}_0)^2 + \tr(\mathbb{1}_L \, \widetilde{\Sigma}_1)^2 \right] + \mathcal{O}\left(\left(\beta J / L \right)^6\right).
\end{split}
\end{align}
As before, $\Sigma_0$ is approximately antisymmetric and so its zero mode is suppressed by $L$. But now the determinant also depends on the zero mode of $\Sigma_1$, which we require to not vanish. Under that assumption, the traces are expected to scale as
\begin{equation}
    \tr(\mathbb{1}_L \, \widetilde{\Sigma}_0)^2 + \tr(\mathbb{1}_L \, \widetilde{\Sigma}_1)^2 \propto c_0 L^2 + c_1 L^4,
\end{equation}
meaning that the first term is still vanishing in the large-$L$ limit, while the second term does not strongly depend on $L$ at all. The presence of zero-modes in the replica-correlations is hence necessary for the periodic determinant to be non-zero at all discretization sizes. Numerical evidence of this is also provided in Figure \ref{fig:two_replica_L_dependence} (with $J = 1$).

But even in this case, the convergence of the determinant is of order $\mathcal{O}(L^{-1/2})$, meaning that for our chosen discretization size of $L = 1000$ there will be significant errors even at relatively low $\beta$. This error will be suppressed by the logarithm occurring in the action, but its existence should be acknowledged regardless as it does affect determinant-dependent solving routines such as the one discussed in Section \ref{sec:summing_over_moments}.

\subsection{Generalized Replica Symmetry}

We can now generalize the arguments made for the $R = 2$ case to arbitrary even $R$. While it is typically not possible to directly verify our choice of replica symmetry using other reliable calculations (except for the aforementioned second R\'enyi entropy discussed in Section \ref{sec:purity}), numerical evidence suggests it to be reasonable. It should be noted though that the choice of generalization made here is not necessarily unique, and that other ways to extend $R = 2$ are possible. We do believe however that at least for $q \geq 4$ our ansatz together with its (signed) permutations is sufficient to retrieve all relevant solutions, in particular those corresponding to a real Gaussian random variable. 

\subsubsection*{Anticirculant Ansatz}
\label{sec:anticirculant}

In the discretized setup, we can decompose the derivative operator $\Partial_{RL}^\mp$ \eqref{eq:discrete_derivative} with either boundary condition in terms of derivatives acting on either the replica space or imaginary time:
\begin{equation}
    \Partial_{RL}^\pm = \frac{1}{2} \Partial_R^\pm \otimes (\Partial_L^- - \Partial_L^+) + I_R \otimes \Partial^+_L.
\end{equation}
From this it follows that $[\Partial^-_{RL}, \Partial_R^- \otimes I_L] = 0$, and since any (dominating) saddle for ordinary SYK also commutes with $\Partial^-_{RL}$ (due to being time-translation invariant) it means that imposing the replica symmetry associated to $\Partial_R^-$ gives $w$-deformed solutions the potential to converge towards such a saddle, at least in some regime. Extending the $R = 2$ case, we hence define the generalized replica symmetry generator
\begin{equation}
    T \coloneqq I_R - \Partial_R^- = \begin{pmatrix} 
        0 & 0 & 0 & \cdots & 0 & -1 \\
        1 & 0 & 0 & \cdots & 0 & 0 \\
        0 & 1 & 0 & \cdots & 0 & 0 \\
        \vdots & \vdots &\vdots & \ddots & \vdots & \vdots \\
        0 & 0 & 0 & \cdots & 0 & 0 \\
        0 & 0 & 0 & \cdots & 1 & 0
    \end{pmatrix},
\end{equation}
and use it to make the following ansatz for any bilocal field $M \in \{ G, \Sigma, \ldots\}$:
\begin{equation}
    M \equiv \sum_{r = 0}^{R - 1} T^r \otimes M_r = \begin{pmatrix} 
        M_0 & -M_{R-1} & \cdots & -M_2 & -M_1  \\
        M_1 & M_0 & -M_{R-1} &  \cdots & -M_2 \\
        \vdots & M_1 & M_0 &\ddots & \vdots\\
        M_{R-2} &  \vdots & \ddots & M_0 & -M_{R-1}  \\
        M_{R-1} & M_{R-2} & \cdots & M_1 & M_0 
    \end{pmatrix}.
\end{equation}
We call (block) matrices with this structure \emph{anticirculant}, as they describe translation-invariant operators with antiperiodic boundary conditions\footnote{Circulant matrices (which have periodic boundary conditions) are common in the literature, whereas their antiperiodic counterparts as presented here are less well known.}. What this means is that the entries $M_{rs}$ of such a matrix only depend on the difference $r-s$ of the row and column indices mod $R$, with entries that are only equivalent mod $R$ having opposing signs:
\begin{equation}
    M_{rs} \sim \sgn(r-s) \, M_{r-s \mod R},
\end{equation}
where $\sgn(0) = 1$. This translation invariance and antiperiodicity also manifests in the spectrum of $T$, which is made up of the odd Fourier modes ($R$th roots of $-1$):
\begin{equation}
    \sigma(T) = \left\{ e^{2 \pi i (s + 1/2) / R} \, | \, s = 0, \ldots, R-1\right\}.
\end{equation}
Like in the $R = 2$ case, it is therefore possible to (block-) diagonalize any anticirculant matrix $M$ by performing a discrete Fourier transform on the individual entries $M_r$:
\begin{equation}
    \widehat{M} = \diag \left( \widehat{M}_0, \ldots, \widehat{M}_{R-1} \right), \quad   \widehat{M}_s = \sum_{r=0}^{R-1} e^{2 \pi i r (s + 1/2) / R} \, M_r.
\end{equation}
Since the determinant of $M$ is left invariant under this transformation, one can also decompose it into a product of determinants of the blocks, such that
\begin{equation}
    \det(M) = \prod_{s = 0}^{R-1} \det \left(\widehat{M}_s\right).
\end{equation}
Furthermore, the Fourier-transformed blocks are (approximately) anti-Hermitian and always occur together with their respective complex conjugate:
\begin{equation}
    \left(\widehat{M}_s\right)^\dagger = - \widehat{M}_s, \quad \left(\widehat{M}_s\right)^* = \widehat{M}_{R-s-1},
\end{equation}
where $\dagger$ symbolizes flipping the imaginary time parameters of the field's complex conjugate in the continuous picture. The second identity implies that any anticirculant $R$-replica matrix is actually uniquely defined by only $R/2$ complex fields (matrices), which all can be inverted independently to significantly speed up the iterative solving of the anticirculant Schwinger-Dyson equations
\begin{align}
\begin{split}
\label{eq:syk_w_schwinger_dyson_discrete_anticirculant}
    \widehat{G}_s &= (1-w) \, (\Partial^- - (\Delta\tau)^2 \, \widehat{\Sigma}_s)^{-1} + w \, (\Partial^+ - (\Delta\tau)^2 \, \widehat{\Sigma}_s)^{-1}, \\
    \Sigma_r(\tau,\tau') &= J^2 \, G_r(\tau,\tau')^{q-1},
\end{split}
\end{align}
or more specifically their analogous discretization\footnote{Same as in the $R = 2$ case, this more general structure is left invariant under the second Schwinger Dyson equation: $G(\tau, \tau')^{q-1} = \sum_{r = 0}^{R-1} T^r \, G_r(\tau, \tau')^{q-1}$}. More details on the exact numerical procedures and runtime complexities for both the unbiased and anticirculant solvers are found in Appendix \ref{sec:unbiased_solver} and \ref{sec:anticirculant_solver} respectively. 

\subsubsection*{Antisymmetry}

For $R > 2$ one has more constraints that the replica structure has to satisfy, namely those imposed by the (approximate) antisymmetry \eqref{eq:G_definition} of $G$ (and $\Sigma$) as a whole. This can be seen from $T T^T = I_R = -T^R$, which implies that
\begin{equation}
    \left(T^r\right)^T = T^{-r} = -T^{R-r}, \quad r = 0, 1, \ldots, R-1.
\end{equation}
Writing $M^T$ to denote switching both replica indices and imaginary times (either continuous or discrete) of some replica field $M$, one then gets the constraint
\begin{align}
\begin{split}
    M^T = \sum_{r = 0}^{R-1} \left(T^r\right)^T \otimes M^T_r = - \sum_{r = 0}^{R-1} T^{R-r} \otimes M^T_r = I_R \otimes M_0^T - \sum_{r = 1}^{R-1} T^r \otimes M^T_{R-r} \stackrel{!}{=} -M,
\end{split}
\end{align}
which is satisfied iff
\begin{equation}
\label{eq:G_replica_antisymmetry_constraint}
    M_r^T = \begin{cases}
        -M_r, \quad & r = 0 \\
        M_{R-r}, \quad & r > 0
    \end{cases}.
\end{equation}
As before, the equality for $r = 0$ is exactly true in the continuous case and only approximately true up to diagonal elements in the discretized case. In either case, any anticirculant and antisymmetric $R$-replica matrix $M$ is hence fully determined by $R/2 + 1$ real entries $M_r$. This leads to further improvements for the numerical procedure laid out in Appendix \ref{sec:anticirculant_solver}.

\subsubsection*{(Signed) Permutation Invariance}
\label{sec:permutations}

As mentioned earlier in this section, the anticirculant replica symmetry chosen here is not the only possible ansatz, but we believe it to be sufficient to probe all possible distinct values the action can take on due to $T$ having non-degenerate eigenvalues. Knowing the number of distinct saddles with the same value for the action is important though to make statements over the statistics of the operator moments considered in this work (see Section \ref{sec:saddle_point_corrections}), as well as narrow down the number of anticirculant solutions one has to keep track of.

Considering the (continuous) action \eqref{eq:syk_w_action} with the second Schwinger-Dyson equation \eqref{eq:syk_w_schwinger_dyson} inserted into it
\begin{align}
\begin{split}
\label{eq:syk_w_action_integrated_out}
    \action^{(R, w)}_\mathrm{SYK}[G, \Sigma] ={}& - \frac{1-w}{2} \ln \det[I_R \otimes \Partial^- - 2 \, \Sigma] - \frac{w}{2} \ln \det[I_R \otimes \Partial^+ - 2 \, \Sigma] \\
    & + \frac{J^2(q-1)}{2q} \sum_{r,s=1}^{R} \int_0^{\beta} d\tau d\tau'  \, G_{rs}(\tau, \tau')^q,
\end{split}
\end{align}
one can see that it for all (even) $q$ it is invariant under both overall sign flips $G \rightarrow -G, \, \Sigma \rightarrow -\Sigma$ (which for the the determinants follows from integration by parts) and signed permutations $P$ of the replica indices:
\begin{equation}
    G(\tau, \tau') \rightarrow (P \, G P^T) (\tau, \tau'), \quad P^T P = I_R, \, P_{rs} \in \{1, 0, -1\}.
\end{equation}
Given such a permutation, the associated new replica symmetry generator $T'$ would then be of the form
\begin{equation}
    T' = P \, T P^T.
\end{equation}

In the case of $q = 2$, the action admits even more symmetries, namely those corresponding to the orthogonal group $\Ortho(R)$. We comment on the explicit implications of this difference later in Section \ref{sec:fourth_moments}, but for now we will restrict ourselves to the subgroup of signed permutations as those are the only symmetries relevant for $q \geq 4$.

Assuming that solutions with anticirculant replica structure are sufficient to probe all possible saddle point values, we are interested in finding representatives from each equivalence class of solutions which result in the same action. As argued above, this is equivalent to understanding in what ways signed permutations can (non-trivially) map between different anticirculant replica structures, or more specifically different powers of $T$. Using a numerical search for small $R$, we found that any such permutation $P$ must satisfy
\begin{equation}
    P \, T P^T \equiv \pm \, T^{2k - 1}, \quad 1 \leq k \leq  R/2, \quad 2k - 1 \neq R / 2,
\end{equation}
meaning $T$ can in general only be mapped to odd powers of itself. The last inequality applies if $R/2$ is odd and arises because in that case $T^{R/2}$ is antisymmetric, a property that is trivially preserved under signed permutations but is not shared by $T$ (unless $R = 2$).

It follows that under the symmetry transformation $T \rightarrow -T$ the associated replica fields $G_r$ transform as
\begin{equation}
    G_r \rightarrow (-1)^r \, G_r,
\end{equation}
whereas under the transformation $T \rightarrow T^{2k-1}$ with $k$ as above they transform as
\begin{equation}
    G_r \rightarrow (-1)^{\lfloor (2k-1) \, r/R \rfloor} \, G_{(2k-1) \, r \mod R}.
\end{equation}
In particular, for $k = R/2$ (which is the only non-trivial case until $R=8$) the permutations transform $G$ such that
\begin{equation}
    G_r \rightarrow (-1)^{r+1} \, G_{R - r} = (-1)^{r+1} \, G^T_r,
\end{equation}
with the second equality following from \eqref{eq:G_replica_antisymmetry_constraint}. Assuming that the off-diagonal replica fields $G_r$ are symmetric, one can therefore freely choose the overall sign shared by all fields with even and odd index respectively.

\subsection{Classification of Initial Values}
\label{sec:initial_value_classification}

Based on these symmetries and some additional (potentially biased) assumptions about the actual replica fields, we can now make an explicit ansatz for the discretized initial values $G_\mathrm{init}$. Inspired by the $R=2$ case \eqref{eq:G_free_decomposition}, we choose the initial replica fields such that\footnote{Subsequent analytical considerations for $R = 2$ have revealed $(G_\mathrm{init})_0(\tau, \tau') = \sgn(\tau - \tau')/ 2 - w (\tau - \tau') / \beta$ and $(G_\mathrm{init})_1(\tau, \tau') = \pm \max((w / (\beta J)^2)^{1/q}, 1/2)$ to lead to better convergence. However, the impact of this improvement on the value of the statistical moments is minimal and only really observable in the regime where $w > 0.5$. It therefore does not affect any of the results stated in Section \ref{sec:applications}, and only slightly affects the observed structure of $G$ as depicted in e.g. Figure \ref{fig:G_R2_saddles}. We expect the same improvements to also apply to $R > 2$, while not affecting the replica structures themselves.}
\begin{equation}
    (G_\mathrm{init})_0 = G_\mathrm{free} \equiv (\Partial^-)^{-1} - I/2, \quad (G_\mathrm{init})_{r>0} \in \{\mathbb{1} / 2, -\mathbb{1} / 2, 0\}.
\end{equation}
In particular, this means that the off-diagonal fields are symmetric, and so \eqref{eq:G_replica_antisymmetry_constraint} demands that one has $(G_\mathrm{init})_r = (G_\mathrm{init})_{R-r}$. All possible initial replica structures can therefore be denoted by $\mathbf{R}_R(s_1, s_2, \ldots, s_{R/2})$ with $s_r \in \{+, -, 0\}$. 

\begin{figure}[p]
    \centering
    \includegraphics[width=\linewidth]{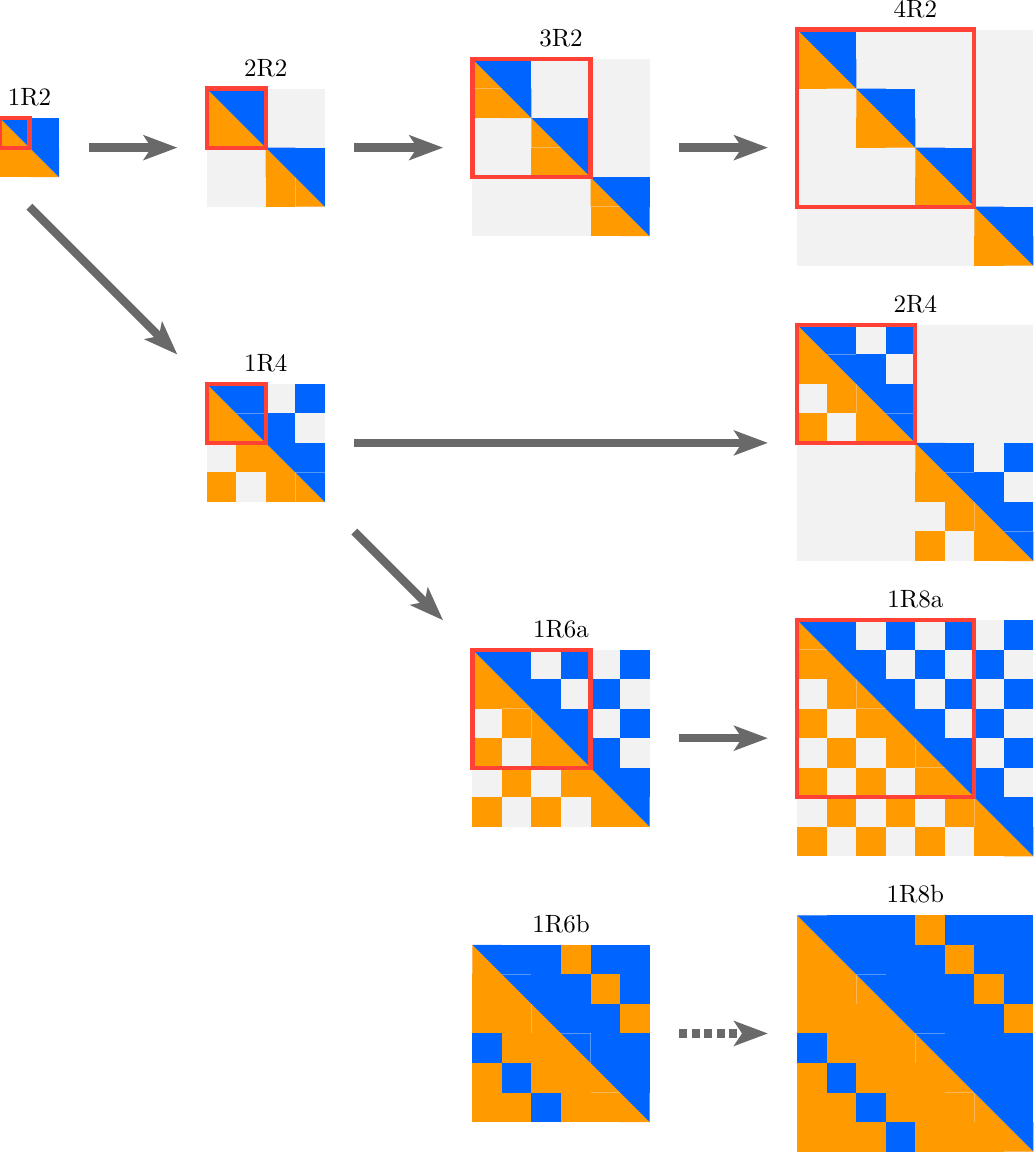}
    \caption{Matrix heatmaps of all distinct (non-diagonal) replica structures ($R = 2, 4, 6, 8$) which are left invariant by the Schwinger-Dyson equations and result in \enquote{well-behaved} saddle points for all $q$ and $\beta$. Orange represents a value of $+1/2$, blue a value of $-1/2$ and gray a value of $0$. Arrows indicate self-similarity relationships between replicas of different sizes.}
    \label{fig:replica_structures}
\end{figure}

While initially there are $3^{R/2} - 1$ such replica structures (ignoring the replica-diagonal case), this number is significantly reduced by identifying those equivalent to each other under (signed) permutations. Using a computer-aided search, we determined all genuinely distinct initial values satisfying our assumptions up to $R = 8$. Furthermore, we also did a heuristic analysis of the solutions resulting from said initial values for $q = 4$ and at small $\beta$ (to avoid potential convergence issues, see Section \ref{sec:w_0_behavior}). A table containing our results can be found in Figure \ref{fig:initial_replica_structures}. What we observed is that many of the initial values seem to converge towards the same solution, if they even converge at all. This suggests that not all possible replica structures are actually realized in the end, with the total number of $R$-replica structures left invariant seemingly being only $R/2$ (excluding the replica-diagonal case at $w = 0$) based on our preliminary results. These initial values and their relations to each other are schematically depicted up to $R = 8$ in Figure \ref{fig:replica_structures}. However, a more systematic study of the claims made here should be conducted in the future.
\begin{figure}[h!tb]
    \centering
    \begin{tabular}{c||c||c||c||c||c||c||c||c||c}
         $R$ & Structure & $G_1$ & $G_2$ & $G_3$ & $G_4$ & $G_5$ & $G_6$ & $G_7$ & Comments \\
         \hline \hline
         2 & $\mathbf{R}_2(+)$ & $\mathbb{1}/2$ & & & & & & &  \\
         \hline
          & $\mathbf{R}_4(0,+)$ & 0 & $\mathbb{1}/2$ & 0 & & & & &  $\cong 2 \times \mathbf{R}_2(+)$ \\
         4 & $\mathbf{R}_4(+,0)$ & $\mathbb{1}/2$ & 0 & $\mathbb{1}/2$ & & & & &  \\
          & $\mathbf{R}_4(+, +)$ & $\mathbb{1}/2$ & $\mathbb{1}/2$ & $\mathbb{1}/2$ & & & & & \makecell{$\leadsto \mathbf{R}_4(+,0)$ \\ limit.\ convergence} \\
         \hline
          & $\mathbf{R}_6(0, 0, +)$ & 0 & 0 & $\mathbb{1}/2$ & 0 & 0 & & & $\cong 3 \times \mathbf{R}_2(+)$ \\
          & $\mathbf{R}_6(0, +, 0)$ & 0 & $\mathbb{1}/2$ & 0 & $\mathbb{1}/2$ & 0 & & & limit.\ convergence \\
          & $\mathbf{R}_6(0, +, +)$ & 0 & $\mathbb{1}/2$ & $\mathbb{1}/2$ & $\mathbb{1}/2$ & 0 & & & $\leadsto \mathbf{R}_6(+, +, -)$ \\
          & $\mathbf{R}_6(+, 0, 0)$ & $\mathbb{1}/2$ & 0 & 0 & 0 & $\mathbb{1}/2$ & & & $\leadsto \mathbf{R}_6(+, 0, +)$ \\
         6 & $\mathbf{R}_6(+, 0, +)$ & $\mathbb{1}/2$ & 0 & $\mathbb{1}/2$ & 0 & $\mathbb{1}/2$ & & & \\
          & $\mathbf{R}_6(+, 0, -)$ & $\mathbb{1}/2$ & 0 & $-\mathbb{1}/2$ & 0 & $\mathbb{1}/2$ & & &limit.\ convergence \\
          & $\mathbf{R}_6(+, +, 0)$ & $\mathbb{1}/2$ & $\mathbb{1}/2$ & 0 & $\mathbb{1}/2$ & $\mathbb{1}/2$ & & & $\leadsto \mathbf{R}_6(+, +, -)$ \\
          & $\mathbf{R}_6(+, +, +)$ & $\mathbb{1}/2$ & $\mathbb{1}/2$ & $\mathbb{1}/2$ & $\mathbb{1}/2$ & $\mathbb{1}/2$ & & & $\leadsto \mathbf{R}_6(+, 0, +)$ \\
          & $\mathbf{R}_6(+, +, -)$ & $\mathbb{1}/2$ & $\mathbb{1}/2$ & $-\mathbb{1}/2$ & $\mathbb{1}/2$ & $\mathbb{1}/2$ & & & \\
         \hline
          & $\mathbf{R}_8(0, 0, 0, +)$ & 0 & 0 & 0 & $\mathbb{1}/2$ & 0 & 0 & 0 & $\cong 4 \times \mathbf{R}_2(+)$ \\
          & $\mathbf{R}_8(0, 0, +, 0)$ & 0 & 0 & $\mathbb{1}/2$ & 0 & $\mathbb{1}/2$ & 0 & 0 & $\leadsto \mathbf{R}_8(+, 0, +, 0)$ \\
          & $\mathbf{R}_8(0, 0, +, +)$ & 0 & 0 & $\mathbb{1}/2$ & $\mathbb{1}/2$ & $\mathbb{1}/2$ & 0 & 0 & $\leadsto \mathbf{R}_8(+, -, +, +)$ \\
          & $\mathbf{R}_8(0, +, 0, 0)$ & 0 & $\mathbb{1}/2$ & 0 & 0 & 0 & $\mathbb{1}/2$ & 0 & $\cong 2 \times \mathbf{R}_4(+,0)$ \\
          & $\mathbf{R}_8(0, +, 0, +)$ & 0 & $\mathbb{1}/2$ & 0 & $\mathbb{1}/2$ & 0 & $\mathbb{1}/2$ & 0 & $\leadsto \mathbf{R}_8(0, +, 0, 0)$ \\
          & $\mathbf{R}_8(0, +, +, 0)$ & 0 & $\mathbb{1}/2$ & $\mathbb{1}/2$ & 0 & $\mathbb{1}/2$ & $\mathbb{1}/2$ & 0 & $\leadsto \mathbf{R}_8(+, +, +, -)$\\
         8 & $\mathbf{R}_8(0, +, +, +)$ & 0 & $\mathbb{1}/2$ & $\mathbb{1}/2$ & $\mathbb{1}/2$ & $\mathbb{1}/2$ & $\mathbb{1}/2$ & 0 & $\leadsto \mathbf{R}_8(+, +, +, -)$ \\
          & $\mathbf{R}_8(0, +, +, -)$ & 0 & $\mathbb{1}/2$ & $\mathbb{1}/2$ & $-\mathbb{1}/2$ & $\mathbb{1}/2$ & $\mathbb{1}/2$ & 0 & $\leadsto \mathbf{R}_8(+, +, +, -)$ \\
          & $\mathbf{R}_8(+, 0, +, 0)$ & $\mathbb{1}/2$ & 0 & $\mathbb{1}/2$ & 0 & $\mathbb{1}/2$ & 0 & $\mathbb{1}/2$ &  \\
          & $\mathbf{R}_8(+, 0, +, +)$ & $\mathbb{1}/2$ & 0 & $\mathbb{1}/2$ & $\mathbb{1}/2$ & $\mathbb{1}/2$ & 0 & $\mathbb{1}/2$ & $\leadsto \mathbf{R}_8(+, 0, +, 0)$ \\
          & $\mathbf{R}_8(+, +, +, 0)$ & $\mathbb{1}/2$ & $\mathbb{1}/2$ & $\mathbb{1}/2$ & 0 & $\mathbb{1}/2$ & $\mathbb{1}/2$ & $\mathbb{1}/2$ & $\leadsto \mathbf{R}_8(+, +, +, -)$ \\
          & $\mathbf{R}_8(+, +, +, +)$ & $\mathbb{1}/2$ & $\mathbb{1}/2$ & $\mathbb{1}/2$ & $\mathbb{1}/2$ & $\mathbb{1}/2$ & $\mathbb{1}/2$ & $\mathbb{1}/2$ & $\leadsto \mathbf{R}_8(+, 0, +, 0)$ \\
          & $\mathbf{R}_8(+, +, +, -)$ & $\mathbb{1}/2$ & $\mathbb{1}/2$ & $\mathbb{1}/2$ & $-\mathbb{1}/2$ & $\mathbb{1}/2$ & $\mathbb{1}/2$ & $\mathbb{1}/2$ &  \\
    \end{tabular}
    \caption{One potential choice for initial values of $G$ whose replica structures are not equivalent under signed permutations, up to $R = 8$. We also comment on if a given initial value can be block-diagonalized, has been observed to converge towards a different equivalence class of replica structures, or does not seem to converge at all under certain circumstances.}
    \label{fig:initial_replica_structures}
\end{figure}

It should also be noted that this selection of initial values is likely not sufficient to probe all possible saddle points. This should not be surprising, as even in ordinary SYK there are infinitely many non-trivial replica solutions when $R > 2$ \cite{arefeva_replica_nondiagonal_solutions_2019}. In future work it might therefore prove interesting to explore a wider class of initial values using (approximate) analytical solutions of the $q = 2$ and large-$q$ cases. But as we will show in Section \ref{sec:fourth_moments}, at least for $q \geq 4$ all $R > 2$ solutions are dominated by $R/2$ copies of the (unique) $R = 2$ solution, and so we don't have to worry about the omission of subleading contributions.

\subsection{Numerical Results}

Our analysis is primarily based on the numerical solver described in Appendix \ref{sec:anticirculant_solver}, which explicitly assumes an anticirculant replica symmetry because it speeds up computations significantly. For $w > 0$, initial values are chosen as shown in Figures \ref{fig:initial_replica_structures} and \ref{fig:replica_structures}, and for $w = 0$ the (dominating) diagonal solution is imposed unless stated otherwise\footnote{The implications of this choice are discussed in Section \ref{sec:w_0_behavior}.}. Setting the general single-replica matrix size to $L = 1000$ and the linear interpolation between iteration steps to $t = 0.01$ has provided fast and (somewhat) accurate convergence up to around $\beta = 50$ (we set $J = 1$ as usual). Quantities are depicted for $q = 2, 4, 6, 8$ unless stated otherwise. We also restrict the maximum inverse temperature of most figures in the main text to $\beta = 20$, as higher values have led to diminishing returns and worse convergence. There are multiple ways one could improve the accuracy of the results in the future, for example -- as mentioned before -- by refining the choice of initial value by means of analytical results, or by studying the asymptotics of the data with regard to increasing $L$. We defer such refinements to future work.

\subsubsection*{Second Moments}
\label{sec:second_moments}

\begin{figure}[p]
    \centering
    \begin{subfigure}[b]{0.32\textwidth}
         \centering
         \includegraphics[width=\textwidth]{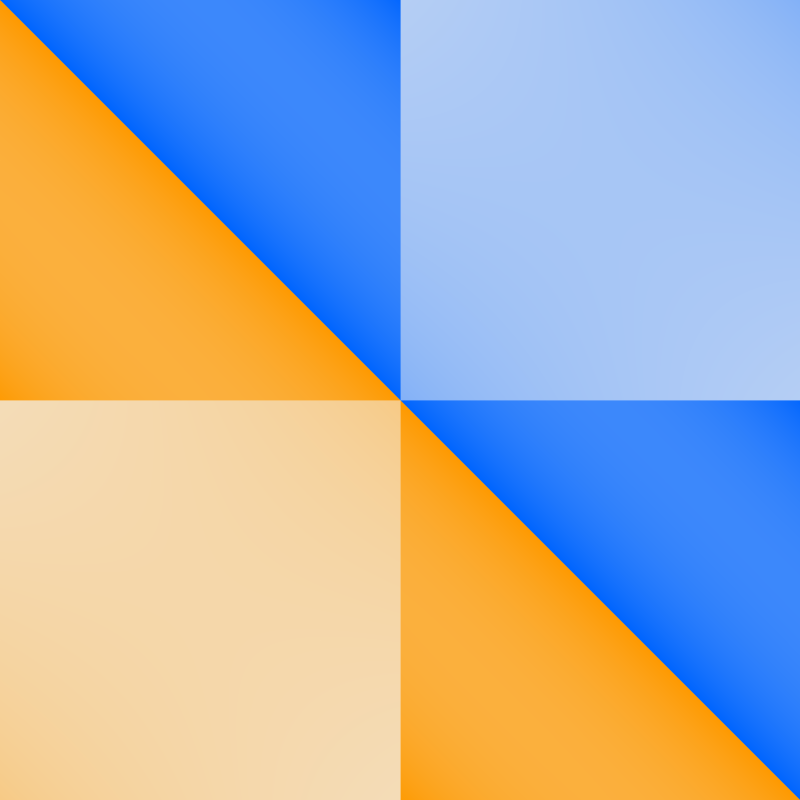}
         \caption{$\beta = 5, w = 0.01$}
     \end{subfigure}
     \hfill
     \begin{subfigure}[b]{0.32\textwidth}
         \centering
         \includegraphics[width=\textwidth]{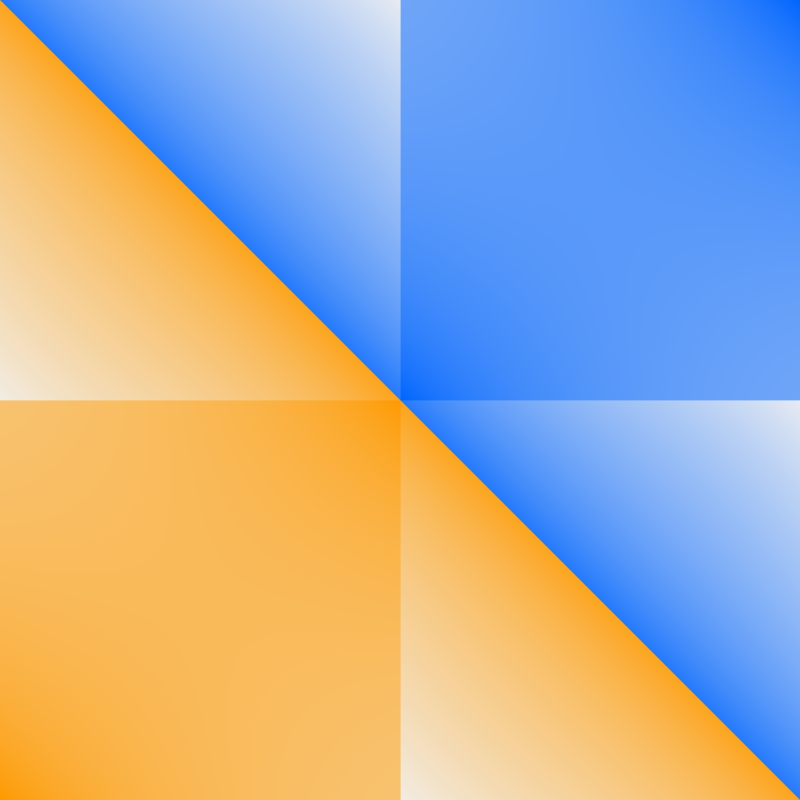}
         \caption{$\beta = 5, w = 0.5$}
     \end{subfigure}
     \hfill
     \begin{subfigure}[b]{0.32\textwidth}
         \centering
         \includegraphics[width=\textwidth]{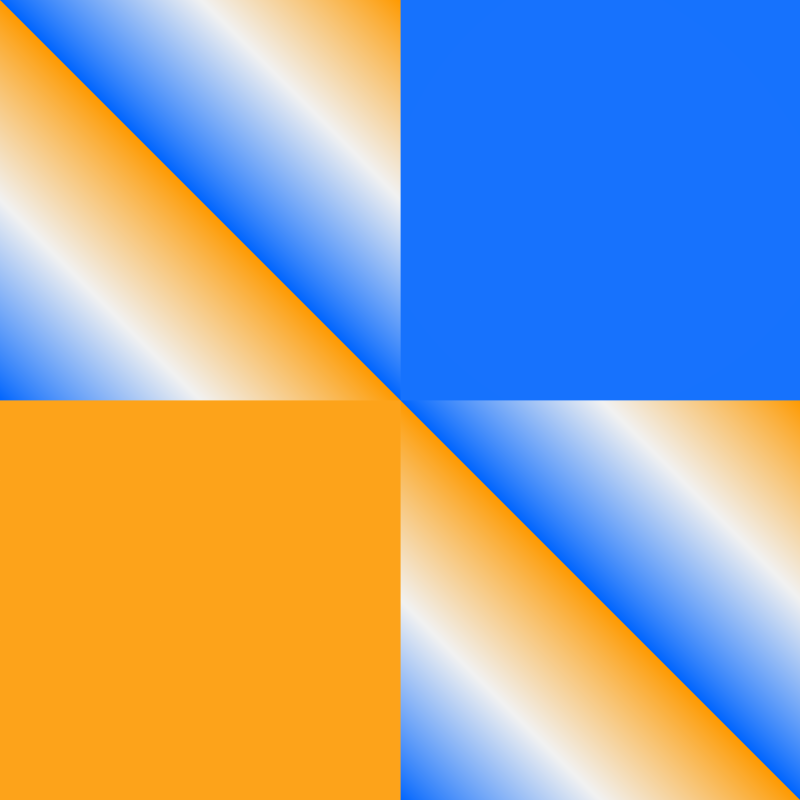}
         \caption{$\beta = 5, w = 1$}
     \end{subfigure}
     \par\smallskip\smallskip
     \begin{subfigure}[b]{0.32\textwidth}
         \centering
         \includegraphics[width=\textwidth]{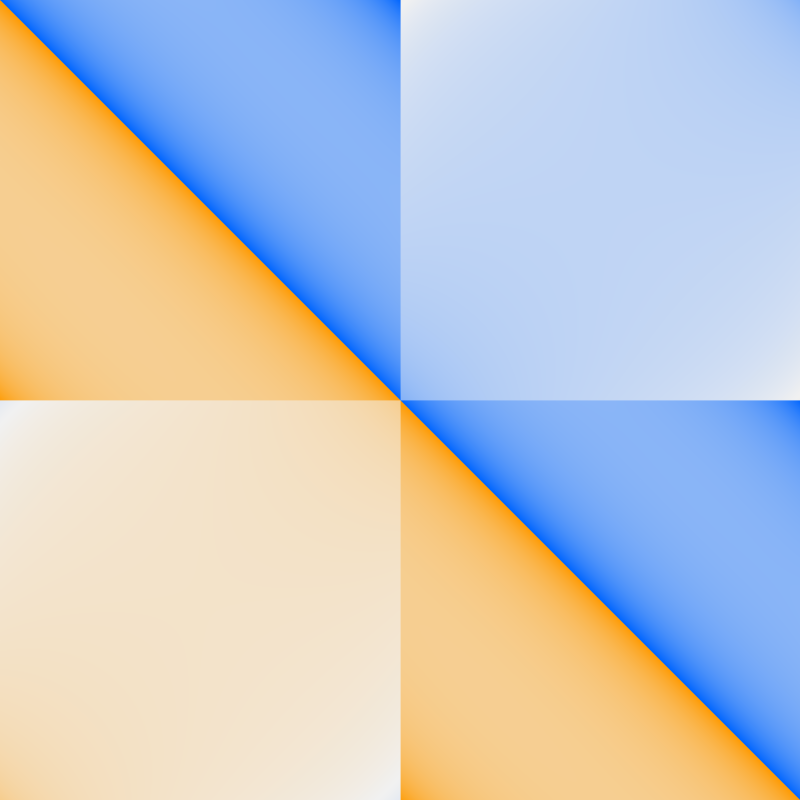}
         \caption{$\beta = 20, w = 0.01$}
     \end{subfigure}
     \hfill
     \begin{subfigure}[b]{0.32\textwidth}
         \centering
         \includegraphics[width=\textwidth]{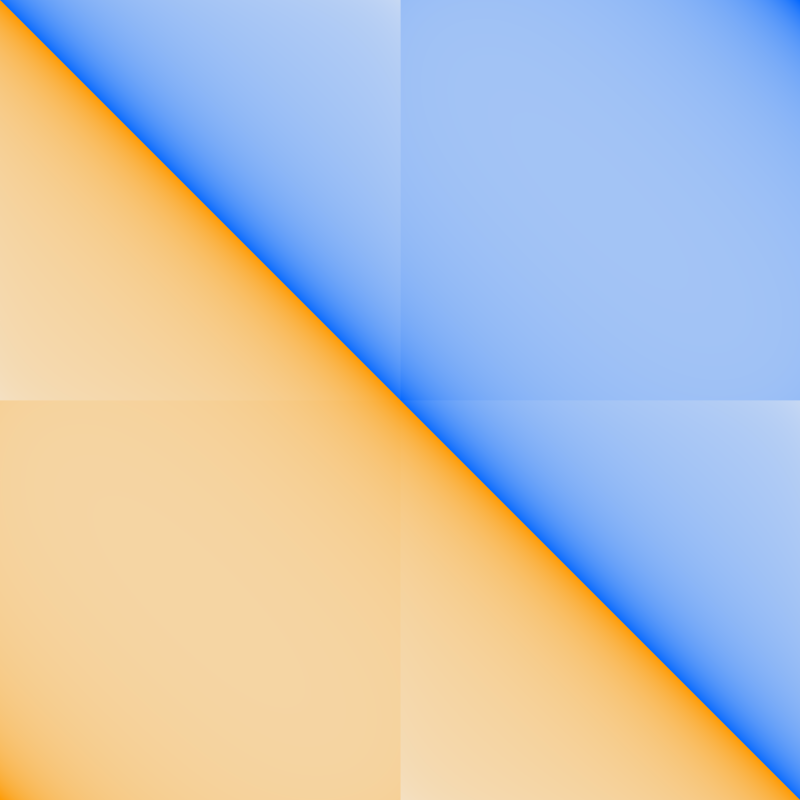}
         \caption{$\beta = 20, w = 0.5$}
     \end{subfigure}
     \hfill
     \begin{subfigure}[b]{0.32\textwidth}
         \centering
         \includegraphics[width=\textwidth]{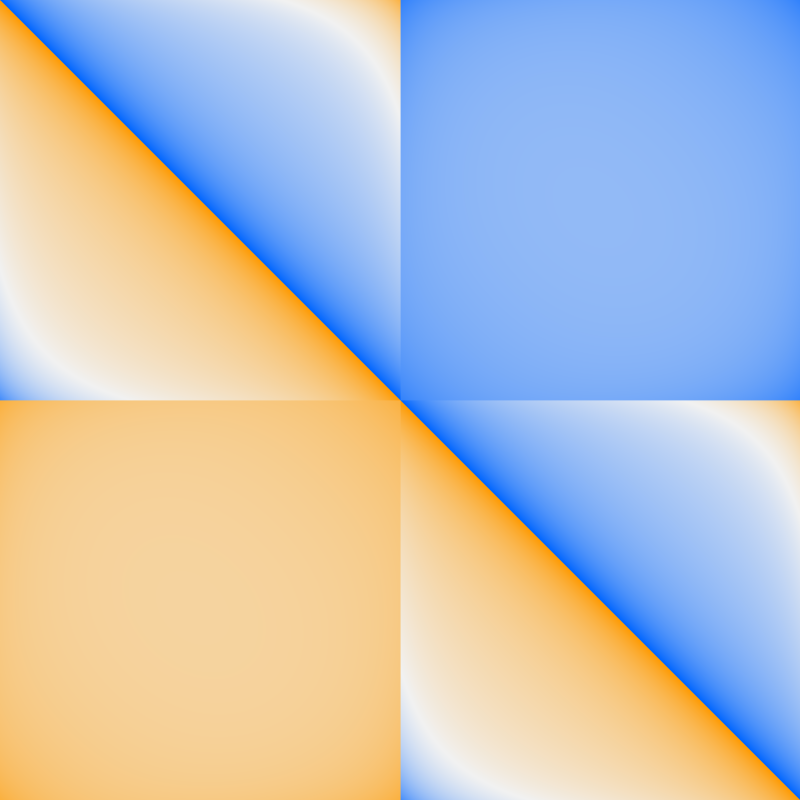}
         \caption{$\beta = 20, w = 1$}
     \end{subfigure}
     \par\smallskip\smallskip
     \begin{subfigure}[b]{0.32\textwidth}
         \centering
         \includegraphics[width=\textwidth]{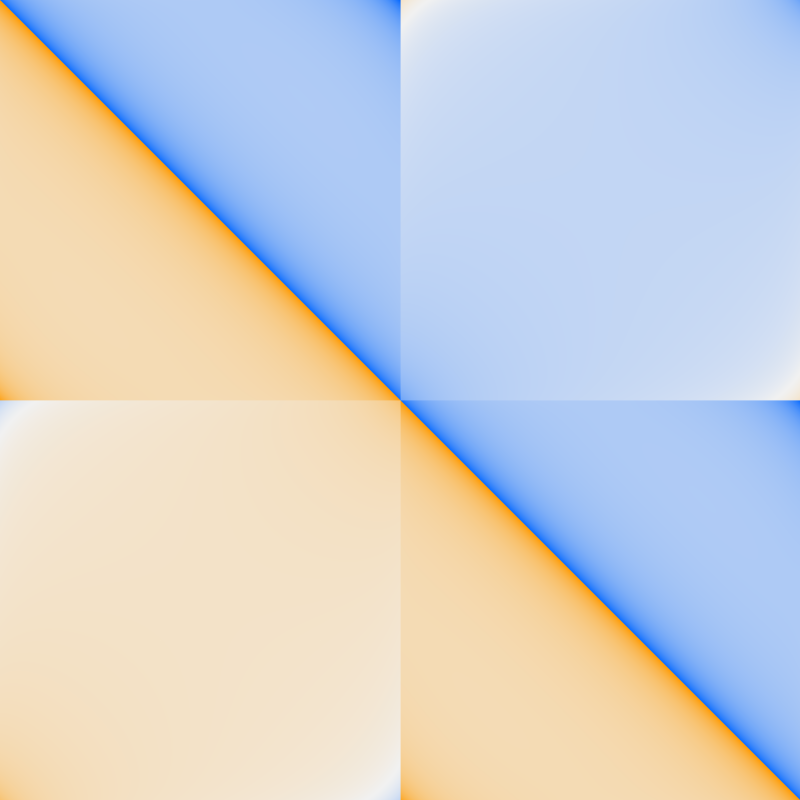}
         \caption{$\beta = 50, w = 0.01$}
     \end{subfigure}
     \hfill
     \begin{subfigure}[b]{0.32\textwidth}
         \centering
         \includegraphics[width=\textwidth]{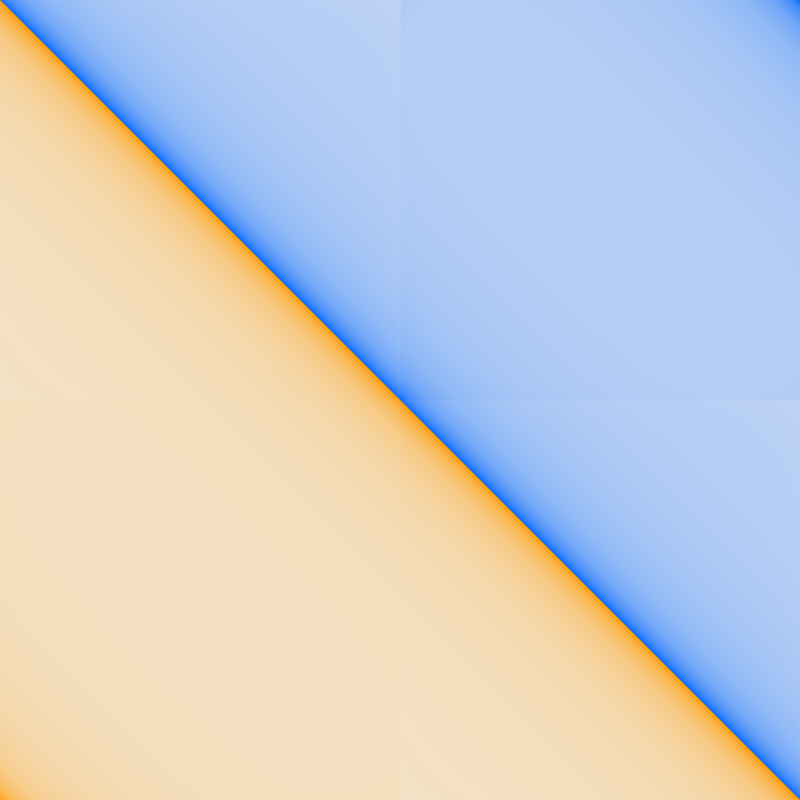}
         \caption{$\beta = 50, w = 0.5$}
     \end{subfigure}
     \hfill
     \begin{subfigure}[b]{0.32\textwidth}
         \centering
         \includegraphics[width=\textwidth]{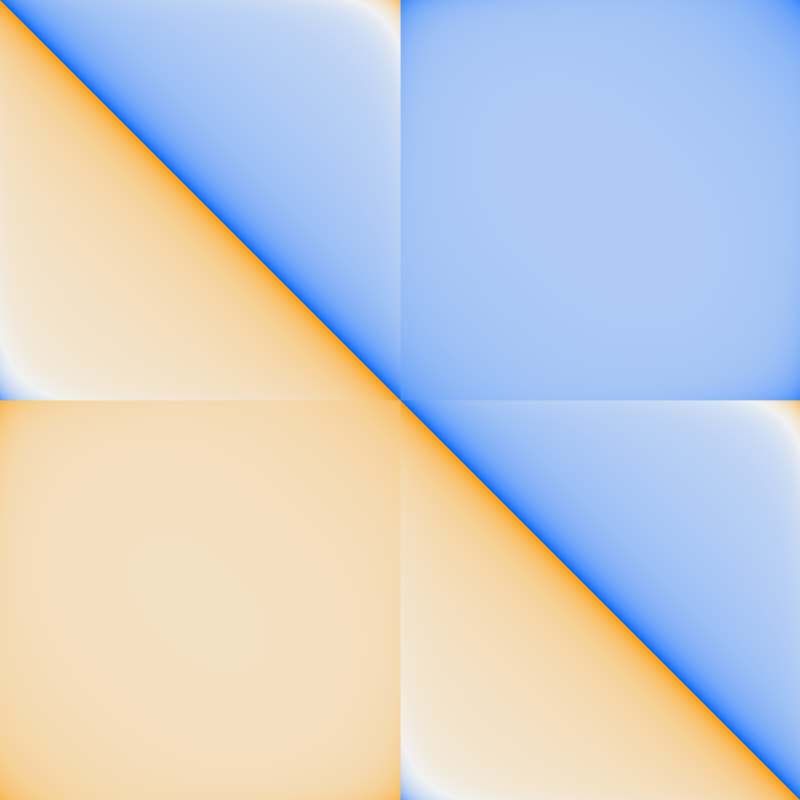}
         \caption{$\beta = 50, w = 1$}
     \end{subfigure}
    \caption{Two-replica saddle point solutions for $G$ with initial value $G_\mathrm{free}$ \eqref{eq:G_free_decomposition}, evaluated for $q = 4$ and $L = 2000$ at $\beta \in \{5, 20, 50\}$ and $w \in \{0.01, 0.5, 1\}$. Of primary significance is that, at large $\beta$ and $w = 0.5$, $G$ converges towards a single-replica saddle of ordinary SYK with doubled inverse temperature $2\beta$.}
    \label{fig:G_R2_saddles}
\end{figure}

We begin our numerical analysis with the case of $R = 2$, where our analysis only yields one (unique) replica structure (unless $w = 0$). Still, there is a vast variety of different possible solutions for $G$ (and $\Sigma$) depending on both $\beta$ and $w$. The matrix structure of some representative solutions of this landscape with (non-diagonal) initial value $G_\mathrm{free}$ \eqref{eq:G_free} are shown for $q = 4$ and $L = 2000$ as heatmap plots in Figure \ref{fig:G_R2_saddles}. Orange indicates a value of $+1/2$, blue a value of $-1/2$, and gray a value of zero\footnote{We point out that a subsequent analytical study revealed that -- while the diagonal replica correlations of the saddle are indeed bounded by $\pm 1/2$ -- the off-diagonal correlations are in fact not, as they instead scale like $(w/(\beta J)^2)^{1/q}$ at high temperatures. Using an improved initial value based on this observation (among others) also leads to a slightly different structure of $G$ as shown in Figure \ref{fig:G_R2_saddles}, especially for $w > 0.5$. However, these observations do not invalidate the general analysis for the given choice of parameters, and will therefore be discussed in more detail in future work.}. There are already multiple interesting observations that can be made with regard to these samples:
\begin{itemize}
    \item The average strength of the cross-replica correlations increases with the relative weight $w$, but actually decreases with increasing inverse temperature $\beta$. In particular, the $\beta = 5$ and $w = 0.01$ case already has comparatively strong correlations, which will only get stronger at higher temperatures. This is because of the existence of $w_\mathrm{crit}$, a certain weight that causes the overall solution to become time-translation invariant across replicas. Its existence is strongly linked to the purity and second R\'enyi entropy of the SYK thermal state, which we discuss in Section \ref{sec:purity}. There we show that $w_\mathrm{crit}$ actually goes towards 0 for small $\beta$.

    \item The solution for $G$ at $\beta = 50$ and $w = 0.5$ also becomes approximately time-translation invariant across replicas, causing it to resemble a single-replica solution of ordinary SYK at $2 \beta$. As is also further explored in Section \ref{sec:purity}, this is a consequence of $w_\mathrm{crit}$ converging towards $0.5$ as $\beta$ grows large.

    \item For larger $\beta$, the resulting cross-replica correlations depend less strongly on the exact value of $w$. In fact, already for $\beta = 20$ and especially for $\beta = 50$ the solutions for $w = 0.01$ are very similar to those obtained by evaluating the non-diagonal initial value directly at $w = 0$. However, saddles obtained this way result in a value of the action that appears to be larger than the dominating diagonal case. We show in Section \ref{sec:w_0_behavior} that this is a consequence of the finite-$L$ numerics, which indicates that the large-$\beta$ and small-$w$ regime needs a more careful analysis in the future.
\end{itemize}

\begin{figure}[h!tb]
    \centering
    \begin{subfigure}[b]{0.49\textwidth}
         \centering
         \includegraphics[width=\textwidth]{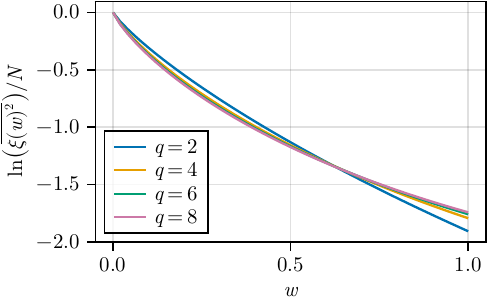}
         \caption{$\beta = 0.5$}
     \end{subfigure}
     \hfill
     \begin{subfigure}[b]{0.49\textwidth}
         \centering
         \includegraphics[width=\textwidth]{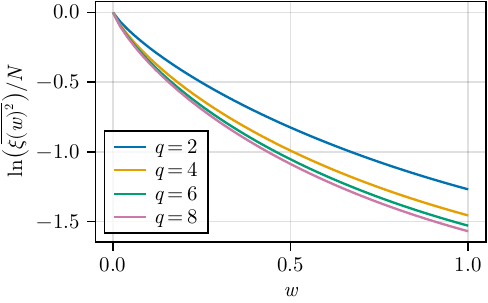}
         \caption{$\beta = 1$}
     \end{subfigure}
     \par\bigskip
     \begin{subfigure}[b]{0.49\textwidth}
         \centering
         \includegraphics[width=\textwidth]{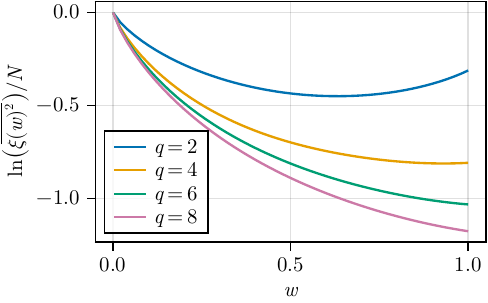}
         \caption{$\beta = 5$}
     \end{subfigure}
     \hfill
     \begin{subfigure}[b]{0.49\textwidth}
         \centering
         \includegraphics[width=\textwidth]{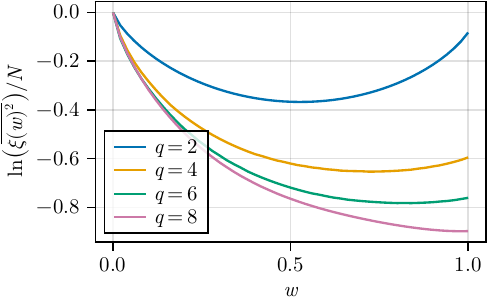}
         \caption{$\beta = 20$}
     \end{subfigure}
    \caption{Logarithm of the disorder-averaged second moments (variance) for $q = 2, 4, 6, 8$ and various representative $\beta$. The $q = 2$ case being symmetric around $w = 0.5$ for larger $\beta$ indicates that $\rho$ becomes a pure state with definite parity in the zero-temperature limit, while for $q > 2$ this does not occur despite the variance rising slightly again.}
    \label{fig:R2_action}
\end{figure}

Using these solutions we are able to evaluate the logarithm of the averaged second moments $\ln(\overline{\xi(w)^2}) / N$ \eqref{eq:averaged_moment_saddle} over a range of relative weights $w$ and for certain representative inverse temperatures $\beta$. The results are shown for $q = 2, 4, 6, 8$ in Figure \ref{fig:R2_action}.

 As expected, there is a significant difference between the behaviors of the Gaussian $q = 2$ case and the others, in particular at low temperatures. There the moments become approximately symmetric around $w = 0.5$ for $q=2$, while only slightly increasing again in the case of $q > 2$. To understand this behavior, it is essential to remind oneself that the case of $w = 1$ corresponds to the insertion of the fermion parity operator $(-1)^F$ \eqref{eq:fermion_parity}. Pure fermionic quantum states $\rho = \ketbra{\psi}$ necessarily have to commute with this operator due to parity superselection rules, which implies that -- like the identity -- it always must appear in the expansion of $\rho$ in terms of Majorana strings. Furthermore, such a state should be invariant under the transformation $\rho \rightarrow (-1)^F \rho$ as it only results in an overall change of phase. But since $(-1)^F$ has the property that it maps any Majorana string to its complement
\begin{equation}
    (-1)^F \majo(a) \propto \majo(a^c),
\end{equation}
this implies that the variance for $w$ and $1 - w$ has to be the same i.e., it has to be symmetric around $w = 0.5$. This argument matches what we observe for $q = 2$ at $\beta = 20$ in Figure \ref{fig:R2_action}, confirming the known fact that the Gaussian case becomes pure at zero temperature \cite{maldacena_remarks_sachdev_ye_2016}. For larger $q$, the variance does eventually increase again, but not enough to become fully symmetric. That is to be expected, as these cases don't converge towards pure states at low temperatures\footnote{This is due to the choice of limits being $N \rightarrow \infty$ and then $T \rightarrow 0$. Doing so \enquote{washes out} the true (unique) finite-$N$ ground state compared to the nearby $\mathcal{O}(e^{N s_0})$ excited states with approximately the same energy, which arise for $q \geq 4$ (for $q = 2$ there are only $\mathcal{O}(1)$ nearby excited states). We revisit this subtlety in the context of the stabilizer R\'enyi entropy discussed in Section \ref{sec:sre}}, meaning that the expectation value for $(-1)^F$ will not be $1$ or $-1$, but something in between and therefore absolutely smaller. It could still happen that the minimum of the variance for $q \geq 4$ will eventually converge towards $w = 0.5$, but the required low-temperature regime can not be probed by our tools as of writing.

\subsubsection*{Fourth Moments}
\label{sec:fourth_moments}

To analyze the different possible replica structures for $R = 4$, we initially restrict ourselves to the cases of $q = 2$ and $q = 4$ as they are representative for the behavior we observed in the case of two replicas. We also note that the $\mathbf{R}_4(+, +)$ structure as listed in Figure \ref{fig:initial_replica_structures} failed to converge in several instances for $q = 4$, even with slower interpolation and modified initial values. And where it did converge, it seemed to exactly match the solution corresponding to $\mathbf{R}_4(+, 0)$ for all $q$, which supports the results of the heuristic analysis of initial values described in Section \ref{sec:initial_value_classification}.

\begin{figure}[h!tb]
    \centering
    \includegraphics[width=0.32\linewidth]{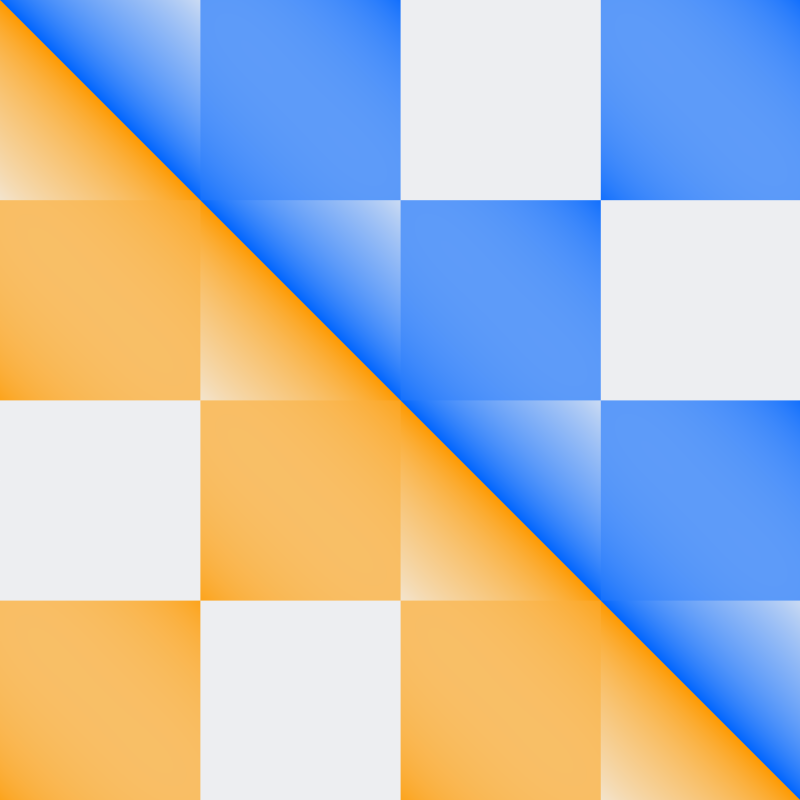}
    \caption{Example saddle for $G$ with 1R4 replica structure, evaluated  for $q = 4$ at $\beta = 5$ and $w = 0.5$. Its structure resembles four pairwise overlapping copies of 1R2 with antiperiodic boundary conditions. Despite this similarity, the actual entries differ noticeably from e.g.\ those of the 2R2 solution evaluated with the same parameters as shown in Figure \ref{fig:2R2_permutations}. } 
    \label{fig:1R4_example}
\end{figure}

\begin{figure}[h!tb]
    \centering
    \begin{subfigure}[b]{0.32\textwidth}
        \includegraphics[width=\linewidth]{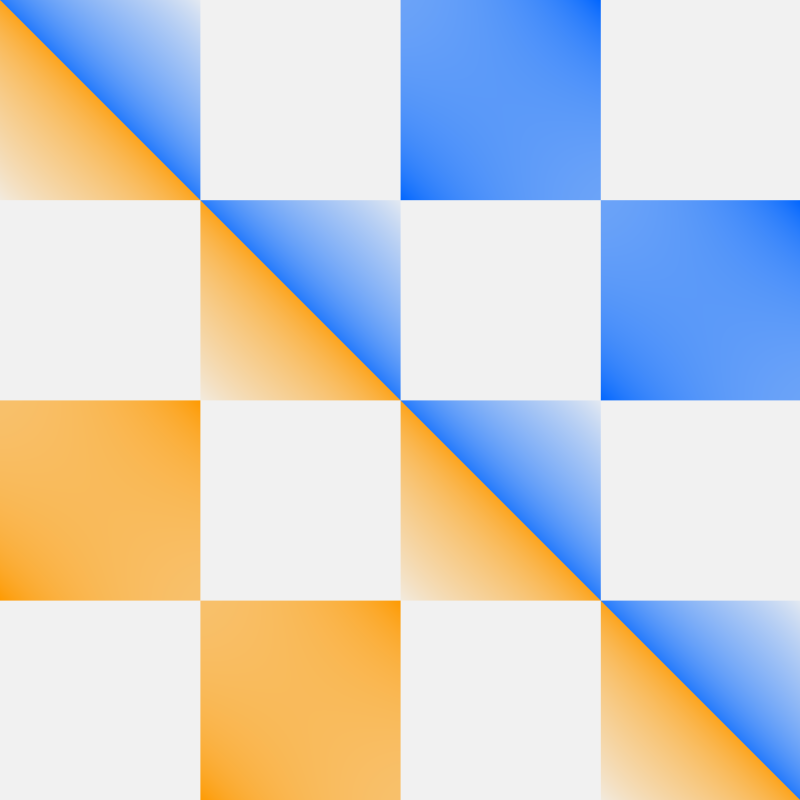}
    \end{subfigure}
    \hfill
    \begin{subfigure}[b]{0.32\textwidth}
        \includegraphics[width=\linewidth]{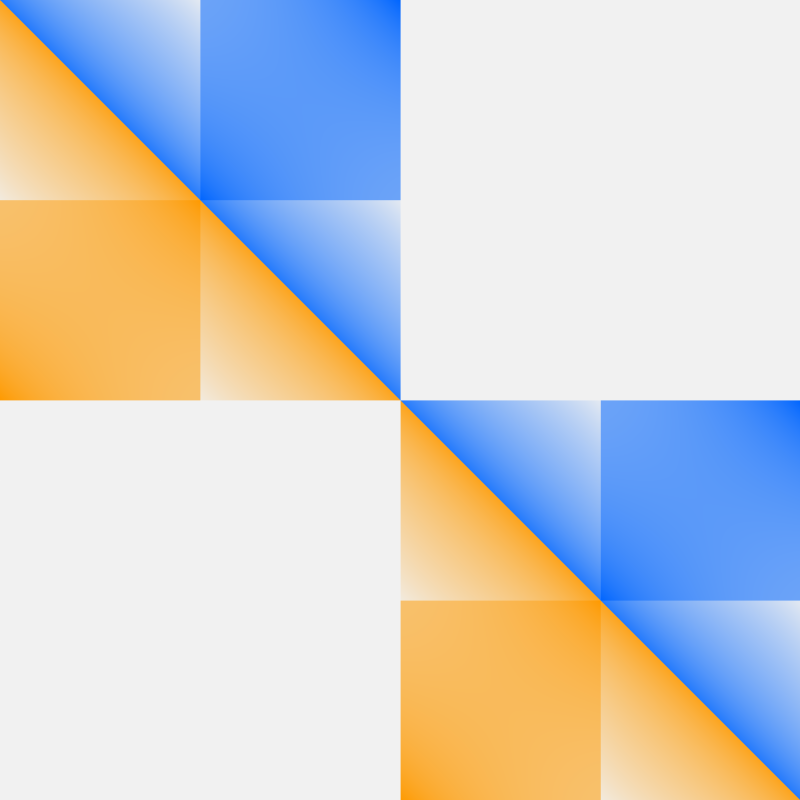}
    \end{subfigure}
    \hfill
    \begin{subfigure}[b]{0.32\textwidth}
        \includegraphics[width=\linewidth]{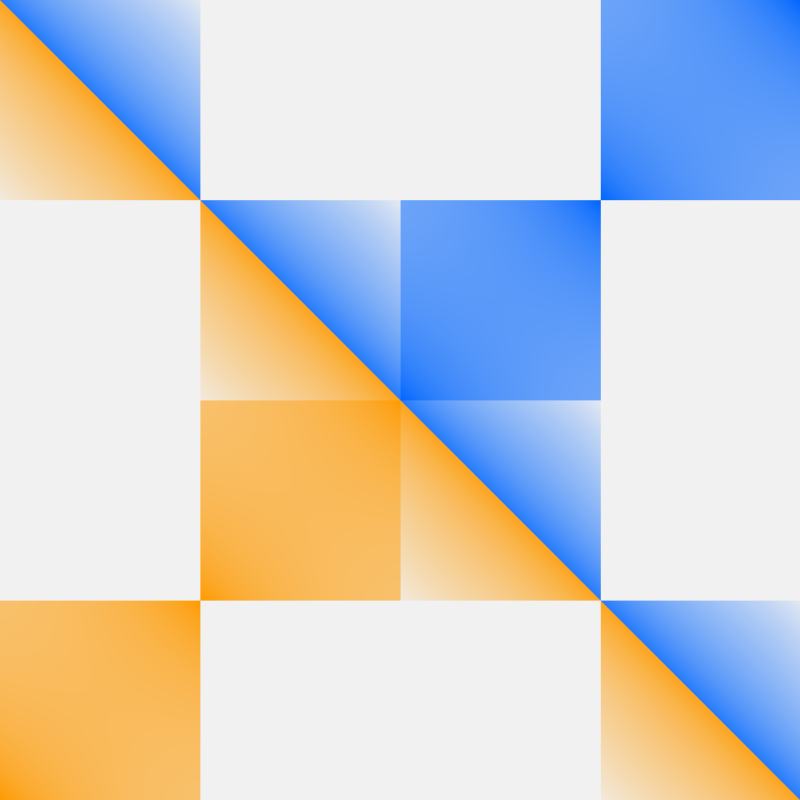}
    \end{subfigure}
    \caption{All three distinct $R=4$ replica saddle points for $G$ which are equivalent to two copies of the $R = 2$ solution up to permutations and therefore result in the same value for $\action^{(4, w)}_\mathrm{SYK}$ (here evaluated with the unbiased matrix solver at $\beta = 5$, $w = 0.5$, $q = 4$ and $L = 1000$). Only the first option has anticirculant replica symmetry. The number of options matches the expected kurtosis of a Gaussian random variable.}
    \label{fig:2R2_permutations}
\end{figure}

The only possible replica structures satisfying our assumptions therefore seem to be $\mathbf{R}_4(0, +)$ and $\mathbf{R}_4(+, 0)$, which we refer to as 2R2 (two two-replica blocks) and 1R4 (one four-replica block, see Figure \ref{fig:1R4_example}) in alignment with Figure \ref{fig:replica_structures}. The reason for the former name is that it is possible to block-diagonalize $\mathbf{R}_4(0, +)$ into two blocks of the (unique) two-replica solution by applying action-preserving permutations. In fact, Figure \ref{fig:2R2_permutations} shows that there are exactly three distinct ways to depict this replica structure (up to signs), which correspond to the three different ways one can Wick-contract the four-replica correlation function in the case that $\xi(w)$ is (at least approximately) described by a Gaussian random variable (this is also referred to as the kurtosis of the probability distribution).

\begin{figure}[h!tb]
    \centering
    \begin{subfigure}[b]{0.49\textwidth}
         \centering
         \includegraphics[width=\textwidth]{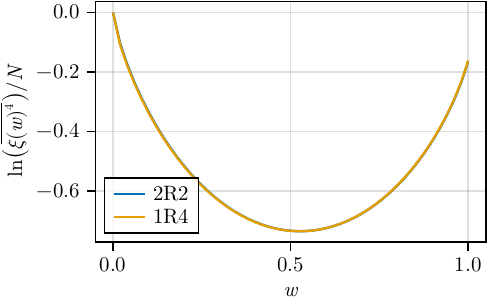}
         \caption{$q = 2$}
     \end{subfigure}
     \hfill
     \begin{subfigure}[b]{0.49\textwidth}
         \centering
         \includegraphics[width=\textwidth]{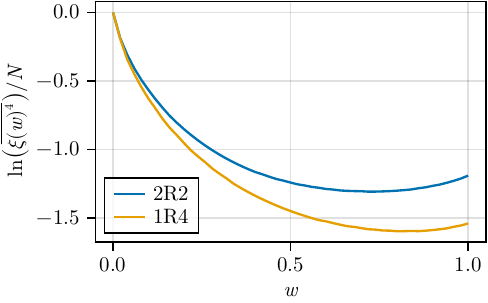}
         \caption{$q = 4$}
     \end{subfigure}
    \caption{Comparison of the fourth moments $\ln(\overline{\xi(w)^4}) / N$ corresponding to the two well-behaved replica structures 2R2 and 1R4 when evaluated at $\beta = 20$. While for $q = 2$ both structures give the same result, for $q = 4$ (and higher) the 2R2 clearly dominates everywhere. This implies that the moments are not Gaussian in the Gaussian $q = 2$ case, and (approximately) Gaussian in the non-Gaussian $q > 2$ cases.}
    \label{fig:R4_comparison}
\end{figure}

In Figure \ref{fig:R4_comparison} we show the logarithm of the averaged fourth moments $\ln (\overline{\xi(w)^4}) / N$ for $q = 2, 4$ at $\beta = 20$ and with both possible replica structures. What immediately becomes apparent is that for $q = 2$ both structures appear to result in the same values for the moment while for $q = 4$ the values are different with the 2R2 solution clearly dominating. The latter behavior (which is the same for other values of $\beta$) implies that for $q > 2$ $\xi(w)$ is indeed well-approximated by a Gaussian random variable due to the same-sign permutations of 2R2 matching the expected kurtosis.

But for the case of $q = 2$ things are more subtle. This is because there the action actually has the aforementioned $\Ortho(4)$ replica symmetry (see Section \ref{sec:permutations}), meaning the saddles are equivalent up to such an orthogonal transformation. This also implies that $\xi(w)$ can not be Gaussian-distributed as the kurtosis depends on the geometry of the exact isometry group. Previous work employing finite-$N$ analysis found the moments to obey a Laplace distribution \cite{bera_non_stabilizerness_sachdev_2025}, but we defer an infinite-$N$ derivation of this result to future work as it would require a different set of tools.

Overall, because the fourth moments are therefore dominated (or accurately represented in the case of $q = 2$) by contributions from the 2R2 saddle point, the resulting (logarithmic)  moments will match those shown in Figure \ref{fig:R2_action}, albeit with an additional factor of 2.

\subsection{Behavior at \textit{w} = 0}
\label{sec:w_0_behavior}

\begin{figure}[h!tb]
    \centering
    \begin{subfigure}[b]{0.49\textwidth}
         \centering
         \includegraphics[width=\textwidth]{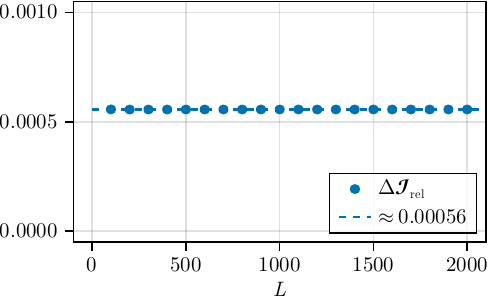}
         \caption{$q = 2$}
     \end{subfigure}
     \hfill
     \begin{subfigure}[b]{0.49\textwidth}
         \centering
         \includegraphics[width=\textwidth]{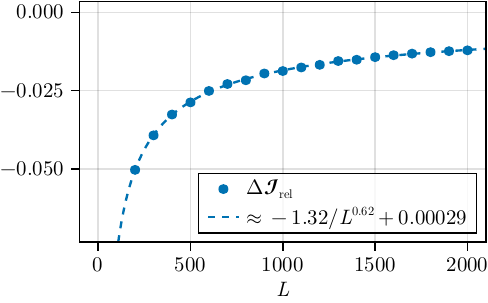}
         \caption{$q = 4$}
     \end{subfigure}
    \caption{$L$-dependence of the relative difference $\Delta \action_\mathrm{rel}$ \eqref{eq:relative_action_difference} between actions corresponding to saddles retrieved from initial values with 2R1 and 1R2 replica structure at $w = 0$ and $\beta = 20$. While in the $q = 4$ case a fit indicates that the difference decreases with $L$ and will therefore vanish in the continuous limit (which is to be expected since $I_\mathrm{1R2}$ appears to be larger than $I_\mathrm{2R1}$), for $q = 2$ the difference is constant and therefore not vanishing.}
    \label{fig:jump_L_dependence}
\end{figure}

One subtlety in the analysis of higher moments -- which does not directly become apparent from the results shown so far -- is that the decision to impose different replica structures at $w = 0$ and $w > 0$ leads to an apparent discontinuity in the moments around that point. This behavior was already observed in the case of $R = 1$ in Section \ref{sec:single_replica_case}, where it arises due to $\overline{\xi(a)}$ vanishing unless $a = 0$ (in which case $\xi(a) = 1$ even before averaging). But for the higher moments displayed in this section, no jump is visible at $w = 0$ due to the limited resolution of the data shown in the figures. It does however become pronounced for larger $\beta$ and higher moments, in particular when also accounting for combinatorial contributions that arise from summing over moments (see Section \ref{sec:sre}). Albeit hidden, we can still analyze the jump in the data in the case of $R = 2$ by considering the relative difference
\begin{equation}
\label{eq:relative_action_difference}
    \Delta \action_\mathrm{rel} = \frac{\action_\mathrm{2R1} - \action_\mathrm{1R2}}{\action_\mathrm{2R1}}
\end{equation}
between the action $\action_\mathrm{2R1}$ of the replica-diagonal (2R1) solution and the action  $\action_\mathrm{1R2}$ of the non-diagonal (1R2) solution, both evaluated for different discretization sizes $L$ at $w = 0$ and $\beta = 20$. The reasoning is that $\action_\mathrm{1R2}$ evaluated at $w = 0$ should roughly match $\action_\mathrm{2R1}$ at $w = 0^+$ for the overall distribution of moments to be continuous there. If that is indeed the case, then the relative difference $\Delta\action_\mathrm{rel}$ should vanish in the continuum limit of $L \rightarrow \infty$. The results of this analysis for $q = 2$\footnote{Note that $1R2$ can not be mapped to $2R1$ under an orthogonal transformation, meaning both replica structures must lead to distinct results even for $q = 2$.} and $q = 4$ are depicted in Figure \ref{fig:jump_L_dependence}.

Let us focus on the case of $q = 4$ first: An important thing to note is that $\Delta \action_\mathrm{rel}$ is actually \emph{negative}, meaning that $-\action_\mathrm{1R2}$ is larger than $-\action_\mathrm{2R1}$. This contradicts the well-known fact that the partition function of ordinary SYK is self-averaging, and so the observed behavior is likely not physical and simply due to finite-$L$ artifacts. This is further supported by the well-fitted trend that $\Delta \action_\mathrm{rel}$ takes as $L$ increases, which seems to tend towards (approximately) zero. In the continuous limit both possible initial values are therefore expected to result in the replica-diagonal solution, meaning that the distribution of moments should be continuous, but possibly with an infinite derivative at $w = 0$. This is also supported by predictions made in the wormhole picture discussed in Section \ref{sec:wormholes}, where the action takes on the approximately entropy-like form $c_1 w \ln(c_2 w)$ \eqref{eq:wormhole_action_prediction}. However, without a more careful analysis at much larger $L$ it is not possible to make more precise statements about this regime.

For $q = 2$, the behavior of $\Delta \action_\mathrm{rel}$ is considerably different: Even though it is still small compared to the action itself, its (positive) value does not depend strongly on $L$ at all. The implication is that the non-zero difference between the two actions is not just a consequence of the finite-$L$ numerics, but a manifest property of the setup. While this does seem to suggest that the moments are indeed discontinuous at $w = 0$, we won't jump to any conclusions here and instead defer to a future in-depth analysis of the $q = 2$ setup as a whole. We do however point out that this observation could be explained in the context of filtered Stabilizer R\'enyi entropies, which are touched on in Section \ref{sec:sre}.

%% file: sections/applications.tex
\section{Applications}
\label{sec:applications}

We are now in a position to apply the toolset developed in the previous two sections to a wide variety of different use cases, of which we consider only a limited selection here. Our analyses are far from exhaustive, and we note a few topics which could reward further study in the outlook in Section \ref{sec:outlook}. 

\subsection{Summing Over Moments}
\label{sec:summing_over_moments}

A technical tool that many results in this section depend on is how to evaluate (disorder-averaged) sums over all distinct absolute statistical moments (also called the $\ell_R$ string-norm):
\begin{equation}
\label{eq:string_R_norm}
    F_R(\rho) \coloneqq \sum_a |\xi(a)|^R.
\end{equation}
However, in the large-$N$ limit we inevitably lose control over selecting individual strings $\majo(a)$ to sum over, as there are $2^N$ such options. Fortunately, we showed in Section \ref{sec:path_integral_derivation} that the saddle associated to each string only depends on the relative weight $w$ of said string, so the same saddle occurs $\binom{N}{w N}$ times in the overall sum. Assuming that $0 < w < 1$, we can use the Stirling approximation to express the binomial coefficient at large $N$ as
\begin{equation}
    \lim_{N \rightarrow \infty} \frac{1}{N} \ln \binom{N}{w N} =   H(w),
\end{equation}
where $H(w)$ is the binary Shannon entropy associated to $w$:
\begin{equation}
    H(w) = - w \ln (w) - (1-w) \ln(1-w).
\end{equation}
Since $w \in \{0,1/N,2/N, \cdots,1\}$ is effectively continuous at large $N$, we can therefore write any averaged sum over \emph{even}\footnote{See Section \ref{sec:robustness} for how to approximate the absolutely dominating term in a sum of odd moments.} moments as an integral over saddle points corresponding to individual $w$, such that
\begin{equation}
    \overline{\sum_{a} \tr \left(\majo(a) \, e^{-\beta H_\mathrm{SYK}}\right)^{R}} \approx \int_0^1 dw \, \exp\left(N \left[H(w) - \action_\mathrm{SYK}^{(R, w)}\right]\right),
\end{equation}
where $\action_\mathrm{SYK}^{(R, w)}$ is the action evaluated at said saddle point. This expression can however still not be evaluated in a straightforward manner since $N$ can't be factorized out when taking the overall logarithm. But as usual when $N$ is large, we expect the dominating regime of this integral to be the one where the exponent has an extremum, meaning\footnote{A priori, one would have to use a total derivative here since the saddle point fields $G$ and $\Sigma$ also implicitly depend on $w$. But applying the chain rule shows that the corresponding contributions to the derivative vanish due to their defining property of extremizing the action.}:
\begin{equation}
    \frac{\partial}{\partial w} \left( H(w) - \action_\mathrm{SYK}^{(R, w)} \right) = 0,
\end{equation}
for some critical weight $w = w_\mathrm{crit}$. This equation can be easily solved for $w_\mathrm{crit}$, yielding
\begin{equation}
\label{eq:w_crit}
    w_\mathrm{crit} = \frac{\pf[I_R \otimes \Partial^+ - \Sigma]}{\pf[I_R \otimes \Partial^+ - \Sigma] + \pf[I_R \otimes \Partial^- - \Sigma]}.
\end{equation} 
This value for the relative weight can then be inserted into the Schwinger-Dyson equations \eqref{eq:syk_w_schwinger_dyson} to yield a new set of equations independent of $w$, and which -- when solved iteratively -- have the potential to directly result in the saddle dominating the overall sum
\begin{align}
\begin{split}
    \overline{\sum_a \tr \left(\majo(a) \, e^{-\beta H_\mathrm{SYK}}\right)^{R}} &\approx \exp\left(N \left[H(w_\mathrm{crit}) - \action_\mathrm{SYK}^{(R, w_\mathrm{crit})} \right]\right) \\
    &\equiv \exp\left(-N \action_\mathrm{crit}^{(R)}\right),
\end{split}
\end{align}
where the effective resummed action can be written as
\begin{align}
\begin{split}
\label{eq:resummed_action}
    \action^{(R)}_\mathrm{crit}[G, \Sigma] ={}& - \ln \left( \pf[I_R \otimes \Partial^- - 2 \, \Sigma] + \pf[I_R \otimes \Partial^+ - 2 \, \Sigma] \right) \\
    & + \frac{1}{2} \sum_{r,s=1}^{R} \int_0^{\beta} d\tau d\tau' \left(\Sigma_{rs}(\tau, \tau') \, G_{rs}(\tau, \tau') - \frac{J^2}{q} \, G_{rs}(\tau, \tau')^q \right).
\end{split}
\end{align}
In fact, one can arrive at this very action and the associated equations directly by rewriting the initial sum over operator moments in \eqref{eq:string_R_norm} as a product of identity and replica parity operators \eqref{eq:replica_parity_operator} (ignoring signs that can be absorbed into the path integral measure):
\begin{equation}
    \sum_{a} \majo(a)^{\otimes_f R} \equiv \prod_{i=1}^N \left(I + (-1)^{F_{R,i}}\right).
\end{equation}
When the path integral factorizes into $N$ mode integrals after disorder-averaging such as in \eqref{eq:syk_path_integral_decoupled}, each term in that product is therefore only associated to a single mode integral and evaluates to $\pf(I_R \otimes \Partial^- - \Sigma) + \pf(I_R \otimes \Partial^+ - \Sigma)$, ultimately leading to \eqref{eq:resummed_action}.

While in theory this resummed approach can directly lead to the dominating saddle point contribution of $F_R(\rho)$ (and in fact was used in \cite{zhang_stabilizer_renyi_entropy_2026} to study the stabilizer R\'enyi entropy discussed in Section \ref{sec:sre}), it does not allow one to analyze subleading moments in the sum. Furthermore, the analysis in Section \ref{sec:single_replica_case} suggests that the presence of determinants in the associated Schwinger-Dyson equations has the potential to impact the numerical stability of the iterative solver due to considerable finite-$L$ effects. These issues can potentially be resolved with a more careful approach, but in this work we therefore restrict ourselves to considering saddles with fixed relative weight $w$. One can still retrieve the solution at $w_\mathrm{crit}$ by either determining the maximum of $H(w) - \action^{(R, w)}_\mathrm{SYK}$ or by finding the root of $w_\mathrm{crit}(w) - w$ (since \eqref{eq:w_crit} is a well-defined quantity that can be evaluated for all $w$). While we restrict ourselves to the former approach (by using golden-section search), we show evidence of how these two approaches lead to the same critical weight in Sections \ref{sec:purity} and \ref{sec:sre}.

\subsection{Second R\'enyi Entropy}
\label{sec:purity}

\begin{figure}[h!tb]
    \centering
    \begin{subfigure}[b]{0.49\textwidth}
         \centering
         \includegraphics[width=\textwidth]{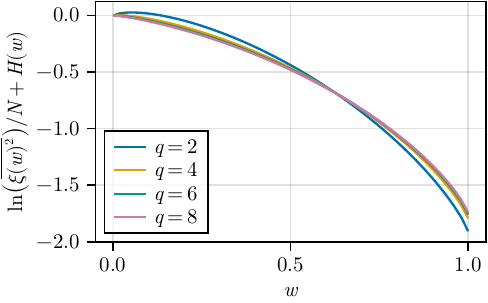}
         \caption{$\beta = 0.5$}
     \end{subfigure}
     \hfill
     \begin{subfigure}[b]{0.49\textwidth}
         \centering
         \includegraphics[width=\textwidth]{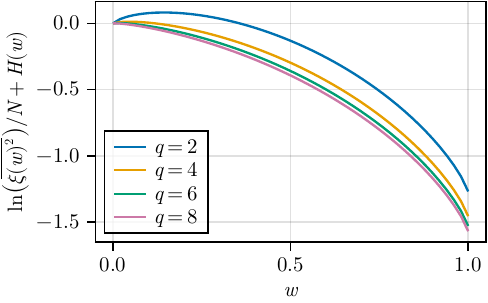}
         \caption{$\beta = 1$}
     \end{subfigure}
     \par\bigskip
     \begin{subfigure}[b]{0.49\textwidth}
         \centering
         \includegraphics[width=\textwidth]{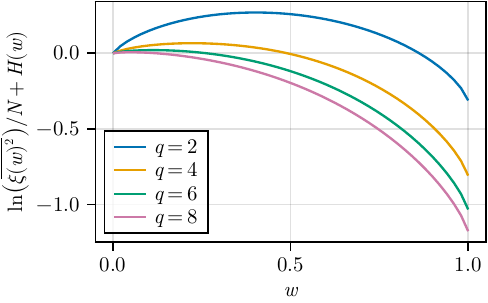}
         \caption{$\beta = 5$}
     \end{subfigure}
     \hfill
     \begin{subfigure}[b]{0.49\textwidth}
         \centering
         \includegraphics[width=\textwidth]{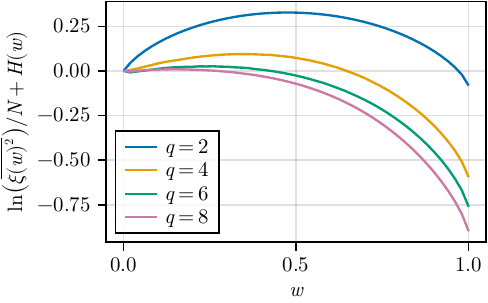}
         \caption{$\beta = 20$}
     \end{subfigure}
    \caption{Combinatorial contributions to $\ln(\overline{\xi(w)^2}) / N$ as a consequence of evaluating it for all Majorana strings with the same relative operator weight $w$. The resulting addition of the binary entropy $H(w)$ causes the maximum of the distribution to be shifted from $w = 0$ to $0 < w_\mathrm{crit} < 1/2$, with smaller $q$ having their maxima closer to $w = 1/2$ at large $\beta$.}
    \label{fig:R2_action_entropy}
\end{figure}

As alluded to before in Section \ref{sec:purity_correlations}, there is a direct connection \eqref{eq:purity_connection} between the sum over second moments of a general quantum state and its purity. Written in terms of the second R\'enyi entropy $S_2(\rho) = -\ln(\tr(\rho^2))$, this is equivalent to
\begin{equation}
    -\ln(\tr(\rho^2)) = \ln(D) - \ln\left( \sum_a \xi(a)^2 \right).
\end{equation}
Evaluating both sides for the SYK Gibbs state using the saddle point approximation for the sum over moments derived as in Section \ref{sec:summing_over_moments}, one finds that for this to be true one has to have
\begin{equation}
    \ln(\overline{Z(2\beta)})/N = \max_{w \in [0,1]}\left( H(w) - \action_\mathrm{SYK}^{(2, w)}(\beta) \right) - \ln(2)/2,
\end{equation}
where $\ln(\overline{Z(2\beta)})/N \equiv - \action_\mathrm{SYK}^{(1, 0)}(2\beta)$ is the partition function of ordinary SYK at inverse temperature $2\beta$. Since that quantity can be evaluated with high accuracy by imposing time-translation invariance, this setup allows us to probe the accuracy of the calculation required to evaluate the right-hand side of that equation, which can then be extrapolated to other applications where a direct comparison is not feasible.

Before we do so however, it is instructive to consider the shape of the distribution $H(w) - \action_\mathrm{SYK}^{(2, w)}$ and its connection to $w_\mathrm{crit}(w)$ \eqref{eq:w_crit} for individual representative values of $\beta$. The results for the former are depicted in Figure \ref{fig:R2_action_entropy} and nicely illustrate how the addition of $H(w)$ shifts the dominating moment from $w = 0$ (as in Figure \ref{fig:R2_action}) to some non-zero weight $w_\mathrm{crit}$ depending on $q$ and $\beta$. As a general rule of thumb, it appears that for smaller $q$ said moment moves faster and closer towards $w = 0.5$ than for larger $q$. In particular for $q = 2$, this aligns with the arguments made in Section \ref{sec:second_moments} about its low-temperature behavior. For $q \geq 4$ it is less clear however if the sum over second moments will also attain a maximum there at sufficiently low temperatures. Probing the relevant regimes requires a numerical precision that the tools at hand are currently unable to achieve in an efficient manner.

\begin{figure}[h!tb]
    \centering
    \begin{subfigure}[b]{0.49\textwidth}
         \centering
         \includegraphics[width=\textwidth]{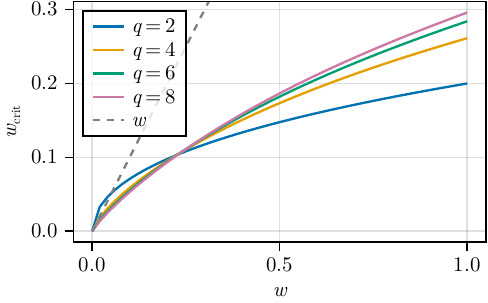}
         \caption{$\beta = 0.5$}
     \end{subfigure}
     \hfill
     \begin{subfigure}[b]{0.49\textwidth}
         \centering
         \includegraphics[width=\textwidth]{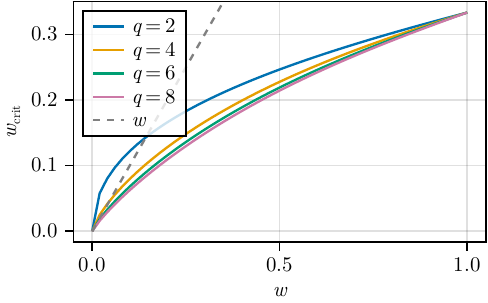}
         \caption{$\beta = 1$}
     \end{subfigure}
     \par\bigskip
     \begin{subfigure}[b]{0.49\textwidth}
         \centering
         \includegraphics[width=\textwidth]{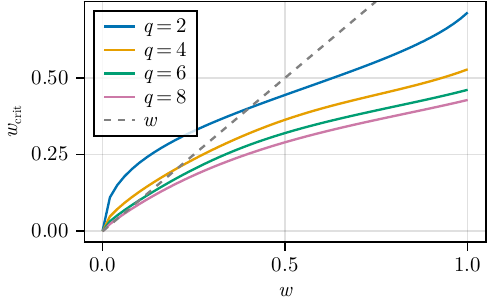}
         \caption{$\beta = 5$}
     \end{subfigure}
     \hfill
     \begin{subfigure}[b]{0.49\textwidth}
         \centering
         \includegraphics[width=\textwidth]{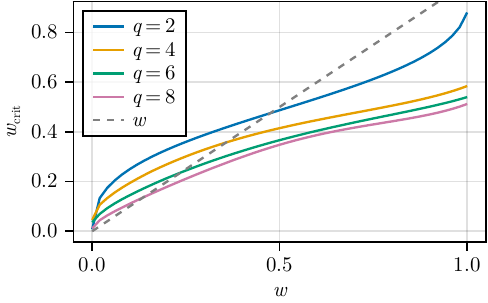}
         \caption{$\beta = 20$}
     \end{subfigure}
    \caption{$w_\mathrm{crit}$ for the case of $R = 2$ evaluated for representative values of $\beta$. The points where $w_\mathrm{crit} = w > 0$ coincide perfectly with the maxima observed in Figure \ref{fig:R2_action_entropy}. However, $w_\mathrm{crit} = 0 = w$ seems to indicate the presence of another extremum, which does not seem to appear in the other figure. This suggests the presence of a discontinuity or an infinite derivative at that point.}
    \label{fig:R2_w_crit}
\end{figure}
 Comparing the maxima in Figure \ref{fig:R2_action_entropy} with the intersections of $w$ and $w_\mathrm{crit}$ in Figure \ref{fig:R2_w_crit} shows that there is almost a perfect one-to-one correspondence, as previously predicted. There is however a slight subtlety to this argument: $w_\mathrm{crit}$ should vanish at $w = 0$ (for the same reasons that are outlined in Section \ref{sec:single_replica_case}), and therefore one also has $w_\mathrm{crit} = w$ there. This does not align with there seemingly only ever being a single extremum visible for each $q$ in Figure \ref{fig:R2_action_entropy}, but does further support the existence of either an infinite derivative or a discontinuity at $w = 0$, as previously discussed in Section \ref{sec:w_0_behavior}. But because this extremum never appears to correspond to the relevant global maximum of $H(w) - \action_\mathrm{SYK}^{(2, w)}$ (except at high/infinite temperature), it can be ignored in this context.

\begin{figure}[h!tb]
    \centering
    \includegraphics[width=0.8\linewidth]{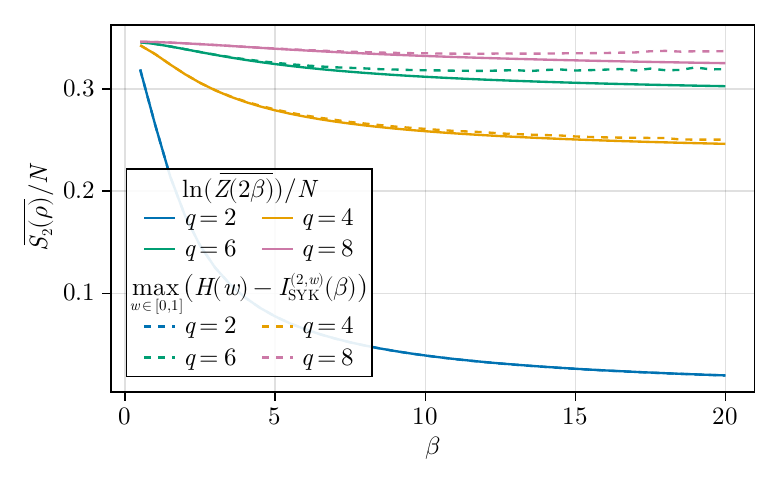}
    \caption{Second R\'enyi entropy derived from both computing $\ln(\overline{Z(2\beta)})/N$ directly (using the time-translation invariant approach with $L_\mathrm{Fourier}=1000000$) and maximizing over $H(w) - \action^{(2,w)}_\mathrm{SYK}(\beta)$ (using the matrix-based approach with $L = 1000$). Both approaches match almost perfectly in the case of $q = 2$, but differ slightly for larger $q$. The relative errors at $\beta = 20$ are $\epsilon_2 = 1.38\%$, $\epsilon_4 = 2.41\%$, $\epsilon_6 = 5.62\%$ and $\epsilon_8 = 3.63\%$. These errors increasing with $\beta$ suggests that they are likely numerical artifacts.}
    \label{fig:purity_renyi}
\end{figure}

To numerically determine the maximum of $H(w) - \action_\mathrm{SYK}^{(2, w)}$ with respect to $w$, we used golden-section search over the interval $[0, 0.5]$ with $\approx 10^{-3}$ accuracy. Each required evaluation of the moments was then performed using the anticirculant solving approach with the usual parameters . To determine $\overline{Z(\beta)}$ and $\overline{Z(2\beta)}$ as accurately as possible, the standard time-translation invariant solving approach \cite{maldacena_remarks_sachdev_ye_2016, bentsen_approximate_quantum_codes_2024} ($L_\mathrm{Fourier} = 1000000$) was employed instead. The two possible ways to compute the R\'enyi entropies $\overline{S_2(\rho)}/N$ are compared for various $q$ in Figure \ref{fig:purity_renyi}.

What we find is that for $q = 2$ both approaches lead to almost identical results for all probed $\beta$, while for $q \geq 4$ there are small but substantial deviations at lower temperatures, which only seem to increase as $q$ grows larger. These errors are almost certainly of a purely numerical nature, and might be related to the iterative nature of solving the Schwinger-Dyson equations \eqref{eq:syk_w_schwinger_dyson_discrete}: Starting with initial cross-replica correlations of $\pm 1/2$ everywhere in $G$ means that taking them to the power of $q - 1$ causes them to decrease significantly when computing the first iteration of $\Sigma$. But since we showed in Section \ref{sec:single_replica_case}  that the lack of zero-mode correlations results in a solution that is highly singular, the initial correlations only having a zero-mode has the potential to lead to significant initial deviations that are only amplified by $\Delta\tau$ growing larger. This would also explain why it does not arise in the Gaussian case, as there this initial decrease in the correlations does not occur.

Overall, the implication is that ideally the discretization size $L$ has to be much larger than $1000$ to be able to probe regimes beyond $\beta = 20$ accurately. This would come with a significant increase in computation time though, and so we leave this for potential future work. There are also other ideas on how the numerics could be made more stable in the future, which we touch on in Section \ref{sec:outlook}. Regardless, this biased data still allows us to make qualitative predications about the setup, even at larger $\beta$.

\subsection{Robustness of Magic}
\label{sec:robustness}

We have just seen that our results for $\overline{\xi(a)^2}$ give a value for the purity that is consistent with direct calculations. We now turn to various measures of magic that can also be related to the moment data. The first measure we consider is known as the robustness of magic.

For an $N$-fermion state $\rho$, we define the fermionic robustness of magic by analogy with its qubit counterpart as
\begin{equation}
    \mathcal{R}(\rho) = \min\left\{ \sum_i |c_i| \, \Bigg\vert \, \rho = \sum_i c_i \, \sigma_i, \, \sigma_i \in \mathrm{STAB} \right\},
\end{equation}
where $\mathrm{STAB}$ is the set of density matrices composed of mixtures of pure fermionic stabilizer states. For general states, the minimizing coefficients $c_i$ will not be positive, and if they are all positive then the state $\rho$ is in $\mathrm{STAB}$ and therefore must have $\mathcal{R}(\rho) = 1$. 

Recalling the $\ell_1$ string norm as defined in \eqref{eq:string_R_norm}:
\begin{equation}
    F_1(\rho) = \sum_a |\tr(\rho \majo(a))|,
\end{equation}
it follows that any pure stabilizer state has $F_1(|\psi\rangle\langle \psi|) = D$, and by convexity any state in $\mathrm{STAB}$ must therefore have $F_1(\rho) \leq D$. Our interest in $F_1(\rho)$ is due to its computation not requiring a convex optimization, but it still being able to lower-bound $\mathcal{R}(\rho)$:
\begin{equation}
    \mathcal{R}(\rho) \geq \frac{F_1(\rho)}{D}.
\end{equation}
This is proven as follows: Let $\{ c_i^\star , \sigma_i^\star\}_i$ be an optimal decomposition of $\rho$ with respect to its robustness $\mathcal{R}(\rho)$. We compute 
\begin{equation}
    F_1(\rho) = \sum_a |\tr(\rho \majo(a))| = \sum_a \left| \sum_i c_i^\star \tr( \sigma_i^\star \majo(a)) \right| \leq D \, \sum_i |c_i^\star | = D \, \mathcal{R}(\rho),
\end{equation}
where the middle inequality follows from $| c_i^\star \tr(\sigma_i^\star \mu) | = | c_i^\star || \tr(\sigma_i^\star \mu) | $ and $\sum_\mu | \tr(\sigma_i^\star \mu) | \leq D$.

We cannot directly carry out the convex optimization in the definition of $\mathcal{R}(\rho)$, but we can easily compute the ensemble average of $F_1(\rho)$ for the SYK thermal state using similar tools as those described in Section \ref{sec:summing_over_moments}. For this calculation, we use our findings that (for $q \geq 4$) the Majorana string expectation value $\xi(a) = \tr(\rho \majo(a))$ is a Gaussian random variable with mean zero and variance $e^{- 2 N \phi(w)}$ (with $\phi(w)$ as in \eqref{eq:phi_w}), which depends only on the relative weight $w$ of the string $\majo(a)$. Using standard properties of such normal-distributed quantities, we have
\begin{equation}
    \overline{|\xi(a)|} = \sqrt{\frac{2}{\pi}} \, e^{- N \phi(w)}.
\end{equation}
Hence, the ensemble average of $F_1(\rho)$ can be approximated by the saddle point
\begin{equation}
    \overline{F_1(\rho)} = \sqrt{\frac{2}{\pi}} \,  \sum_w \binom{N}{wN} \, e^{- N \phi(w)} \approx \sqrt{\frac{2}{\pi}} \,  \int_0^1 dw \, e^{N [H(w) -\phi(w)]},
\end{equation}
which can be numerically evaluated by finding the extremum of $H(w) - \phi(w)$.

Note that -- even without the Gaussian assumption -- our knowledge of the second and fourth moments of $\xi(a)$ suffices to give a lower bound on $\overline{F_1(\rho)}$ of the same order, differing only in the $\mathcal{O}(1)$ numerical coefficient.

\begin{figure}[h!tb]
    \centering
    \includegraphics[width=0.8\linewidth]{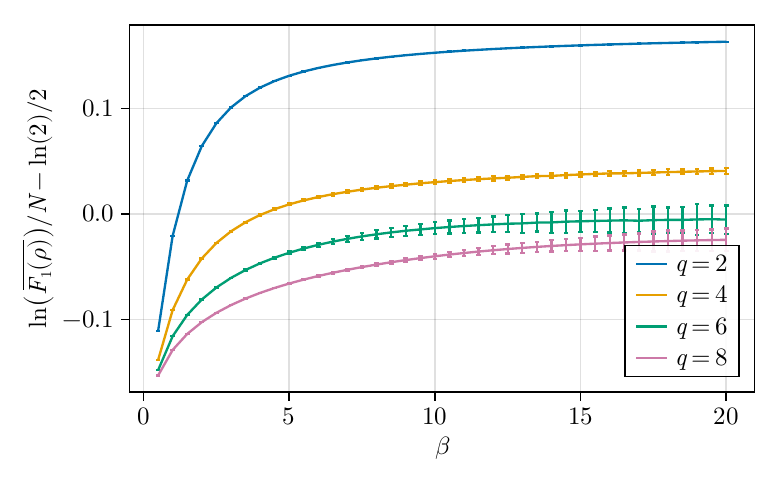}
    \caption{Logarithm of the robustness of magic lower bound versus inverse temperature for $q=2$ through $8$. The error bars indicate numerical uncertainty estimates based on the observed deviations in the second R\'enyi entropy shown in Figure \ref{fig:purity_renyi}.}
    \label{fig:norm}
\end{figure}

In Figure \ref{fig:norm} we depict $\ln(\overline{F_1(\rho)})/N - \ln(2)/2$ for different $q$s as a function of $\beta$, the normalization by $N$ allowing us to drop the aforementioned subleading numerical prefactors that differ between $q = 2$ and $q \geq 4$. The error bars are an estimate of the uncertainty due to finite-$L$ errors and other numerical artifacts, and were obtained by repurposing the error in the purity/second R\'enyi entropy shown in Figure \ref{fig:purity_renyi} to estimate the error in the action. It is clear that both $q=2$ and $q=4$ have exponentially large robustness of magic at sufficiently low temperature. Larger $q$s do not reliably reach positive values in this range of temperatures.

\begin{figure}[h!tb]
    \centering
    \includegraphics[width=0.8\linewidth]{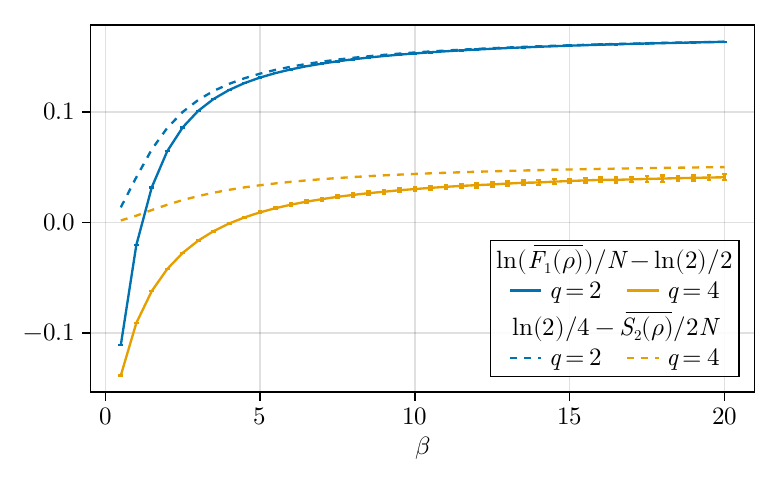}
    \caption{Normalized $\ell_1$ string norm compared to the purity formula derived from the $w_{\text{crit}} \to 1/2$ model culminating in \eqref{eq:F1_lowT_model}. For $q = 2$ we see great agreement at larger $\beta$, while for $q = 4$ further analysis at lower temperatures is required.}
    \label{fig:norm_vs_purity}
\end{figure}

It is possible to model the $q=2$ and $q=4$ curves in Figure \ref{fig:norm} as follows: We observed empirically that the purity saddle point weight $w_\mathrm{crit}$ is close to $1/2$ at low temperature in these cases. If $w_\mathrm{crit}^{(1)}$ and $w_\mathrm{crit}^{(2)}$ are the weights which control the linear norm and purity respectively, then we have $w_\mathrm{crit}^{(2)} \leq w_\mathrm{crit}^{(1)}$. Hence, $w_\mathrm{crit}^{(1)}$ is close to $1/2$ if $w_\mathrm{crit}^{(2)}$ is. In this situation, we may estimate the purity and the linear norm by simply setting $w_\mathrm{crit}^{(1,2)} \to 1/2$. Using $H(1/2) = \ln(2)$, the purity is approximately 
\begin{equation}
    D \, \overline{\tr(\rho^2)} \approx e^{N[\ln(2) - 2 \phi(1/2)]}, 
\end{equation}
which can be solved for $\phi(1/2)$ in terms of the second Renyi entropy per particle $s_2 \coloneqq \overline{S_2(\rho)}/N$:
\begin{equation}
    \phi(1/2) = \frac{1}{4} \ln(2) + \frac{s_2}{2}.
\end{equation}
Similarly, the disorder-average of the linear norm $F_1(\rho)$ is approximately 
\begin{equation}
    \overline{F_1(\rho)} \approx e^{N [ \ln(2) - \phi(1/2) ]}.
\end{equation}
Re-expressing this in terms of the purity and dividing by $D$ gives
\begin{equation}
    \frac{\overline{F_1(\rho)}}{D} \approx e^{N \left[ \ln(2) / 4 - s_2/2 \right]}.
    \label{eq:F1_lowT_model}
\end{equation}
This implies that the state is magical since $s_2 < \ln(2)/2 = s_{2,\max}$ at low temperature. Furthermore, it suggests that $q = 2$ will have a larger amount of magic at low temperatures compared to $q \geq 4$ due to the lack of residual entropy in the former.  We can compare this approximation to the data in Figure \ref{fig:norm_vs_purity}, where we see excellent agreement for $q=2$ and reasonable agreement for $q=4$ at the lowest temperatures we were able to access with reasonable accuracy.

\subsection{Stabilizer R\'enyi Entropy}
\label{sec:sre}

Another notion of magic that has gained popularity in recent years is the stabilizer R\'enyi entropy (SRE) \cite{leone_stabilizer_renyi_entropy_2022, leone_stabilizer_entropies_are_2024}, which we define for both pure and mixed fermionic states $\rho$ as the classical R\'enyi entropy
\begin{equation}
\label{eq:sre_definition}
    M_{\alpha} (\rho) \coloneqq \frac{1}{1-\alpha} \ln \left( \sum_{a} \prob^{\alpha}_{\rho}(a) \right) - \ln(D \tr(\rho^2)),
\end{equation}
associated to the probability of measuring the state $\rho$ in the basis of Hermitian Majorana strings $\majo(a)$\footnote{Note that the existence of this probability distribution is a direct consequence of \eqref{eq:purity_connection}.}:
\begin{equation}
\label{eq:sre_probabilities}
    \prob_{\rho}(a) \coloneqq \frac{\tr(\rho \majo(a))^2 }{D \tr(\rho^2)} \geq 0, \quad \sum_{a} \prob_{\rho}(a) = 1.
\end{equation}
The unusual shift by $-\ln(D \tr(\rho^2))$ ensures the SRE's defining property of being zero if and only if $\rho$ is of the form
\begin{equation}
\label{eq:state_stabilizer_sum}
    \rho \equiv \frac{1}{D} \sum_{\majo \in \mathcal{S}} (-1)^{c_\rho(\majo)} \majo,
\end{equation}
where $\mathcal{S}$ is a fermionic stabilizer (i.e. a subgroup of mutually commuting Majorana strings that commute with $(-1)^F$), and positive otherwise. Furthermore, it is invariant under certain \enquote{free} operations such as (parity-preserving) Cliffords.

What matters for this analysis is that $M_\alpha(\rho)$ is only a rigorous measure of magic if $\rho$ is a pure state ($\tr(\rho^2) = 1$), since generic statistical ensembles of stabilizer states are not necessarily of the form \eqref{eq:state_stabilizer_sum}, and therefore have a non-zero SRE despite not being magical\footnote{A proper extension of the SRE to mixed states can be retrieved from the convex roof construction \cite{leone_stabilizer_entropies_are_2024}}. Regardless, the SRE is a quantity that can be computed for the SYK Gibbs state using the tools developed by us in this paper, and so we proceed for now. Later we compare our results to those retrieved by computing the SRE for generic statistical ensembles of random stabilizer states, which allows us to make statements about its ability to detect magic in the case of the SYK thermal state.

\begin{figure}[h!tb]
    \centering
    \begin{subfigure}[b]{0.49\textwidth}
         \centering
         \includegraphics[width=\textwidth]{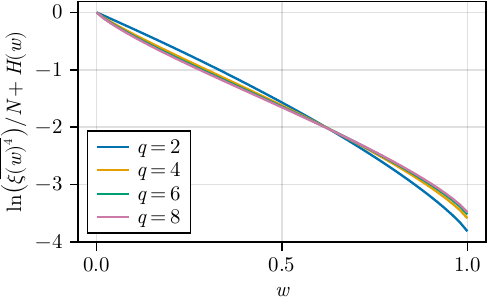}
         \caption{$\beta = 0.5$}
     \end{subfigure}
     \hfill
     \begin{subfigure}[b]{0.49\textwidth}
         \centering
         \includegraphics[width=\textwidth]{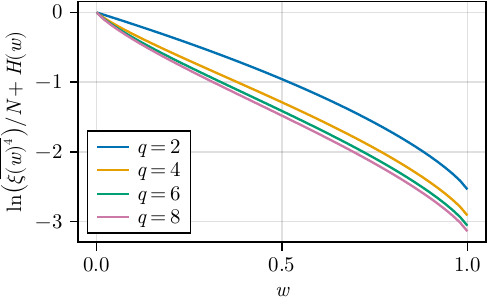}
         \caption{$\beta = 1$}
     \end{subfigure}
     \par\bigskip
     \begin{subfigure}[b]{0.49\textwidth}
         \centering
         \includegraphics[width=\textwidth]{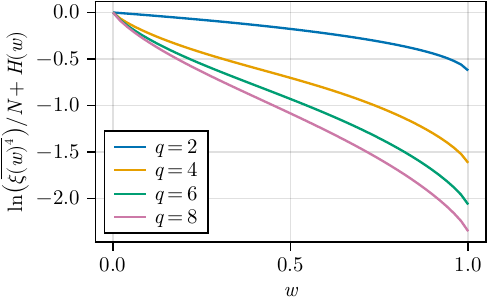}
         \caption{$\beta = 5$}
     \end{subfigure}
     \hfill
     \begin{subfigure}[b]{0.49\textwidth}
         \centering
         \includegraphics[width=\textwidth]{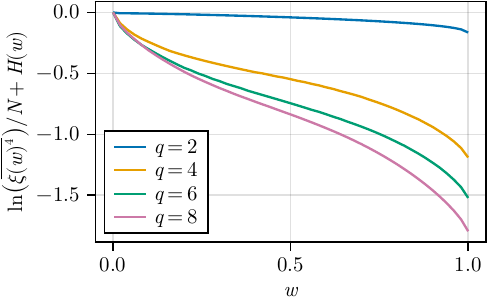}
         \caption{$\beta = 20$}
     \end{subfigure}
    \caption{Combinatorial contributions to $\ln(\overline{\xi(w)^4}) / N = 2 \ln(\overline{\xi(w)^2}) / N$ resulting from the addition of the binary entropy function $H(w)$. Compared to the second moments, almost all maxima are now at or close to $w = 0$ for $\beta \leq 20$ as a result of the 2R2 replica structure doubling the downward slope of the fourth moment for a single operator string.}
    \label{fig:2R2_action_entropy}
\end{figure}

To relate the stabilizer R\'enyi entropy $M_{\alpha}$ to the summing over statistical moments, we can express it more concisely in terms of the $\ell_{2\alpha}$ string norm $F_{2\alpha}(\rho)$ \eqref{eq:string_R_norm} of a given state $\rho$:
\begin{equation}
\label{eq:sre_ratio}
    M_{\alpha} (\rho) = \frac{1}{1-\alpha} \ln \left( \frac{F_{2\alpha}(\rho)}{D \tr(\rho^2)} \right).
\end{equation}
Using the fact that we found the dominating four-replica saddle to be given by 2R2, in the simplest nontrivial case ($\alpha = 2$), the dominating SRE saddle can therefore be computed by finding the maximum of $H(w) - 2 \, \action^{(2,w)}_\mathrm{SYK}$. Before we do so, we again first consider individual instances of this distribution for different choices of $\beta$. The results are shown in Figure \ref{fig:2R2_action_entropy}.

\begin{figure}[h!tb]
    \centering
    \begin{subfigure}[b]{0.49\textwidth}
         \centering
         \includegraphics[width=\textwidth]{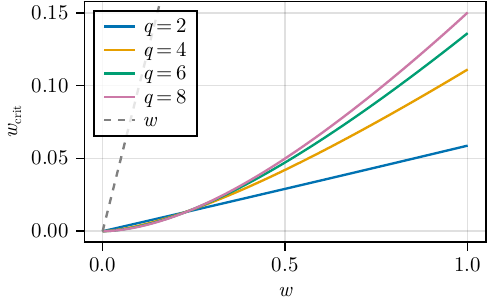}
         \caption{$\beta = 0.5$}
     \end{subfigure}
     \hfill
     \begin{subfigure}[b]{0.49\textwidth}
         \centering
         \includegraphics[width=\textwidth]{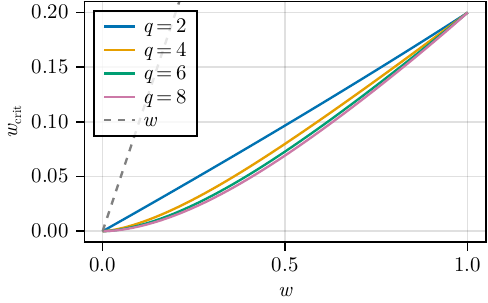}
         \caption{$\beta = 1$}
     \end{subfigure}
     \par\bigskip
     \begin{subfigure}[b]{0.49\textwidth}
         \centering
         \includegraphics[width=\textwidth]{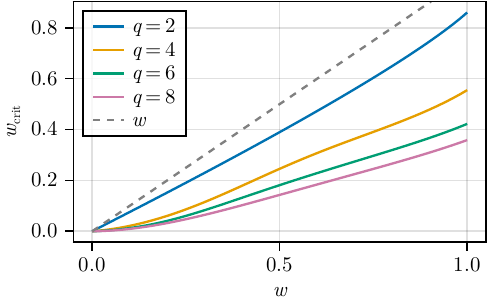}
         \caption{$\beta = 5$}
     \end{subfigure}
     \hfill
     \begin{subfigure}[b]{0.49\textwidth}
         \centering
         \includegraphics[width=\textwidth]{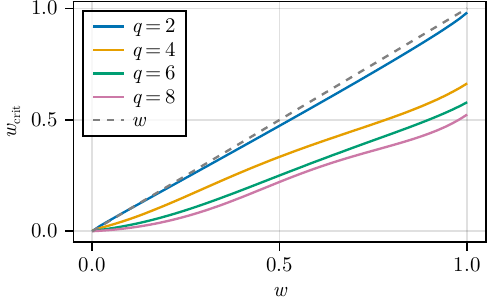}
         \caption{$\beta = 20$}
     \end{subfigure}
    \caption{$w_\mathrm{crit}$ for the case of $R = 4$ using the 2R2 replica structure. It too indicates that all maxima for the inverse temperatures depicted here can be found at or close to $w = 0$. The $q = 2$ case is ambiguous though because $w$ and $w_\mathrm{crit}$ overlap for small $w$. }
    \label{fig:2R2_w_crit}
\end{figure}

Compared to the two-replica case, the maxima for all $q$ seem to remain at or around $w = 0$ for the inverse temperatures being considered, although the maximum for $q = 2$ is harder to determine from just looking at the plot. This is further confirmed by also considering the intersections of $w_\mathrm{crit}(w)$ and $w$,  which are shown in Figure \ref{fig:2R2_w_crit}. Again, they only seem to intersect when both vanish, although the $q = 2$ case is again harder to read off due to both lines almost completely overlapping. In fact, both Figures together imply that one seems to get $\ln(\overline{\xi(w)^4}) \approx - N H(w)$ for the Gaussian case at low temperatures.

This overall change in behavior should not be surprising, as the slope of the fourth moments is twice as steep as that of the second moments due to the 2R2 replica structure dominating. But since the distribution of moments originates at 0 for $w = 0$ and is monotonically increasing until $w = 1/2$, the contribution by the identity is amplified more and more for increasing values of $R$. On the other hand, the binary entropy $H(w)$ remains a $\mathcal{O}(1)$ quantity for any choice of $R$, and therefore it is to be expected that -- for sufficiently large $R$ -- it is not able to get the combined global maximum to move away from $w = 0$ anymore.

The same argument can also be made by just considering the implicit upper bound imposed on the second moments through \eqref{eq:purity_connection} and by inserting $\xi(0) = \tr(I \rho) = 1$. Assuming the moments only depend on the absolute weight $W$, this implies:
\begin{equation}
    \sum_{\majo \neq I} \tr \left(\majo \rho \right)^2 \equiv \sum_{W \neq 0} \binom{N}{W} \xi^2(W) = D \tr(\rho^2) - 1.
\end{equation}
By further identifying $\sum_{W} \binom{N}{W} = D$, we arrive at the following inequality:
\begin{equation}
    \sum_{W \neq 0} \binom{N}{W} \xi^2(W) = \sum_{W \neq 0} \binom{N}{W} \tr(\rho^2) + \tr(\rho^2) - 1 < \sum_{W \neq 0} \binom{N}{W} \tr(\rho^2),
\end{equation}
under the condition that $\tr(\rho^2) < 1$. For random states such as the SYK Gibbs state, it is then reasonable to make the assumption that the non-identity moments hence must satisfy $\xi(W) \lesssim \tr(\rho^2) < 1$. Of course the continuity of the observed moment distributions in Figure \ref{fig:R2_action} indicates that this assumption must break down when going towards $W = 0$ (or $W = N$), but it still seems to hold for the majority of the moments. This can be seen by comparing their values with the relevant upper bound given by (minus) the second R\'enyi entropy as captured in Figure \ref{fig:purity_renyi}. Ultimately, this implies that all mixed state moments away from $W = 0$ are bound to get smaller and smaller as $R$ increases, which is in line with the path integral results we discussed previously.

\begin{figure}[h!tb]
    \centering
    \begin{subfigure}[b]{0.49\textwidth}
         \centering
         \includegraphics[width=\textwidth]{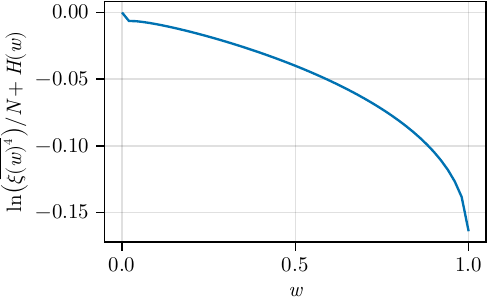}
         \caption{$q = 2$}
     \end{subfigure}
     \hfill
     \begin{subfigure}[b]{0.49\textwidth}
         \centering
         \includegraphics[width=\textwidth]{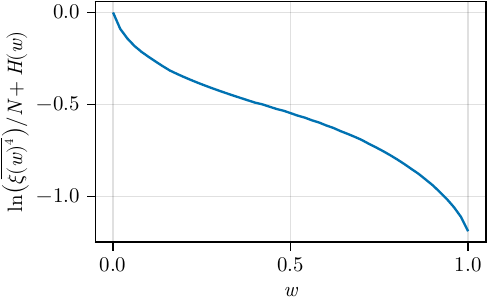}
         \caption{$q = 4$}
     \end{subfigure}
    \caption{Comparison of the combinatorically weighted fourth moments $\ln(\overline{\xi(w)^4}) / N + H(w)$ (for 2R2) at $\beta = 20$ and for $q = 2$ and $q = 4$. While for $q = 4$ the moment distribution still appears to be continuous, for $q = 2$ the discontinuity initially predicted in Section \ref{sec:w_0_behavior} is clearly visible at $w = 0$, with a second local extremum following shortly after.}
    \label{fig:2R2_action_entropy_comparison}
\end{figure}

While these arguments hold well for the case of $q \geq 4$, where the state is guaranteed to have residual entropy at all temperatures, for the Gaussian case we actually observe slightly different behavior, which is best seen by comparing it directly to the $q = 4$ case at $\beta = 20$. This is done in Figure \ref{fig:2R2_action_entropy_comparison}, and again highlights the apparent discontinuity/infinite derivative in the distribution of the $q = 2$ moments compared to their seemingly continuous $q \geq 4$ counterparts (see Section \ref{sec:w_0_behavior}).

Either way, all $q$ appear to realize their global maximum at $w = 0$, meaning that the identity term always dominates in the disorder-averaged stabilizer R\'enyi entropy $\overline{M_\alpha(\rho)}$ for $\alpha = 2$ (and higher). One can therefore approximate it in terms of just the averaged second R\'enyi entropy of $\rho$:
\begin{equation}
\label{eq:sre_purity_approximation}
    \overline{M_\alpha(\rho)} \approx \frac{\ln(D) - \overline{S_2(\rho)}}{\alpha - 1}.
\end{equation}
Note the similarity to the approximation of $\ln(\overline{F_1(\rho)})$ in Section \ref{sec:robustness}, which indicates that the case of $q = 2$ also has the largest SRE at low temperatures\footnote{It is also the only case for which the SRE can actually act as a magic monotone, at least in the zero-temperature limit where the state becomes approximately pure.}, compared to the cases with residual entropy.

The identity term dominating in the computation of the SRE is something that has already been encountered in the context of other systems that admit a large-$N$ limit \cite{turkeshi_pauli_spectrum_nonstabilizerness_2025, haug_efficient_witnessing_testing_2026, collura_non_stabilizerness_fermionic_2026}. This has led to the introduction of a \enquote{filtered} version of $M_\alpha(\rho)$ that intentionally excludes the terms corresponding to the identity and the parity operator $(-1)^F$ (in the case of fermionic systems). This is because their contributions are essentially the same for all states, thus hiding subleading corrections that are more indicative of the specific state at hand.

Based on the analysis in Section \ref{sec:w_0_behavior}, we expect that for $q \geq 4$  removing the identity term (which corresponds to excluding $w = 0$) does not change the overall result due to the distribution seemingly being continuous and its maximum (or rather supremum) therefore still occurring at $w = 0$. The filtered SRE would therefore also be of the form \eqref{eq:sre_purity_approximation} in this case. For $q=2$ however, removing the $w = 0$ \enquote{discontinuity} could actually lead to a new global maximum at some $w > 0$ that shifts the saddle away from the identity term. The resulting change of the saddle point value is only minuscule though -- at least for $\beta = 20$ as can be seen in Figure \ref{fig:2R2_action_entropy_comparison}. So while the filtered SRE does seem to lead to a different result for $q = 2$, which can also be computed using our tools, the result is still not expected to differ much from \eqref{eq:sre_purity_approximation} because the other terms in that expression are much larger in comparison. Either way, the $q = 2$ case of this setup requires a more careful analysis in general, which we plan to address in future work.

\subsection{Holographic Wormholes}
\label{sec:wormholes}

We now show that a simple quantum gravitational model can capture the properties of the two-replica ($R=2$) SYK saddle point discussed above when $\beta J \gg 1$ and $q =4$. The model consists of JT gravity coupled to matter. There are many reviews available that explain the content of JT gravity and its relation to the SYK model, e.g. \cite{sarosi_ads_2_holography_syk_2018, rosenhaus_introduction_syk_model_2019, mertens_solvable_models_quantum_2023}. Here we consider two boundaries to represent the two replicas and work out a model of the inter-replica correlation, $G_{21}(\tau,\tau')$. We first present the gravity calculation of the analogous object and then discuss the comparison between the SYK and gravity computations. 

The field content consists of a metric $g_{\mu\nu}$, a scalar dilaton $\Phi$, a heavy particle, and $N$ light fermions. The gravity action is
\begin{equation}
    I_{\text{JT}}=-S_0 \chi -\frac{1}{2}\int d^2 x \sqrt{g} \Phi (\tilde{R}+2) + \text{(boundary terms)} + m \ell_H + I_{\text{matter}}, \label{eq:JT}
\end{equation}
where $\tilde{R}$ is the Ricci scalar, $m$ is the heavy particle mass, $\ell_H$ is the length of the heavy particle's worldline, and $I_{\text{matter}}$ denotes the action of the light fermions. We remark that the full bulk dual of SYK is not understood, however, it should include the metric-dilaton sector plus the $N$ light fermion fields, since each of the $N$ fermions in SYK has the same scaling dimension. Thus the above theory may be regarded as a (possibly uncontrolled) truncation of the full bulk dual of SYK.

 The solution of interest is a Euclidean wormhole spacetime stabilized by a heavy matter particle. The connectedness of the wormhole reflects the non-trivial inter-replica correlation, $G_{21} \neq 0$. The heavy matter particle is the dual of the Majorana string insertion, and it is heavy because we assume the Majorana string has large weight, $W \propto N$. With this choice, the heavy particle exerts a significant effect on the geometry, in fact allowing the wormhole solution to exist. We then model $G_{21}(\tau,\tau')$ by studying boundary-to-boundary correlations of the $N$ light fermion fields, treated in the geodesic approximation.

The calculation begins with the double trumpet solution in JT gravity sourced by the heavy particle.\footnote{If the $N$ light fermions do not couple to the dilaton, then the solution still obeys $\tilde{R}=-2$ locally. Thus we ignore the fermions when determining the classical solution, though their 1-loop effects may become significant when the wormhole throat $b$ shrinks.} We largely follow the notation in \cite{usatyuk_closed_universes_two_2024}. The metric is
\begin{equation}
    ds^2 = d\rho^2 + b^2 \cosh^2 \rho \, d\sigma^2
\end{equation}
where $\sigma \sim \sigma+1$ and the heavy particle sits at $\sigma=-1/2$. The boundaries are $\rho \to \pm \infty$. The corresponding dilaton profile is
\begin{equation}
    \Phi = \frac{\pi\phi_r}{\beta} \cosh \rho \cosh b \sigma.
\end{equation}

In these formulae, $b$ measures the width of the wormhole at its midpoint at $\rho=0$ and $\phi_r$ is a constant that is related to the boundary value of $\Phi$. In the regime where $b\gg 1$, they are related by
\begin{equation}
    b = 2 \ln \frac{m \beta}{\pi \phi_r} \label{eq:b}
\end{equation}
where $m$ is the mass of the heavy particle. 

The formula \eqref{eq:b} is derived as follows. The heavy particle provides a singular source for the dilaton which leads to a jump in its normal derivative across the particle's location, $\sigma=\pm 1/2$. For a circle of metric $ds^2 = b^2 \cosh^2 \rho \, d\sigma^2$, the unit normal is 
\begin{equation}
    n =  \frac{1}{b \cosh \rho} \partial_\sigma.
\end{equation}
The values of the normal derivative at $\sigma = \pm 1/2$ are
\begin{equation}
    n(\Phi)_{\sigma=\pm 1/2} = \pm \frac{ \pi \phi_r}{\beta} \sinh \frac{b}{2},
\end{equation}
so the jump is
\begin{equation}
    \Delta n(\Phi) = 2 \frac{\pi \phi_r}{\beta} \sinh \frac{b}{2} \approx \frac{\pi \phi_r}{\beta} e^{b/2} = m,
\end{equation}
where the final equality imposes an equation of motion. Working in the limit where $b \gg 1$ gives \eqref{eq:b}.

The bulk is cutoff at a surface defined by $\rho=\rho_c(\sigma)$ obeying
\begin{equation}
    \Phi(\rho_c(\sigma),\sigma) = \frac{\phi_r}{\epsilon}.
\end{equation}
One may think of $\epsilon$ as a short-time cutoff in the theory. Assuming $\rho_c$ is large, we have
\begin{equation}
    e^{\rho_c} =  \frac{2 \beta}{\pi \epsilon} \frac{1}{\cosh b \sigma}.
\end{equation}

The map between $\sigma$ and the boundary time $\tau$ is obtained by integrating the proper time along the boundary,
\begin{equation}
    \frac{\tau(\sigma)}{\epsilon} = \int_{-1/2}^{\sigma} d \sigma' b \cosh \rho_c(\sigma').
\end{equation}
Using our expression for $\rho_c$ in terms of $\sigma$ gives
\begin{equation}
    \frac{\tau}{\beta} = \frac{b}{\pi} \int_{-1/2}^{\sigma} d \sigma' \frac{1}{\cosh b \sigma'}.
\end{equation}
The integral is standard,
\begin{equation}
   \frac{ \tau(\sigma)}{\beta} = \frac{2}{\pi} \arctan e^{b \sigma} - \frac{2}{\pi} \arctan e^{-b/2},
\end{equation}
and in the large $b$ limit this gives
\begin{equation}
   \tau(-1/2) = 0, \quad \tau(0) = \beta/2, \quad \tau(1/2)=\beta.
\end{equation}

For comparison with the SYK data where $b$ is finite but large, it is useful to invoke the corrected formula
\begin{equation}
    \frac{\tau(\sigma)}{\beta} = \frac{\arctan e^{b \sigma} - \arctan e^{-b/2}}{\arctan e^{b/2} - \arctan e^{-b/2}}. \label{eq:t_sigma}
\end{equation}
This formula is obtained by correcting the relationship between the dilaton prefactor and $\beta$ and $b$ and the relationship between $b$ and $m$.

We now turn to the calculation of geodesics. It is convenient to use an embedding space approach. The double trumpet geometry is a quotient of the hyperbolic disk which can be embedded into a $(2+1)d$ space with metric
\begin{equation}
    ds^2_{2+1}=-(dX^0)^2 + (dX^1)^2 + (dX^2)^2.
\end{equation}
In terms of the coordinates used above, the embedding is
\begin{align}
\begin{split}
    X^0 &= \cosh \rho \cosh b \sigma, \\
    X^1 &= \sinh \rho, \\
    X^2 &=\cosh \rho \sinh b \sigma,
\end{split}
\end{align}
which satisfies
\begin{equation}
    - (X^0)^2 + (X^1)^2 + (X^2)^2 = - 1.
\end{equation}

In this embedding framework, the distance $d$ between two points $P$ and $Q$ in the double trumpet obeys
\begin{equation}
    \cosh d = - X(P)\cdot X(Q).
\end{equation}
We consider two points, one on either boundary:
\begin{align}
\begin{split}
    P: (\rho &=\rho_c(\sigma), \, \sigma) \\
    Q: (\rho &= -\rho_c(\sigma'), \, \sigma').
\end{split}
\end{align}
The inner product of the corresponding embedding coordinates is
\begin{equation}
    - X(P) \cdot X(Q) = \cosh \rho_c(\sigma) \cosh \rho_c(\sigma') \cosh b (\sigma - \sigma') + \sinh \rho_c(\sigma) \sinh \rho_c(\sigma').
\end{equation}
When $\rho_c$ is large, $d$ is also large, thus giving the simplified formula
\begin{equation}
    \frac{e^d}{2} = \frac{e^{\rho_c(\sigma)} e^{\rho_c(\sigma')}}{4} (\cosh b (\sigma-\sigma') +1).
\end{equation}
The distance is then
\begin{equation}
    d = \rho_c(\sigma) + \rho_c(\sigma') + \log \cosh^2 \frac{b(\sigma-\sigma')}{2}.
\end{equation}

There are also winding geodesics that wrap around the double trumpet $n$ times. The corresponding distance is 
\begin{equation}
    d_n(\sigma,\sigma') = \rho_c(\sigma) + \rho_c(\sigma') + \log \cosh^2 \frac{b(\sigma-\sigma' - n)}{2}.
\end{equation}
When $n$ is large, $d_n \sim n b + \cdots$ in accord with the fact that a highly wound geodesic wraps the $\rho=0$ circle many times.

\begin{figure}[h!tb]
    \centering
    \includegraphics[width=0.8\linewidth]{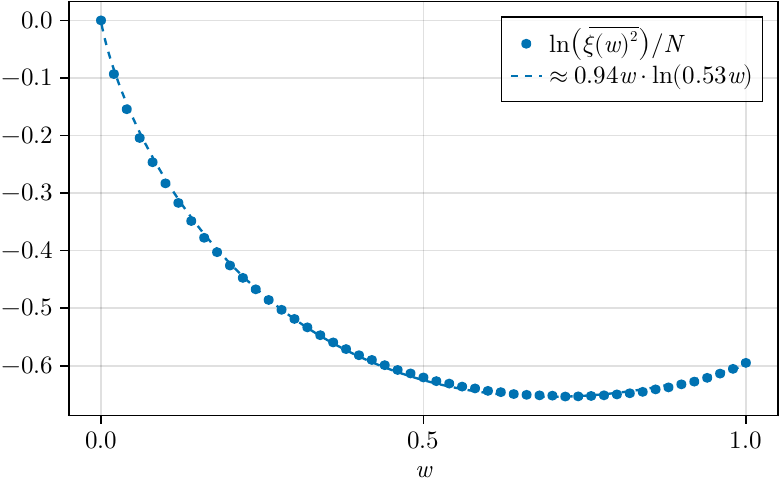}
    \caption{Comparison of the SYK data for the moments $\ln(\overline{\xi(w)^2})/N$ (evaluated for $q = 4$ at $\beta = 20$ and with $L = 1000$) as a function of weight to the trend predicted by the classical action of the wormhole model. As discussed in the text, the wormhole model prediction is $c_1 w \ln (c_2 w)$ for constants $c_1$ and $c_2$. A fit of the data according to this trend results in good agreement for $c_1 \approx 0.94$ and $c_2 \approx 0.53$.}    
    \label{fig:wh_action}
\end{figure}

With this information about geodesics, it is now possible to calculate the classical action of the double trumpet saddle. Specifically, the $w$-dependent piece of this action gives the wormhole prediction for $\ln \overline{\xi^2}$. This prediction can then be compared to microscopic SYK data by taking $m \propto w$. The relevant part of the JT action \eqref{eq:JT} is $-m \ell_H$ where $m$ is the heavy particle mass and $\ell_{H}$ is the length of the heavy particle worldline, 
\begin{equation}
    \ell_{H} = d_{n=0}(\tau=\tau'=0) = 2 \left( -\frac{b}{2} + \ln \frac{ 4 \beta}{\pi \epsilon} \right) = 2 \ln \frac{4 \phi_r}{\epsilon m}.
\end{equation}
With $m \propto w$, the wormhole model predicts an action going as 
\begin{equation}
\label{eq:wormhole_action_prediction}
    c_1 w \ln (c_2 w) + \text{(independent of $w$)}
\end{equation}
for constants $c_{1,2}$. One could try to be more predictive, but there are also additional contributions, such as from 1-loop determinants of matter fields, that would require further analysis. However, the basic $w$-dependence appears robust, so we simply fit $c_{1,2}$ to SYK data as shown in Figure \ref{fig:wh_action}. The agreement is quite good over the whole range of $w$. It is also worth noting that including $(1-w)$ and $(1-w)\ln(1-w)$ terms further improves the fit.

The geodesic data also allows an evaluation of the correlations of quantum fields propagating on the double trumpet background. In brief, a boundary operator of scaling dimension $\Delta$ is dual to bulk field of mass determined by $\Delta$. By solving for the 2-point function of this bulk field on the given background spacetime, one can then determine the boundary correlations by taking the points to the boundary. 

Here we use a simplified geodesic approximation to the bulk 2-point function. Let $\mathcal{G}(\tau,\tau')$ denote the two-boundary correlator. With periodic boundary conditions, we have
\begin{equation}
    \mathcal{G}_P(\tau,\tau') = \mathcal{N}_P \sum_n e^{- \Delta d_n(\sigma(\tau),\sigma'(\tau'))},
\end{equation}
and with anti-periodic boundary conditions,
\begin{equation}
    \mathcal{G}_{AP}(\tau,\tau') = \mathcal{N}_{AP} \sum_n (-1)^n e^{- \Delta d_n(\sigma(\tau),\sigma'(\tau'))},
\end{equation}
where
\begin{equation}
    \sigma(\tau) = \frac{1}{b} \ln \tan \left( [\arctan e^{b/2} - \arctan e^{-b/2}] \frac{\tau}{\beta} + \arctan e^{-b/2} \right)
\end{equation}
follows from \eqref{eq:t_sigma}. See also \cite{lin_revisiting_second_order_2026} for further discussion of correlations on the double trumpet.

In general, we may consider a mixture of periodic and anti-periodic boundary conditions. This is motivated by viewing the microscopic $G_{21}$ as a sum over $N$ flavors of bulk fermions of which $N_P$ have periodic boundary conditions and $N_{AP} = N-N_P$ have antiperiodic boundary conditions,
\begin{equation}
    G_{21} \sim \frac{N_P}{N} \mathcal{G}_P + \frac{N_{AP}}{N}  \mathcal{G}_{AP} \equiv \mathcal{G}_f.
\end{equation}
We introduce $f= N_P/N$ as a convenient short-hand. At the special point $f=1/2$, all the odd winding geodesics cancel. One can see an emergent time-translation symmetry at this special point, similar to the emergent time translation symmetry that occurs at $w=w_{\text{crit}}$ in the purity calculation.

\begin{figure}[h!tb]
    \centering
    \includegraphics[width=\linewidth]{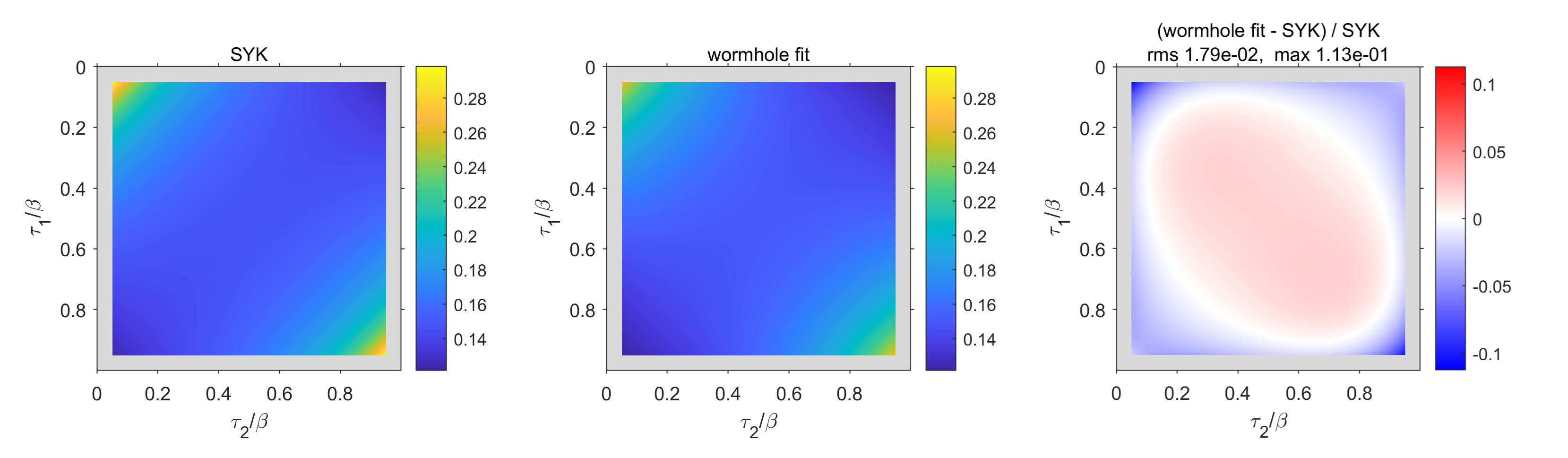}
    \caption{(Left) The plot shows the value of $G_{21}(\tau,\tau')$ in color as a function of $\tau/\beta$ and $\tau'/\beta$. The SYK parameters are $\beta=20$, $w=.25$, $L=2000$. (Middle) The same configuration now showing the wormhole prediction obtained by fitting to the SYK data excluding $\delta \tau = 1 = .05 \beta$ (in units with $J=1$) around the heavy operator (grey boundary). The fit values are $b=4.7$ and amplitude $= .42$. (Right) The relative difference between (Left) and (Middle). The maximum deviation is around 12\%.}
    \label{fig:wh_fit}
\end{figure}

To compare with the SYK data for $G_{21}(\tau,\tau')$, we set $\Delta = 1/q$ and then adjust $b$ and the overall amplitude to fit the data. We set $f=w$ since for a given $\mu$ of weight $W$, the $N-W$ fermions not in $\mu$ maintain anti-periodic boundary conditions around the thermal circle while the $W$ fermions in $\mu$ now acquire periodic boundary conditions. When fitting, we also exclude a small window of size $1/J$ around the heavy operator on the grounds that this is the most UV sensitive region. Figure \ref{fig:wh_fit} shows an example of the comparison for $\beta=20$ and $w=.25$ with the SYK data in the left of Figure \ref{fig:wh_fit} and the wormhole fit in the middle of Figure \ref{fig:wh_fit}. The residuals are shown on the right where we see a root-mean-square error of 1.8\% and a maximum error of about 12\%. 

The agreement is quite good even though we are (a) using a truncation of the full bulk dual of SYK, (b) not very deep in the low-energy regime ($\beta =20$), and (c) have numerical errors from the finite grid size and convergence at the 5\% level (see Figure~\ref{fig:purity_renyi}). We found that the fit can be improved by the ad hoc procedure of letting $f$ float instead of fixing it by $w$, but this additional freedom is not well-motivated. Overall, this example is representative of the quality of the fit we can achieve at this value of $\beta$, and we expect the fit would be improved by going to larger $\beta$ where the simple gravity model considered here should better match SYK.


%% file: sections/outlook.tex
\section{Outlook}
\label{sec:outlook}

There are a number of promising directions for future work. In the short term, we have work in progress extending our results to thermofield double states and to the more general coupled SYK models considered in \cite{sasieta_baby_universes_thermal_2025}, for which the thermal 1-point functions considered here arise in a particular limit. Another near-term goal is increased numerical stability, perhaps by integrating out zero-modes analytically \cite{halder_disorder_averaged_framework_2025}, and pursuing additional numerical optimizations to reach larger $L$ and $\beta$ values. One could also investigate how the bulk model might be improved to better capture small $\beta$ values, and it would be enlightening to study the large $q$ limit where analytic expressions might be available.

We also have work in progress on the $q=2$ case. We already discussed how the $q=2$ action has an enhanced symmetry based on $O(R)$ rotations between the different replicated fermions in the path integral formulation. By carefully accounting for the space of saddle points and the action of the symmetry group on this space, one can perform a more systematic accounting of the moments of the 1-point functions and determine the precise numerical prefactors, i.e. the ratio of $\overline{\xi^{2k}}/\overline{\xi^2}^k$. One goal is to precisely characterize the non-Gaussian distribution governing the thermal 1-point functions and connect it to finite $N$ formulas and Wick's theorem.

Another direction concerns a tensor network model of SYK called NoRA, which we recently introduced in \cite{bettaque_nora_tensor_network_2024}. It would be interesting to understand if the detailed microstructure of operator expectation values could be reproduced within that framework. It would also be interesting to develop tensor network models of the dual gravity physics \cite{antonini_baby_universe_fine_2025, sasieta_baby_universes_thermal_2025} using NoRA.

Finally, we note that one can also give explicit single instance formulas for the thermal 1-point functions when the number of operators involved is small. Indeed, for $q=2$ one can obtain any thermal one-point function in terms of the basic 2-point function using Wick's theorem, and there are explicit determinant formulas that give the same result in compact form \cite{collura_non_stabilizerness_fermionic_2026}. In the case of $q=4$, one can show diagrammatically that $\tr(\sqrt{\rho} \chi_i \sqrt{\rho} \chi_j) \propto \sum_{lmn} J_{i lmn} J_{j lmn}$ and that $\tr(\rho \chi_i \chi_j \chi_k \chi_l) \propto J_{ijkl}$ (with analogous formulas for larger $q$). Operators with more fermions can be treated similarly by summing over all diagrams that contribute, however, the complexity of the calculation quickly grows untenable as the number of fermions increases. Nevertheless, it would be interesting to analyze the leading diagrams statistically to connect to our path integral results.

\emph{Acknowledgments:} We thank Martin Sasieta for useful discussions. We acknowledge support for this work from the DOE Quantised 2.0 program through the GeoFlow consortium grant DE-SC0019380 (VB, BS) and the Heising-Simons Foundation through grant 2024-4849 (BS). BS acknowledges useful discussions with Claude Opus 4.5, however, all calculations and numerical results in the paper were carried out and checked by the authors and the text is not AI generated. VB would also like to thank Mountain Dew Baja Caba Citrus\texttrademark \, for its caffeinated support throughout writing this paper.

%% file: appendices/numerics.tex
\section{Numerical Implementation}
\label{app:numerical_implementation}

In this appendix, we discuss the details of our numerical implementation, including the discretization of the time variable, the iterative solver, and the assumed structure of the solution space.

\subsection{Matrix Discretization}
\label{sec:discretization}

To solve the Schwinger-Dyson equations \eqref{eq:syk_w_schwinger_dyson} numerically, they have to be discretized one way or another. The most straightforward approach with no additional assumptions about the possible solutions (such as time-translation invariance)  is demoting $\Sigma_{rs}$ and $G_{rs}$ from (anti-) symmetric functions in $\tau$ to (anti-) symmetric $L \times L$  matrices. This is accomplished by splitting the intervals $[0, \beta)$ being integrated over in \eqref{eq:syk_path_integral_decoupled} and \eqref{eq:syk_path_integral_integrated_out} into $L$ bins with width $\Delta\tau = \beta / L$ and applying the trapezoid rule 
\begin{equation}
\label{eq:trapezoid rule}
    \int_{k \Delta\tau}^{(k+1) \Delta\tau} d\tau f(\tau) \approx \Delta\tau \cdot \frac{f(k \Delta\tau) + f((k+1) \Delta\tau)}{2}  \quad \forall k = 0, \ldots, L - 1
\end{equation}
Assuming that $f$ is periodic (i.e.\ $f(\beta) = f(0)$), then each term $f(k \Delta\tau)$ will appear exactly twice in the overall discretized integral, resulting in
\begin{align}
\begin{split}
    \int_0^\beta d\tau f(\tau) &= \sum_{k=0}^{L-1} \int_{k \Delta\tau}^{(k+1) \Delta\tau} d\tau f(\tau) \\
    &\approx \Delta\tau \sum_{k=0}^{L-1} \frac{f(k \Delta\tau) + f((k+1) \Delta\tau)}{2} \\
    &= \Delta\tau \sum_{k=0}^{L-1} f(k \Delta\tau).
\end{split}
\end{align}
The same rules apply if $f$ has two arguments and is periodic in both of them, except that we now have two sums and an additional factor of $\Delta\tau$:
\begin{equation}
\label{eq:two_argument_discretization}
    \int_0^\beta d\tau d\tau' f(\tau, \tau') \approx (\Delta\tau)^2 \sum_{k,l = 0}^{L-1} f(k\Delta\tau, l\Delta\tau).
\end{equation}

Since every integrand in \eqref{eq:syk_path_integral_decoupled} and \eqref{eq:syk_path_integral_integrated_out} is periodic in both $\tau$ and $\tau'$, we can directly apply these general rules to each of them, starting with the kinetic term of a single (decoupled) Majorana mode with either boundary condition as seen in \eqref{eq:syk_path_integral_decoupled}:
\begin{align}
\begin{split}
\label{eq:syk_discretized_kinetic_term}
    \int_0^{\beta} d\tau d\tau' \chi^{(r)}(\tau) \left[ \delta_{rs} \, \Partial(\tau, \tau') \right] \chi^{(s)}(\tau') &= \int_0^{\beta} d\tau \, \chi^{(r)}(\tau) \left(\partial_{\tau} \chi^{(r)}\right)(\tau) \\
    &\approx \Delta\tau \sum_{k=0}^{L-1} \chi^{(r)}(k \Delta\tau) \left(\partial_{\tau} \chi^{(r)}\right)(k \Delta\tau) \\
    &\equiv \Delta\tau \sum_{k=0}^{L-1} \chi^{(r)}_k \left(\partial_{\tau} \chi^{(r)}\right)_k
\end{split}
\end{align}
However for this term we also have to choose a discretization of the imaginary time derivative. For reasons that will become evident shortly, we do \emph{not} choose a symmetric difference quotient, but an asymmetric one:
\begin{equation}
    \left(\partial_{\tau} \chi^{(r)}\right)_k \approx \frac{\chi^{(r)}_{k} - \chi^{(r)}_{k-1}}{\Delta\tau}.
\end{equation}
The reason why this seems problematic is because when inserting it into \eqref{eq:syk_discretized_kinetic_term} one would get products of $\chi^r_k$ which would vanish due to it being a Grassmann number. However, we will ignore this issue for now and keep said products in the expressions. Rearranging the terms in \eqref{eq:syk_discretized_kinetic_term} then gives
\begin{align}
\begin{split}
    \Delta\tau \sum_{k=0}^{L-1} \chi^{(r)}_k \left(\partial_{\tau} \chi^{(r)}\right)_k &= \sum_{k=0}^{L-1} \chi^{(r)}_k \left(\chi^{(r)}_k - \chi^{(r)}_{k-1}\right) \\
    &= \chi^{(r)}_0 \left(\chi^{(r)}_0 \pm \chi^{(r)}_{L-1}\right) + \sum_{k=1}^{L-1} \chi^{(r)}_k \left(\chi^{(r)}_k - \chi^{(r)}_{k-1} \right) \\
    &\equiv \left(\chi^{(r)}\right)^T \left[\delta_{rs} \, \Partial^{\mp}\right] \chi^{(s)},
\end{split}
\end{align}
where in penultimate line we identified $\chi^r_{-1} \equiv \mp \chi^r_{L-1}$ depending on if the fermion experiences odd or even boundary conditions respectively, and in the last line we rewrote the whole sum as a matrix equation involving $\chi^{(r)} \equiv (\chi^{(r)}_0, \chi^{(r)}_1, \ldots, \chi^{(r)}_{L-1})^T$ and 
\begin{equation}
\label{eq:discrete_derivative_2}
    \Partial^\mp  = \begin{pmatrix} 
        1 & 0 & 0 & \cdots & 0 & \pm 1 \\
        -1 & 1 & 0 & \cdots & 0 & 0 \\
        0 & -1 & 1 & \cdots & 0 & 0 \\
        \vdots &\vdots &\vdots & \ddots & \vdots & \vdots \\
        0 & 0 & 0 & \cdots & 1 & 0 \\
        0 & 0 & 0 & \cdots & -1 & 1
    \end{pmatrix}.
\end{equation}
Note that $\Partial^\mp$ is not a skew-symmetric matrix and therefore does not have a Pfaffian associated to it, another pathology resulting from our choice of discretization. However, by instead evaluating the square root of the determinant corresponding to the antiperiodic derivative $\Partial^-$ we get\footnote{For the periodic derivative one always has $\pf(\Partial^+) \sim \sqrt{\det \Partial^+ } = 0$ due to the presence of the zero mode.}
\begin{equation}
    \pf(\Partial^-) \sim \sqrt{\det(\Partial^-) } = \sqrt{2},
\end{equation}
regardless of matrix size. This coincides with the number of degrees of freedom expected from a single Majorana mode, as two such modes form the representation of a two-dimensional Hilbert space. Had we chosen a symmetrized discretization of this derivative in form of an antisymmetric matrix, we would instead have gotten $\pf(\Partial^-) = 2$, which corresponds to two Majorana modes. This is known as \emph{fermion doubling} and is a direct consequence of the Nielsen-Ninomiya theorem \cite{nielsen_absence_neutrinos_lattice_1981}. To avoid these additional unphysical degrees of freedom, we have to break one of the theorem's assumptions, namely the Hermiticity of the discretized kinetic term. One price we pay is that we have to use $\sqrt{\det M}$ instead of $\pf(M)$ for any matrix $M$ representing an antisymmetric function in the continuum. Since $\pf(M)^2 = \det(M)$, this means we lose information about the overall sign of the Pfaffian. Fortunately this sign can be safely ignored, and one can also assume $\det(M)$ to always be positive due to the number $R$ of replicas being even.

Compared to $\Partial^\mp$, discretizing the Gaussian self-energy term $\chi^{(r)}(\tau) \Sigma_{rs}(\tau, \tau') \chi^{(s)}(\tau')$ in \eqref{eq:syk_path_integral_decoupled} is straightforward using \eqref{eq:two_argument_discretization}. Both terms together can then be written as a matrix expression over both discretized time steps and replica indices:
\begin{align}
\begin{split}
    & \sum_{r,s = 1}^R \int_0^{\beta} d\tau d\tau' \chi^{(r)}(\tau) \left[\delta_{rs} \, \Partial(\tau, \tau') - \Sigma_{rs}(\tau, \tau') \right] \chi^{(s)}(\tau') \\
    \approx{}&  \sum_{r,s = 1}^R \left(\chi^{(r)}\right)^T \left[\delta_{rs} \, \Partial^{\mp} - (\Delta \tau)^2 \, \Sigma_{rs}\right] \chi^{(s)} \\
    \equiv{}& \chi^T \left[I_R \otimes \Partial^{\mp} - (\Delta \tau)^2 \, \Sigma\right] \chi.
\end{split}
\end{align}
Repeating these steps for the integral in \eqref{eq:syk_path_integral_integrated_out} and integrating out the Majorana modes as before results in the discretized action
\begin{align}
\begin{split}
    I^{(R, w)}_\mathrm{SYK}[G, \Sigma] \approx{}& - \frac{1-w}{2} \ln \det\left(I_R \otimes \Partial^- - (\Delta\tau)^2 \, \Sigma\right) - \frac{w}{2} \ln\det\left(I_R \otimes \Partial^+ - (\Delta\tau)^2 \, \Sigma\right) \\
    & + \frac{(\Delta\tau)^2}{2} \sum_{i,j = 1}^{RL} \left( \Sigma_{ij} \, G_{ij} - \frac{J^2}{q} (G_{ij})^q \right),
\end{split}
\end{align}
where the indices $i, j$ sum over both discretized times and replicas. Varying this expression with regard to $\Sigma$ and $G$ then results in the corresponding Schwinger-Dyson equations:
\begin{align}
\begin{split}
\label{eq:syk_w_schwinger_dyson_discrete_2}
    -G^T &= (1-w) \, (I_R \otimes \Partial^- - (\Delta\tau)^2 \, \Sigma)^{-1} + w \, (I_R \otimes \Partial^+ - (\Delta\tau)^2 \, \Sigma)^{-1}, \\
    \Sigma_{ij} &= J^2 \, (G_{ij})^{q-1}.
\end{split}
\end{align}
Note that we have $-G^T$ instead of just $G$ on the left-hand side of the first equation. The transpose arises due to $\partial_M \ln \det(M) = (M^{-1})^T$, and is usually removed in the continuous Schwinger-Dyson equations by imposing antisymmetry as in \eqref{eq:G_definition}. However, by breaking the antisymmetry of $\Partial^\mp$ we also broke the antisymmetry of $G$ (and $\Sigma$), which is best seen in the case of the free Green's function
\begin{equation}
    G_\mathrm{free} \sim (\Partial^-)^{-1} = \frac{1}{2}\begin{pmatrix}
        1 & -1 & -1 & \cdots & -1 & -1 \\
        1 & 1 & -1 & \cdots & -1 & -1 \\
        1 & 1 & 1 & \cdots & -1 & -1 \\
        \vdots & \vdots & \vdots & \ddots & \vdots & \vdots \\
        1 & 1 & 1 & \cdots & 1 & -1 \\
        1 & 1 & 1 & \cdots & 1 & 1
    \end{pmatrix}
\end{equation}
associated to \eqref{eq:discrete_derivative_2} being antisymmetric up to non-vanishing diagonal elements. Leaving the transpose in \eqref{eq:syk_w_schwinger_dyson_discrete_2} untouched is therefore necessary to ensure mathematical consistency, and has also proven itself to lead to improved numerical accuracy.

\subsection{Unbiased Iterative Solving}
\label{sec:unbiased_solver}

We first discuss the usual way to solve \eqref{eq:syk_w_schwinger_dyson_discrete_2} iteratively, which is equivalent to the ones found in \cite{maldacena_remarks_sachdev_ye_2016}. Starting with some suitable initial $RL \times RL$ matrix $G_\mathrm{init}$ (e.g.\ those discussed in Section \ref{sec:initial_value_classification}), an interpolation parameter $t \in (0, 1)$, an interpolation scaling factor $b > 1$, and a relative error tolerance $\varepsilon > 0$, one can perform the following algorithm:
\begin{enumerate}
    \item Set $G \gets G_\mathrm{init}$ and use it to compute $\Sigma_\mathrm{SD}$ and then $G_\mathrm{SD}$ using \eqref{eq:syk_w_schwinger_dyson_discrete_2}.
    \item Compute the relative difference $\delta = \|G_\mathrm{SD} - G\|_F / \|G\|_F$ (where $\|M\|_F = \sqrt{\sum_{ij} |M_{ij}|^2}$ is the Frobenius norm).
    \item Compute the interpolation $G_t \gets (1 - t) \, G + t \, G_\mathrm{SD}$.
    \item Compute the new relative difference $\delta_\mathrm{new} = \|G_t - G\|_F / \|G\|_F$
    \item Check if $\delta_\mathrm{new} < \varepsilon$. If that is the case, return $G_t$ (and $\Sigma$ computed from $G_t$ using \eqref{eq:syk_w_schwinger_dyson_discrete_2}).
    \item Check if $\delta_\mathrm{new} > \delta$. If that is the case, set $t \gets t/b$ and go back to Step 3.
    \item Update $\delta \gets \delta_\mathrm{new}$, $G \gets G_t$, and recompute $\Sigma_\mathrm{SD}$ and $G_\mathrm{SD}$ with the new $G$ using \eqref{eq:syk_w_schwinger_dyson_discrete_2}.
    \item Go back to Step 3.
\end{enumerate}

While for ordinary SYK ($w = 0$) an interpolation parameter of $t = 0.5$ is largely sufficient, for $w > 0$ a value of $t = 0.01$ was necessary to achieve proper convergence. For both cases, having $L = 1000$, $b = 2$ and $\varepsilon = 10^{-5}$ seems to provide a reasonable tradeoff between speed and accuracy for the temperature regime $\beta \in (0, 50]$, although choosing $\varepsilon \lesssim (\beta J/L)^2$ only leads to corrections that are not physically relevant. We also aborted the solver after 1000 iterations, which usually was enough to have $\delta$ get sufficiently close to $\varepsilon$ for our choice of parameters. This only proved to be an issue for $q = 2$ though, which still turned out to be the case that converged the best.

The biggest downside of this iterative algorithm is that each iteration step requires the inversion of two $RL \times RL$ matrices, each of which has $\mathcal{O}((RL)^3)$ runtime complexity. While inverting a single-replica $1000 \times 1000$ matrix is still reasonably fast, going to $R = 2$ already severely impacts the performance and $R = 4$ could not be probed in reasonable time at all with our available resources.

\subsection{Iterative Solving with Anticirculant Replica Structure}
\label{sec:anticirculant_solver}

To significantly reduce the impact on the iteration runtime that choosing large $R$ has, we can exploit the anticirculant replica symmetry proposed in Section \ref{sec:anticirculant}. As discussed there, choosing this ansatz allows one to block-diagonalize any multi-replica expression
\begin{equation}
    M = \sum_{r = 0}^{R-1} T^r \otimes M_r
\end{equation}
by performing a Fourier transform over the replica indices:
\begin{equation}
\label{eq:fourier_transform}
    \widehat{M} = \diag \left( \widehat{M}_0, \ldots, \widehat{M}_{R-1} \right), \quad   \widehat{M}_s = \sum_{r=0}^{R-1} e^{2 \pi i r (s + 1/2) / R} \, M_r.
\end{equation}
The resulting complex matrices are (approximately) anti-Hermitian and always occur together with their respective complex conjugate:
\begin{equation}
    \left(\widehat{M}_s\right)^\dagger = - \widehat{M}_s, \quad \left(\widehat{M}_s\right)^* = \widehat{M}_{R-s-1}.
\end{equation}
Together with the fact that the overall antisymmetry of the replica matrix \eqref{eq:G_replica_antisymmetry_constraint} implies that $M_r^T = M_{R-r}$ for $r > 0$, we therefore only need to keep track of $R/2 + 1$ many $L \times L$ real matrices, and only have to invert $R/2$ complex ones after performing a Fourier transform.

The solving routine in Appendix \ref{sec:unbiased_solver} mostly remains the same, except that now we have $G \equiv (G_0, G_1, \ldots G_{R/2})$ and $\Sigma \equiv (\Sigma_0, \Sigma_1, \ldots \Sigma_{R/2})$ respectively. To compute $\Sigma_{SD}$, we first perform the Fourier transform \eqref{eq:fourier_transform} of $I_R \otimes \Partial^\mp - (\Delta\tau)^2 \, \Sigma$, and then use the first Schwinger-Dyson equation in \eqref{eq:syk_w_schwinger_dyson_discrete_anticirculant}. To transform back to real space, the inverse Fourier transform
\begin{equation}
    M_r = \frac{1}{R} \sum_{s = 0}^{R - 1} e^{-2\pi i r(s + 1/2) / R} \, \widehat{M}_s
\end{equation}
is employed. The second Schwinger-Dyson equation can be computed for each (real) entry individually. The Frobenius norm of any replica matrix $M$ is also given by
\begin{equation}
    \|M\|_F = \sqrt{ R \sum_{r = 0}^{R-1} \|M_r\|^2_F }.
\end{equation}

Overall, instead of a $\mathcal{O}((RL)^3)$ runtime for each replica matrix inversion, we only have to perform $2 L^2$ fast Fourier transforms -- each with runtime $\mathcal{O}(R \ln(R))$ -- and $R/2$ (complex) matrix inversions with runtime $\mathcal{O}(L^3)$. The overall complexity is therefore $\mathcal{O}(R L^3 + R \ln(R) L^2) = \mathcal{O}(R L^3)$, which is a significant improvement.

%% file: appendices/ensembles.tex
\section{Comparison to Other Ensembles }
\label{app:ensembles}

Here we present some useful comparisons for the results in the main text. We consider two ensembles of mixed states, a uniform mixture of $K$ random stabilizer states and a uniform mixture of $K$ Haar random states, and calculate relevant properties, including purity, $1$-norm, and stabilizer Renyi entropy. These are somewhat crude comparisons in that we do not attempt to mimic the full weight distribution of the operator statistics, but these ensembles do have important elements in common with the SYK-derived ensemble, including a statistical permutation symmetry acting on the fermions. For simplicity, we work in the large $N$ limit and drop subleading factors in $N$ and the dimension $D$. Moreover, since the second Renyi entropy of the SYK thermal state is sub-maximal when $\beta > 0$, we will assume $K \ll D$ in the computations below.

For the 1-norm, which lower bounds the robustness of magic, we always find a trivial bound for the random stabilizer mixture (as had to be the case) whereas the random Haar mixture yields a bound that is always non-trivial, indicating magic. On the other hand, for the stabilizer Renyi entropy with index $2$, we find that both random mixtures give the same value provided the entropy is large enough. Hence, while this large value of the SRE is consistent with magic, it does not guarantee it. One can also apply a depolarizing channel of some strength $\gamma$ which has the effect of damping strings of weight $W$ by a factor of $e^{- \gamma W}$. This extra knob allows further fine-tuning of the weight distribution, especially if we also consider probabilistic mixtures of different $\gamma$s. We tried to use this extra degree of freedom to more closely mimic the SYK thermal weight distribution using states in the stabilizer polytope, but we could not find a way to capture the detailed statistical properties when $s_2 \coloneqq- \frac{1}{N} \ln \tr(\rho^2) < \frac{1}{2} \ln 2$.

\subsection{Mixture of Random Stabilizer States}

In this case each density matrix in the ensemble is a mixture of $K$ randomly chosen pure stabilizer states. These can be obtained by choosing a random Clifford unitary and acting it on a fixed initial state. Denote an instance as
\begin{equation}
    \rho = \frac{1}{K} \sum_\alpha | \psi_\alpha \rangle \langle  \psi_\alpha | .
\end{equation}

\subsubsection*{Purity} Starting from the definition of the purity,
\begin{equation}
    \tr(\rho^2) = \frac{1}{K^2} \sum_{\alpha \beta} |\langle \psi_\alpha | \psi_\beta \rangle|^2,
\end{equation}
the fact that Cliffords form a 2-design gives, at large $D$,
\begin{equation}
    \tr(\rho^2) \approx \frac{1}{K} + \frac{1}{D}.
\end{equation}
Hence, so long as $K \ll D$, the 2nd Renyi entropy per particle is
\begin{equation}
    s_2 = \frac{\ln K}{N}.
\end{equation}

\subsubsection*{Moments of Quantum Expectation Values} The random variable of interest is the quantum expectation value,
\begin{equation}
    \xi = \frac{1}{K} \sum_\alpha \xi_\alpha, \quad \xi_\alpha = \langle \psi_\alpha | \mu | \psi_\alpha\rangle.
\end{equation}
Each $|\psi_\alpha\rangle$ is a random stabilizer state and a given non-identity string is in the stabilizer group with probability $1/D$. If a string is in the stabilizer group, then its expectation value is $\pm 1$ with equal probability. Thus
\begin{equation}
    P(\xi_\alpha = 0) = 1 - \frac{1}{D}, P(\xi_\alpha=1) = \frac{1}{2D}, P(\xi_\alpha =-1) = \frac{1}{2D}.
\end{equation}
And different states in the mixture give independent contributions. Of course, the identity string and fermion parity are special.

Since the $\xi_\alpha$ are independent, it immediately follows that all odd moments of $\xi$ vanish. Therefore we consider just the even moments of $\xi$. The even moments of $\xi_\alpha$ are
\begin{equation}
    \overline{\xi_\alpha^{2k}} = \frac{1}{D},
\end{equation}
independent of $k$.

Using the vanishing of the odd moments and the independence of the $\xi_\alpha$, the second moment of $\xi$ is
\begin{equation}
    \overline{\xi^2} = \frac{1}{K^2} \sum_\alpha \overline{\xi_\alpha^2} = \frac{1}{K^2} \sum_\alpha \frac{1}{D} = \frac{1}{KD}.
\end{equation}
Similarly, the fourth moment of $\xi$ is
\begin{equation}
    \overline{\xi^4} = \frac{3}{K^4} \sum_{\alpha \neq \beta} \overline{\xi_\alpha^2} \, \overline{\xi_\beta^2} + \frac{1}{K^4} \sum_\alpha \overline{\xi_\alpha^4}.
\end{equation}
At large $K$ this is 
\begin{equation}
    \overline{\xi^4} \approx \frac{3}{K^2 D^2} + \frac{1}{K^3 D}.
\end{equation}
Provided $K \ll D$, the second term dominates.

\subsubsection*{1-Norm} The ensemble average of $|\xi|$ can be obtained as follows. With $K \ll D$, the most likely outcome is that $\xi = 0$. The two largest probabilities are
\begin{align}
\begin{split}
    P(\xi =0) &= (1-1/D)^K + \cdots \\
    P(|\xi|=1/K) &= K (1-1/D)^{K-1}/D + \cdots.
\end{split}
\end{align}
To leading order in $K/D$, we therefore have
\begin{equation}
    \overline{|\xi|} = \frac{1}{D}.
\end{equation}
As usual, the identity is special, since $\xi(0)=1$. 

Overall, the linear norm is thus
\begin{equation}
    \overline{F_1} = \sum_a \overline{|\xi(a)|} = 1 + \frac{D^2-1}{D} \approx D.
\end{equation}

\subsubsection*{Stabilizer Renyi Entropy} The probabilities entering the SRE are 
\begin{equation}
    \prob_\rho(a) = \frac{\xi(\mu(a))^2}{D \tr(\rho^2)}.
\end{equation}
We proceed by averaging numerator and denominator separately. The quantity of interest for $M_2$ is $
    \sum_a \overline{ \prob_\rho(a)^2}$, which boils down the fourth moment computed above. We find
\begin{equation}
    \sum_a \overline{ \prob_\rho(a)^2} = \sum_a  \frac{\frac{1}{K^3 D}}{\frac{D^2}{K^2}} = \frac{1}{K D}.
\end{equation}
However, the identity operator is special. It's contribution is 
\begin{equation}
    \prob_\rho(0)^2 = \frac{K^2}{D^2}.
\end{equation}
Hence, the full sum is
\begin{equation}
    \sum_a \overline{\prob_\rho(a)^2} = \frac{K^2}{D^2} + \frac{1}{KD}.
\end{equation}
For $K \ll D^{1/3}$, this sum is controlled by the second term, but for $D^{1/3} \ll K \ll D$, the sum is controlled by the first, identity, term. 

\subsection{Mixture of Random Haar States}

We now consider the analogous calculations for a mixture of $K$ Haar random states and continue to work in the regime $K \ll D$. In this regime, the purity is identical to the mixture of random stabilizer states, so we have $K = e^{N s_2}$. However, the Majorana string expectation values are now Gaussian with $\overline{\xi_\alpha^2} = \frac{1}{D}$.

\subsubsection*{Moments of Quantum Expectation Values} The second moment of $\xi$ is
\begin{equation}
    \overline{\xi^2} = \frac{1}{K^2} \sum_\alpha \overline{\xi_\alpha^2} = \frac{1}{KD},
\end{equation}
just as in the stabilizer case. The fourth moment is
\begin{equation}
    \overline{\xi^4} = \frac{3}{K^2 D^2}
\end{equation}
because $\xi$ is Gaussian. This is very different from the stabilizer case. In particular, the kurtosis,
\begin{equation}
    \kappa = \frac{\overline{\xi^4}}{\overline{\xi^2}^2}, 
\end{equation}
is very different in the two ensembles.

\subsubsection*{1-Norm} The Gaussianity of $\xi$ gives
\begin{equation}
    \overline{|\xi|} = \frac{1}{\sqrt{KD}}.
\end{equation}
Hence the linear norm is 
\begin{equation}
    \overline{F_1} = 1 + \frac{D^2-1}{\sqrt{KD}} \approx \sqrt{\frac{D^3}{K}}.
\end{equation}
The robustness of magic therefore obeys
\begin{equation}
   \overline{\mathcal{R}} \geq \frac{\overline{F_1}}{D} = \sqrt{\frac{D}{K}} \gg 1,
\end{equation}
so the state is always far outside the stabilizer polytope.

\subsubsection*{Stabilizer Renyi Entropy} Once again, the probabilities entering the SRE are 
\begin{equation}
    \prob_\rho(a) = \frac{\xi(\mu(a))^2}{D \tr(\rho^2)},
\end{equation}
and we continue to average numerator and denominator separately. $M_2$ is determined by $
    \sum_a \overline{ \prob_\rho(a)^2}$, which boils down the fourth moment computed above. We find
\begin{equation}
    \sum_a \overline{ \prob_\rho(a)^2} = \sum_a  \frac{\frac{3}{K^2 D^2}}{\frac{D^2}{K^2}} = \frac{3}{ D^2}.
\end{equation}
However, the identity operator is again special. Its contribution is 
\begin{equation}
    \prob_\rho(0)^2 = \frac{K^2}{D^2}.
\end{equation}
Hence, the full sum is
\begin{equation}
    \sum_a \overline{\prob_\rho(a)^2} = \frac{K^2}{D^2} + \frac{3}{D^2},
\end{equation}
which is always dominated by the identity contribution for $K \gg 1$.